\documentclass[aps,pre,reprint,onecolumn,superscriptaddress, longbibliography, 12pt]{revtex4-1}

\usepackage{amssymb,amsmath}
\usepackage[table]{xcolor}
\usepackage{graphicx}
\usepackage{mathtools}
\usepackage[export]{adjustbox}
\usepackage{overpic}
\usepackage[colorlinks=true,
 linkcolor=black,
 urlcolor=blue,
 citecolor=blue]{hyperref}
\usepackage{nicefrac}
\usepackage{stackrel}
\usepackage{multirow}
\usepackage{lipsum} 
\usepackage[normalem]{ulem}
\usepackage{pbox}
\newcolumntype{C}[1]{>{\centering\let\newline\\\arraybackslash\hspace{0pt}}m{#1}}
\usepackage{float}
\usepackage{verbatim}

\usepackage{enumitem}
\usepackage{bbold}
\usepackage{lineno}
\usepackage{framed,color}
\definecolor{shadecolor}{rgb}{0.85,0.80,0.80}
\definecolor{myorange}{RGB}{253, 184, 99}
\definecolor{mypurple}{RGB}{178, 171, 210}
 \usepackage{multirow}

\usepackage{float}
  
\allowdisplaybreaks

\begin{document}
\title{Dispersal-induced instability in complex ecosystems}
\author{Joseph W. Baron*}
\affiliation{Department of Physics and Astronomy, School of Natural Sciences, The University of Manchester, Manchester M13 9PL, United Kingdom}
\affiliation{Instituto de F{\' i}sica Interdisciplinar y Sistemas Complejos IFISC (CSIC-UIB), 07122 Palma de Mallorca, Spain \\ \textsuperscript{*}Corresponding author: J.W.B (email: josephbaron@ifisc.uib-csic.es)}

\author{Tobias Galla}
\affiliation{Department of Physics and Astronomy, School of Natural Sciences, The University of Manchester, Manchester M13 9PL, United Kingdom}
\affiliation{Instituto de F{\' i}sica Interdisciplinar y Sistemas Complejos IFISC (CSIC-UIB), 07122 Palma de Mallorca, Spain \\ \textsuperscript{*}Corresponding author: J.W.B (email: josephbaron@ifisc.uib-csic.es)}

\maketitle

\section*{Abstract}

In his seminal work in the 1970s, Robert May suggested that there is an upper limit to the number of species that can be sustained in stable equilibrium by an ecosystem. This deduction was at odds with both intuition and the observed complexity of many natural ecosystems. The so-called stability-diversity debate ensued, and the discussion about the factors contributing to ecosystem stability or instability continues to this day. We show in this work that dispersal can be a destabilising influence. To do this, we combine ideas from Alan Turing's work on pattern formation with May's random-matrix approach. We demonstrate how a stable equilibrium in a complex ecosystem with trophic structure can become unstable with the introduction of dispersal in space, and we discuss the factors which contribute to this effect. Our work highlights that adding more details to the model of May can give rise to more ways for an ecosystem to become unstable. Making May's simple model more realistic is therefore unlikely to entirely remove the upper bound on complexity.

\section*{Introduction}

\noindent Providing a firm counterpoint to the view that greater ecosystem complexity promoted stability \cite{odum, elton, macarthur, paine, landi, mccann}, Robert May used a simple statistical model \cite{may, may71} to argue that increasing the number of species in an ecosystem could in fact reduce stability. By analysing the eigenvalues of a randomly constructed community matrix, May deduced the following criterion for stability \cite{may} 
\begin{linenomath}
\begin{align}\label{eq:may}
\sigma^2 N C < 1.
\end{align}
\end{linenomath}
In this criterion $\sigma^2$ is the variance in the inter-species interactions, $N$ is the number of species and $C$ is the connectance (the probability that any given pair of species interact with one another). May's result shows that stability is decided by the product $c=\sigma^2 N C$. We will call this quantity the `complexity' of the ecosystem.

May's model suggests that more complex ecosystems tend towards instability. For a fixed variance of interactions, there is an upper bound to the number of species and food web connections that the ecosystem can sustain. This idea quickly became controversial and May's work sparked the so-called complexity-stability (or diversity-stability) debate, which continues to this day \cite{landi, mccann, namba}. 

Since the 1970s, the discussion around stability has been made more precise and subtle. It is now understood that there are a number of senses in which an ecosystem can be unstable and, indeed, there are a number of ways one can define diversity \cite{landi, justus}. An ecosystem can be unstable with respect to, for example, the introduction of new species, the extinction of existing species, or environmental changes. May's work specifically addresses stability with respect to fluctuations in species abundance.

In order to understand the influence of the many aspects of real ecosystems on stability, May's fairly austere model has since been augmented and improved upon. Features not captured by May's initial model include food-web structure (e.g., trophic levels, modularity, and nestedness) \cite{grilli, may71, allesinatang1, allesinatang2, hutchinson2019, pilosof2017}, the feasibility of the equilibrium \cite{stone, gibbs}, nonlinearities and alternative interpretations of `interaction strength' \cite{fyodorov2016,gross2009,berlow2004}, and variability of the environment and of species' susceptibility to environmental change \cite{chesson, mcnaughton1977, thebault, tilman1999}. In many models of complex ecosystems however, only the total abundance of each species is considered without appreciating how the members of that species are distributed in space \cite{may, grilli, may71, allesinatang1, allesinatang2,stone, gibbs, berlow2004, gross2009, chesson, mcnaughton1977, thebault, tilman1999}. In such models there is no notion of space and hence no dispersal. In this work we explicitly include the effects of diffusive dispersal in space and study the effects on stability. 

Dispersal may intuitively be expected to be a homogenising and stabilising influence. As demonstrated recently in Ref.~\cite{gravel}, it can indeed stabilise equilibria in spatially heterogeneous ecosystems. Perhaps counter-intuitively, the insight of Turing's seminal work \cite{turing} was that dispersal can also destabilise a dynamical system. Such an instability has been studied in meta-population \cite{hanski, hanski1999} predator-prey models with small numbers of species \cite{hassell1991, hassell1994, levinsegel} and numerically for food webs on networks \cite{brechtel2018}. We combine Turing's idea with May's random-matrix approach to show that a similar destabilising effect can be seen in models of complex ecosystems.

In order not to obscure the key effects at work, we opt to modify May's paradigmatic model sparingly. This allows us to highlight the consequences of the inclusion of dispersal. We suppose that the abundances of the species rest in a steady, homogeneous equilibrium. In order to study stability, we examine the Jacobian matrix governing perturbations in species abundance about this equilibrium. Like May, we ask what statistical properties are required of the Jacobian matrix in order for the ecosystem to return to equilibrium when perturbed. 

Unlike May however, we allow for trophic structure in our model. It is the combination of dispersal and trophic structure which gives rise to the Turing-type instability. For the sake of mathematical simplicity, we confine our model to only two broad trophic classifications: predator and prey species. The two groups of species are distinguished by statistical differences in their interactions. Our approach can be generalised to more complicated food web structures.  

We now postulate the form of the Jacobian matrix central to our problem, $\mathbf{M}_q$. The elements of this matrix describe how spatial disturbances of wavelength $\lambda = 2\pi/q$ in the abundance of one species affect the abundances of the other species; $q$ is known as the wavenumber \cite{crosshohenberg}.

\begin{figure}[t]
	\centering
	\includegraphics[scale = 0.9]{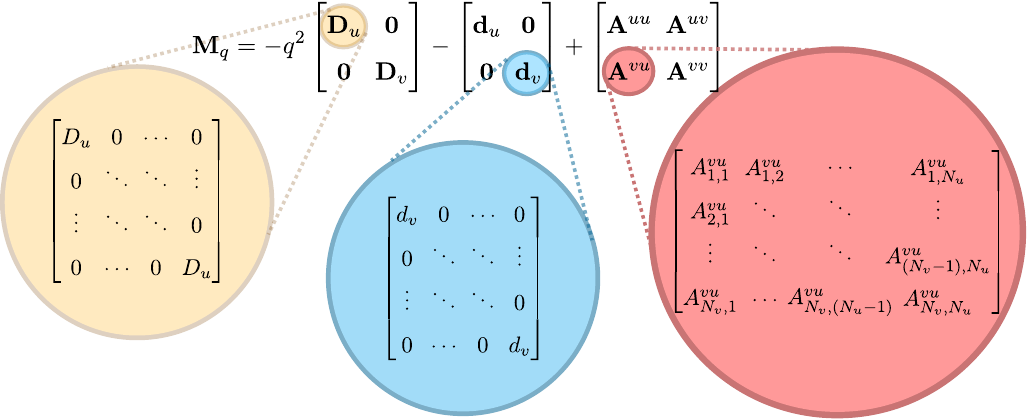}
	\caption{ The structure of the stability matrix $\mathbf{M}_q$. The matrix is composed of three parts: a diffusion matrix $\mathbf{D}$, a self-interaction matrix $\mathbf{d}$, and an interaction matrix $\mathbf{A}$ with entries drawn at random from a probability distribution. Each matrix is split into blocks due to the trophic structure of the community -- we use the subscript $u$ to denote species which have mostly prey-like interactions and $v$ for species with mostly predator-like interactions. The approach can be generalised for more complicated block structures. The stability of the non-spatial ecosystem is described by the matrix for $q=0$, or equivalently by setting $D_u=D_v=0$ (see Methods, and Section S1 in the Supplementary Material).}
	\label{fig:mymatrix}
\end{figure}

Because of the trophic structure of the community, the matrix $\mathbf{M}_q$ has a block structure where each block has different statistics. Similarly structured random matrices have been used in previous literature \cite{grilli2017, allesina2015}. The matrix $\mathbf{M}_q$ is comprised of three terms: a diffusion term, an intra-species interaction term and an inter-species interaction term.  We write
\begin{linenomath}
\begin{align}\label{eq:mq}
\mathbf{M}_q &= -q^2\mathbf{D} - \mathbf{d} + \mathbf{A}.
\end{align}
\end{linenomath}
The diffusion coefficients for prey and predator species are $D_u$ and $D_v$ respectively (Fig.~\ref{fig:mymatrix}). The interaction matrix $\mathbf{A}$ is modelled as having elements drawn from a correlated Gaussian ensemble, although other distributions may be used to obtain the same results (see Section S7 in the Supplementary Material). Further details on the structure of $\mathbf{M}_q$ and how one arrives at this form are given in Fig. \ref{fig:mymatrix} and in the Methods section.

The problem of analysing stability reduces to finding the eigenvalue spectrum of the matrix $\mathbf{M}_q$. If the eigenvalues of this matrix all have negative real parts, then the equilibrium is stable with respect to disturbances of wavenumber $q$. Else, it is unstable. In order for the equilibrium to be stable on the whole, all eigenvalues of $\mathbf{M}_q$ must have negative real parts for all values of $q$.

\begin{figure}[t]
	\centering 
    \includegraphics[scale = 0.15]{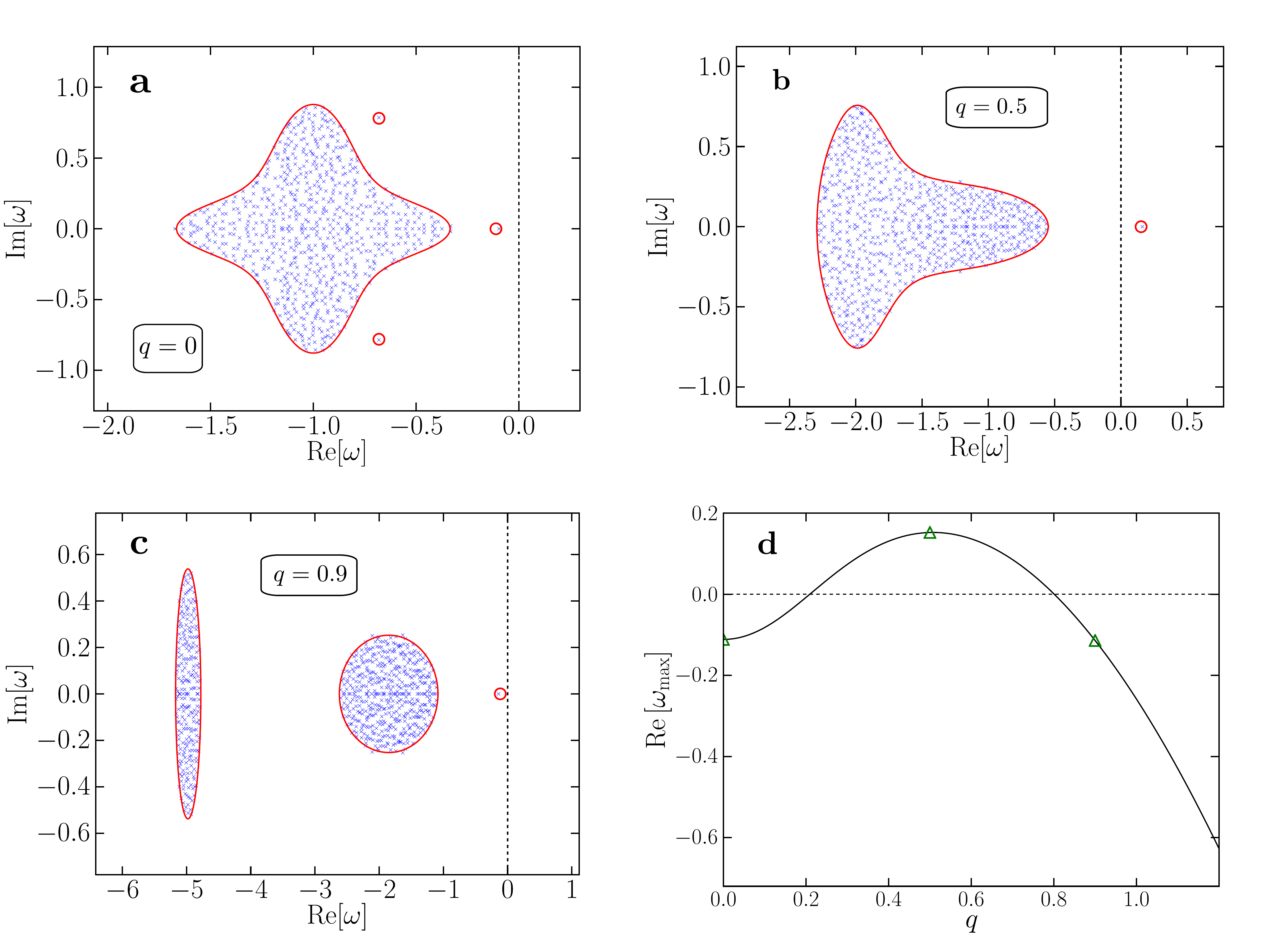}

	\caption{Eigenvalue spectra of the stability matrices in Fig. \ref{fig:mymatrix}. In May's original model, the eigenvalues all lay uniformly within a circle in the complex plane. In our model, the circle is warped and split into more complicated shapes. Also, a small number of outlier eigenvalues now stray from the bulk. Eigenvalues of computer-generated random matrices $\mathbf{M}_q$ are shown as blue crosses. They are compared to theoretical predictions for the boundary of the bulk region (red solid line) and theoretical predictions for the outliers (open red circles). For $q=0$, all of the eigenvalues have negative real part (panel {\bf a})-- the equilibrium is stable in the non-spatial ecosystem. For disturbances with larger wavenumber $q$, one of the eigenvalues crosses the real axis (panel {\bf{b}}). The equilibrium is unstable against such perturbations. If the wavenumber $q$ is larger still, the rightmost eigenvalue returns to the negative half-plane (panel {\bf{c}}). This is characteristic of a Turing instability \cite{crosshohenberg} -- the equilibrium is unstable with respect to disturbances of a finite range of wavelengths. This is shown in panel {\bf{d}}, where we plot the real part of the rightmost eigenvalue as a function of $q$. The equilibrium is unstable against perturbations of wavenumber $q$ whenever $\mbox{Re}[\omega_{\rm max}]>0$. The green triangles mark the wavenumbers from panels {\bf{a}}--{\bf{c}}.  }
	\label{fig:turinginstability}
\end{figure}

If the number of species in the ecosystem is large, then the eigenvalue spectrum is dependent only on the statistics of $\mathbf{M}_q$ and not on its specific entries. Using random-matrix theory and ideas from statistical physics, we are able to deduce a mathematical expression for the support of the eigenvalue spectrum of $\mathbf{M}_q$. That is, we can find the region in the complex plane in which the eigenvalues sit and, most importantly, whether or not they have positive real parts. Examples are shown in Fig. \ref{fig:turinginstability}. 
 \newcommand\figsize{0.16}
 
With this analytical approach we can calculate what properties of $\mathbf{M}_q$ make the equilibrium unstable. Thus, we can deduce how May's upper bound on ecosystem complexity is modified by the inclusion of dispersal and trophic structure. Most crucially, we show that equilibria which would be stable without spatial effects can be destabilised by dispersal. We find that this dispersal-induced instability is possible not only in a linear model but also in a non-linear system where the equilibrium is arrived at dynamically and hence is feasible by construction.

\bigskip

\section*{Results}

\noindent {\bf Eigenvalue spectra.} We show some example eigenvalue spectra of the matrix $\mathbf{M}_q$ in Fig. \ref{fig:turinginstability}. The vast majority of the eigenvalues group into a `bulk' region, with the exception of a few outliers. These outliers cannot be ignored -- the excursion of even one eigenvalue across the imaginary axis to the positive real side makes the equilibrium unstable. 

Using the statistical properties of $\mathbf{M}_q$ we are able to calculate mathematically the bulk regions to which most of the eigenvalues are confined and the locations of any outliers.  In Figs. \ref{fig:turinginstability}a --\ref{fig:turinginstability}c we show that these calculations agree very well with the spectra of computer-generated random matrices. We can therefore predict what community properties lead to stability or instability.

As Fig. \ref{fig:turinginstability} demonstrates, it is possible to find circumstances under which the model community is destabilised by the inclusion of dispersal. Figure \ref{fig:turinginstability}a shows the eigenvalue spectrum for the model without spatial effects ($q=0$, or equivalently $D_u = D_v = 0$, see Methods). All eigenvalues in Fig. \ref{fig:turinginstability}a have negative real parts so we conclude that the equilibrium is stable for the model ecosystem without dispersal. In Fig. \ref{fig:turinginstability}b we take into account dispersal and show the eigenvalue spectrum for a non-zero wavenumber. All other parameters are the same as in Fig. \ref{fig:turinginstability}a. Now, an outlier eigenvalue strays over the imaginary axis, demonstrating that the equilibrium is unstable. For perturbations with higher wavenumbers the outlier returns to the negative half-plane (Fig. \ref{fig:turinginstability}c) -- the equilibrium is stable with respect to perturbations of higher wavenumber.  Figure \ref{fig:turinginstability}d shows the real part of the rightmost eigenvalue ($\mbox{Re}[\omega_{\rm max}]$) as a function of the wavenumber $q$, highlighting the set of wavenumbers against which the equilibrium is unstable.

To understand the role of trophic structure in dispersal-induced instability, let us consider momentarily a model without statistical distinction between predator and prey species (like May's original model \cite{may}), and with a common diffusion coefficient $D$ for all species. The rightmost eigenvalue would have the following dependence on wavenumber: $\mbox{Re}[\omega_{\rm max}] = \omega_0 - D q^2$, where $\omega_0$ is rightmost eigenvalue of the community stability matrix without dispersal ($q = 0$). In this simple case, $\mbox{Re}[\omega_{\rm max}]$ is a purely decreasing function of $q$ -- if $\omega_0<0$ then $\mbox{Re}[\omega_{\rm max}]<0$ for all values of $q$. There can be no dispersal-induced instability here. The inclusion of trophic structure in our model leads to a maximum in $\mbox{Re}[\omega_{\rm max}]$ at a non-zero value of $q$ (Fig. \ref{fig:turinginstability}d) and, consequently, a finite band of non-zero wavenumbers against which the equilibrium is unstable. This is the essence of why dispersal in combination with tropic structure can promote instability.

In Turing's original work on chemical reaction systems \cite{turing}, instability with respect to perturbations of a finite band of wavenumbers [as in Fig. \ref{fig:turinginstability}d] signalled the formation of stable periodic patterns. The exact shape of these patterns is usually determined by non-linearities in the differential equations describing the reactions. Our model is valid only in the vicinity of the supposed equilibrium and, similar to May \cite{may}, we have not specified the nature of any non-linearities. We therefore do not speculate for now about what might happen after the system has departed from the fixed point. We merely point out here the dispersal-induced instability of the equilibrium about which we have linearised.

\vspace{4em}

\begin{figure}[t]
	\centering
	\includegraphics[scale = 0.45]{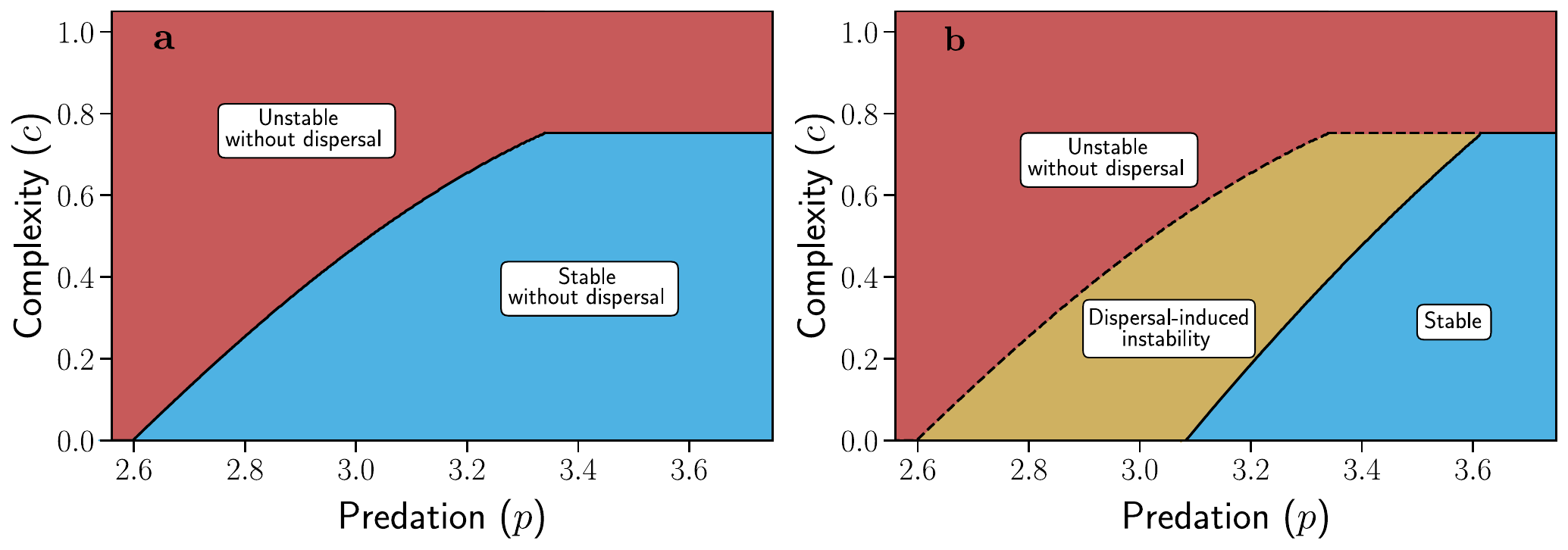}
	\caption{Dispersal as a destabilising influence. Panel {\bf a} shows the stability diagram for the non-spatial ecosystem model with trophic structure. The solid line is the upper bound on complexity $c$ for the equilibrium to be stable. Increasing complexity leads to instability. There is also a lower bound on the amount of predation $p$ required for the community to be stable (if $p$ is too small, then the equilibrium is unstable even for small values of the complexity). Stability for the spatial ecosystem with dispersal is shown in panel {\bf b}. In the blue region the equilibrium is stable. In the yellow region the equilibrium would be stable in a non-spatial model but dispersal in space induces instability. Crossing from the blue into the yellow region in panel {\bf b} the ecosystem undergoes a Turing instability. In the red region in {\bf b} the equilibrium is unstable both in the non-spatial and in the spatial ecosystem. }
	\label{fig:phasediagram}
\end{figure}

\noindent {\bf Modifiying May's bound: stability with and without dispersal.} In order to further appreciate the effect that the inclusion of dispersal has on stability, we first consider the conditions under which the non-spatial ecosystem becomes unstable (see Methods). This enables us to study how May's bound on the complexity $c$ changes for our model, which has distinct predator and prey species. A stability plot is shown in Fig.~\ref{fig:phasediagram}a. The horizontal axis shows the average degree of predation $p = CN\mu_{vu}$ (see Methods). The vertical axis is the complexity parameter $c$. The solid line indicates the upper bound on the complexity: below the line the equilibrium is stable, above this line it is unstable. 

We see from Fig. \ref{fig:phasediagram} that greater predation $p$ increases the amount of complexity $c$ that can be sustained in stable equilibrium by the ecosystem. Notably, in order to have stability at all in the model with trophic structure there is a lower bound on the predation parameter $p$.

If we now include dispersal, the stability diagram changes [Fig.~\ref{fig:phasediagram}b]. In particular the upper bound on the complexity can become lower than in the non-spatial system. This is because a new type of instability is now possible -- the Turing instability. Thus, there are instances in which the model is stable without dispersal but unstable when dispersal is introduced (yellow area  in Fig.~\ref{fig:phasediagram}b labelled `dispersal-induced instability').  

There are no situations in which an unstable equilibrium is stabilised by the combination of dispersal and trophic structure alone. However, previous work \cite{gravel} has shown how the combination of spatial heterogeneity (in inter-species interactions) with dispersal can be stabilising. We comment on the effect of including spatial heterogeneity in our model in the Discussion and in the Supplementary Material (Section S14). 

\vspace{2em}
\noindent{\bf How does complexity affect the Turing instability?} So far we have concerned ourselves with the effects of dispersal on complex ecosystems. We now ask the reverse question: Spatial instability and pattern formation have been found in `simple' models of ecosystems with a small number of species \cite{neubert, patternformationreview, vegetationpatterns, levindispersion}. What are the effects of complexity on this Turing instability?

In general, Turing instabilities in simple systems typically occur when the diffusion coefficients of the `activator' and `inhibitor' components are quite disparate -- this is known as the `fine-tuning issue' with the Turing mechanism \cite{murraybook, barongallastochastic}. The activating components in our system are the prey species, while predators play the role of the inhibitors. We would like to know what the effects of complexity are on the threshold ratio $D_v/D_u$ at which the Turing instability occurs. To answer this question, we compute this threshold for different values of the complexity parameter $c$.

The case $c=0$ can be achieved by setting the width $\sigma$ of the distribution for the elements of the matrix $\mathbf{A}$ to zero. The matrix elements within a block are then identical to each other. The model thus reduces to a simple two-species predator-prey system. Increasing $c$ from zero introduces heterogeneity between the species within the blocks.

\begin{figure}[t]
	\centering
	\includegraphics[scale = 0.5]{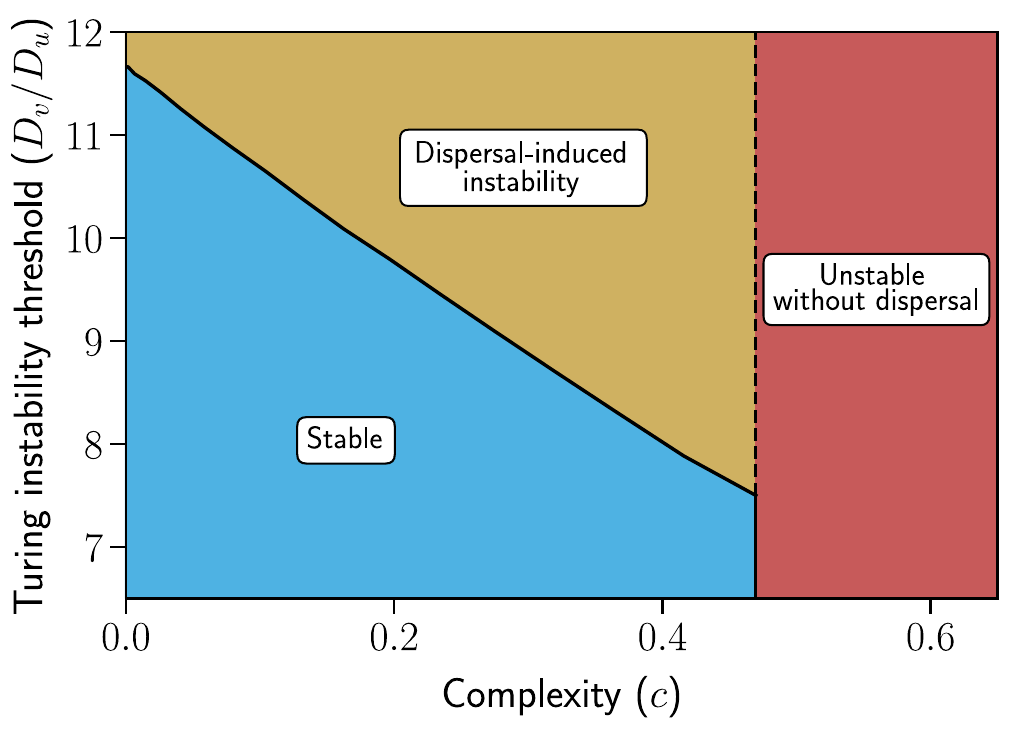}
	\caption{ Complexity reduces the ratio of diffusion coefficients required for Turing instability. In the yellow region the community is unstable due to Turing instability. Conversely, a high value for the ratio of diffusion coefficients $D_v/D_u$ reduces the complexity that can be sustained in stable equilibrium. For sufficiently high complexity the non-spatial community becomes unstable (red area to the right of the vertical line). The equilibrium of the spatial ecosystem is then also unstable, irrespective of the diffusion coefficients.}
	\label{fig:turingvssigma}
\end{figure}

We find that complexity can lower the ratio of diffusion coefficients required for Turing instability (Fig.~\ref{fig:turingvssigma}). For more complex ecosystems the Turing instability therefore sets in more easily. So, not only can complexity decrease the stability of non-spatial model ecosystems, it can also reduce stability in spatial models.  Conversely, increasing the ratio of diffusion coefficients $D_v/D_u$ reduces the complexity that can be sustained in stable equilibrium. As can be seen in Fig.~\ref{fig:turingvssigma}, the upper bound on $c$ is lower when disparity of dispersal rates is large.

\vspace{3em}

\ 
\noindent {\bf Spatial instability in a non-linear model with complexity.} The linear analysis we have focused on so far, although informative, has drawbacks. It only deals with the dynamics in the vicinity of a homogeneous equilibrium and it tells us nothing about how the ecosystem behaves in the long-run if the equilibrium is unstable. Further, one could object that the linear model is somewhat contrived and that it does not capture how a `real' ecosystem constructs itself in equilibrium. 

\begin{figure}[t]
	\centering
	\includegraphics[scale = 0.22]{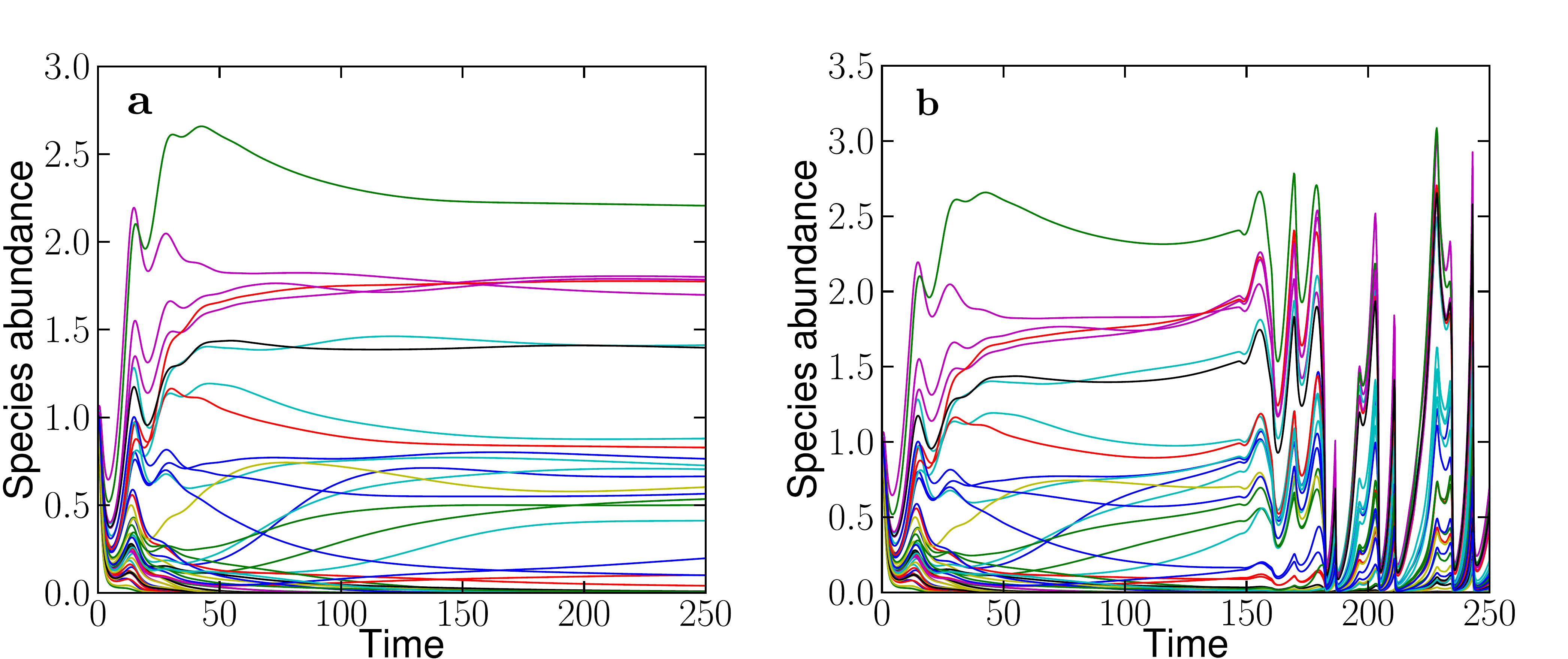}
	\caption{Dispersal can induce instability of feasible equilibria. The prey species abundances in the non-spatial Levin-Segel ecosystem in panel (a) converge to a stable feasible equilibrium. In panel (b) we simulate the ecosystem with the same paramaters as in (a), but allowing for dispersal in space. Volatile behaviour is found in the spatial ecosystem, even though the non-spatial system would approach a stable equilibrium. }
	\label{fig:stabilitypreynonlinear}
\end{figure}

We now present simulation data of a complex ecosystem obeying non-linear Levin-Segel-type dynamics \cite{levinsegel} (see Methods). In  Fig. \ref{fig:stabilitypreynonlinear} we demonstrate that a dispersal-induced instability can occur in this model as well. Figure \ref{fig:stabilitypreynonlinear}a shows a realisation of the dynamics without dispersal. The model ecosystem converges to an equilibrum composed only of surviving species. By definition this is a feasible equilibrium \cite{grilli2017}. 

Figure~\ref{fig:stabilitypreynonlinear}b shows the same model (with the same interaction matrix $\mathbf{A}$) but now with dispersal. The abundances do not settle in the long run. Instead, they display quite erratic behaviour. We stress that this is different to the typical behaviour seen in two-species systems with a Turing instability. In such simple systems, the abundance of each species converges to a constant value eventually. The fixed-point value varies across space, generating a periodic pattern. The complex nature of the interactions in our model leads to the more complicated dynamics in Fig. \ref{fig:stabilitypreynonlinear}b known as diffusion-induced chaos \cite{kuramoto1978, pascual1993}. 
 
\begin{figure}[t]
	\centering
	\includegraphics[scale = 0.27]{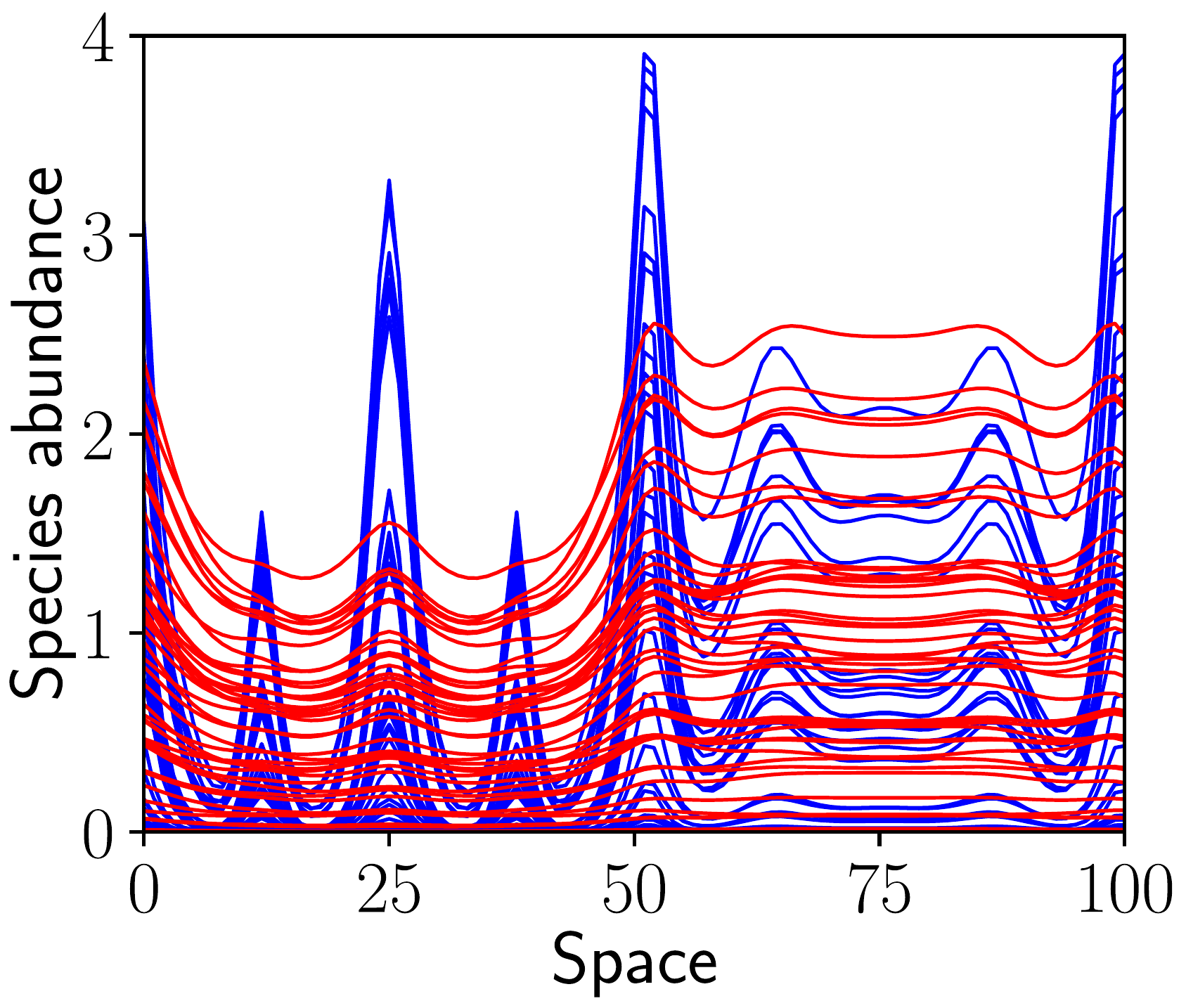}
	\caption{Distribution of species in space as a result of dispersal-induced instability. The figure shows a snapshot of all $N=100$ species abundances in a complex Levin-Segel ecosystem. The model parameters are the same as in Fig.~\ref{fig:stabilitypreynonlinear}b, which are given in full in the Supplementary Material. Predator species abundances are shown as red lines, prey species as blue lines. The most abundant species varies from place to place. This structure is not a static pattern as in conventional reaction-diffusion systems with Turing instability. Instead the abundances change with time. }
	\label{fig:patterns}
\end{figure}

With that being said, at any point in time one can take a snapshot of the spatial profile of the species abundances. An example is shown in Fig.~\ref{fig:patterns}. One finds some spatial structure for the abundances of prey species (green lines) but this differs from classic Turing patterns, which are typically more periodic and regular. We note that the quickly diffusing predator species (red lines) have more smoothly undulating spatial profiles than the slowly diffusing prey species in this example.
 
\vspace{4em}

\section*{Discussion}
\noindent The random-matrix approach to modelling ecosystems has developed substantially since May's original work. We contribute to this development with a model of a complex ecosystem which includes trophic structure and dispersal in space. These additional features change May's bound on complexity. How the bound changes tells us about the influence of the new model components on stability. For example, we find that predation acts as a stabilising influence (Fig.~\ref{fig:phasediagram}, and Sections S5 and S14 of the Supplementary Material). 

Inspired by Turing's mechanism for pattern formation \cite{turing}, we show how dispersal can be a destabilising factor in a complex ecosystem. An equilibrium which would be stable in the non-spatial model can become unstable through dispersal. Such an instability has been suggested as a mechanism for the observed patterns in natural ecosystems with small numbers of species involved \cite{Rietkerk2005, Rietkerk2008, vandeKoppel2008, meron2012, Liu2014, Karig2018}. Our work extends the consideration to larger number of species and suggests the possibility of dispersal-induced instability in more complex communities.

Conversely, we observe that increased complexity can lower the threshold for the diffusion coefficients required for a Turing instability. So, not only does complexity reduce stability in a non-spatial model as was May's conclusion \cite{may}, it also destabilises spatial models with dispersal. This is interesting especially in light of the so-called `fine-tuning' issue with the Turing mechanism. The ratio of diffusion coefficients required for a Turing instability is usually large, making it hard to find experimental examples of Turing pattern formation \cite{barongallastochastic, crosshohenberg,lengyel1992, castets1990}. Based on our findings, we speculate that dispersal-induced instability may be easier to observe in complex dynamical systems.  

To demonstrate that dispersal-induced instability can also be seen in model ecosystems where the equilibrium is arrived at more naturally, we performed simulations of a complex ecosystem with Levin-Segel dynamics (Fig.~\ref{fig:stabilitypreynonlinear}). We found that this model also exhibits a dispersal-induced instability. Because of the complex nature of the interactions, the system does not reach a stable patterned state, which is normally characteristic of Turing instabilities in models with fewer species. Instead, one sees persistently volatile dynamics known as diffusion-induced chaos \cite{kuramoto1978, pascual1993}.

So as to highlight the destabilising effect of the combination of dispersal and trophic structure as clearly as possible, we were parsimonious with the inclusion of additional detail in the model. It is possible to relax some of the restrictions that we imposed without sacrificing the ability to perform the mathematical analysis. For example, variation in the autoregulation coefficients $d_\alpha$ and the dispersal coefficients $D_\alpha$ between species can be taken into account (in a similar way to Ref. \cite{barabas2017}). In Section S9 of the Supplementary Material we show that the possibility of dispersal-induced instability remains despite this additional complication.

Provided that relatively mild conditions are met, we also demonstrate that our results still apply when the random matrix elements are drawn from a non-Gaussian distribution (see Section S7 in the Supplementary Material). More precisely, we show that the eigenvalue support only depends on the first and second moments of the distribution of the random matrix elements $A_{ij}^{\alpha\beta}$. This feature of the random matrix model is known as universality \cite{taovu2010, taovukrishnapur2010}. This is interesting from a theoretical point of view and also allows one to constrain the signs of the interaction coefficients within the different blocks in the interaction matrix (see Section S8 in the Supplementary Material). This means that it is possible to enforce strict loss-gain interactions between prey and predator species and to eliminate effects such as intraguild predation. Again, the possibility for dispersal-induced instability remains.

In spatial models of ecosystems, dispersal is often implemented as migration between discrete patches rather than by diffusion in a continuous domain \cite{hanski, hanski1999}. The set of interspecies interactions in the different patches can then be conceptualised as the edges of a multi-layer network \cite{hutchinson2019, pilosof2017}. Our results, which were derived using diffusive dispersal in continuous space, also continue to hold in such meta-ecosystems (see Sections S1 C and S13 of the Supplementary Material).

A patched landscape is also a particularly convenient way to include spatial heterogeneity in the model. A recent study by Gravel et al. \cite{gravel} demonstrated that dispersal can stabilise equilibria in complex ecosystems with spatial heterogeneity (we replicate these findings with our method in Section S11 of the Supplementary Material). This study did not include trophic structure, which is a key aspect of our model. If spatial heterogeneity is combined with trophic structure, both the stabilising mechanism reported in \cite{gravel} and the dispersal-induced instability at the focus of our work can be seen in the same model (see Section and S14 in the Supplementary Material and in particular Fig. S13).
	
The stabilising effect in Ref. \cite{gravel} and the destabilising mechanism we present can be viewed as somewhat separate. Dispersal-induced instability is associated with the outliers in the eigenvalue spectrum. The basis for the stabilising mechanism in Ref. \cite{gravel} is a reduction in the bulk eigenvalue spectrum (see Supplementary Material Sections S11 and S12). Which one of the effects takes precedence depends on the circumstances. A community with trophic structure and a significant predator-to-prey ratio of dispersal rates will be more likely to exhibit dispersal-induced instability. A community with significant spatial variation in interactions and consistently high dispersal rates across all species will be more likely to be subject to the stabilising effects of dispersal (for further discussion see Section S14 of the Supplement and Fig. S14 in particular). 

One criticism levelled at May's model is that it is too simple and that perhaps through the inclusion of further aspects of natural ecosystems, the upper bound on the complexity could be eliminated. Our results do not support this hypothesis. Like May, we also find that there is always an upper limit on the complexity $c =\sigma^2 NC$ that an ecosystem can stably sustain, even when stabilising factors such as predation and spatial heterogeneity are taken into account. Other recent studies using random-matrix approaches arrive at similar conclusions. For example, Allesina and Tang take into account more realistic food-web structures and still find upper limits on the number of interconnected species \cite{allesinatang1, allesinatang2}. May's result therefore generalises to models capturing more aspects of `real' communities in ecology. This prediction is supported by the observation of `diversity regulation' in some ecosystems \cite{osullivan2019, gotelli2017, magurran2018, brown2001, parody2001}. 

A final observation that we wish to convey is that making models for complex ecosystems more detailed introduces the opportunity for new types of instability. In May's original model, for example, the mean of the community matrix elements was zero. As a consequence, any one species is equally likely to benefit or suffer from the presence of another species. Mathematically, all eigenvalues then reside within one bulk region, and it is this bulk region that determines stability. Mutualism can be introduced through interaction coefficients which are positive on average, and competition through a negative average interaction \cite{allesinatang1}. This leads to additional outlier eigenvalues, which can make an equilibrium unstable even though it would otherwise be stable. Introducing trophic structure can generate complex-conjugate pairs of outliers (Fig. \ref{fig:turinginstability}a), allowing further opportunity for instability (Section S5 of the Supplementary Information). Dispersal, finally,  leads to the possibility of a Turing-type instability. Overall, adding more details to the model of May tends to give rise to more ways in which an equilibrium can become unstable.

\section*{Methods}

\noindent {\bf Linear model.}  We imagine that we find the ecosystem at an homogenenous equilibrium. Our model is concerned with the dynamics of small perturbations of the species abundances about this fixed point. The stability of the homogeneous fixed point is determined by whether or not these perturbations decay or increase with time. We write $u_i(x,t)$ and $v_j(x,t)$ for the perturbations of the prey and predator species abundances respectively at position $x$ and time $t$. These are the deviations away from the fixed point. There are $N_u$ prey species and $N_v$ predator species with $N = N_u + N_v$ species in total. We define the constants $\gamma_u = N_u/N$ and $\gamma_v = N_v/N$.~

We assume that prey species diffuse at rate $D_u$ and predator species at rate $D_v$ in a spatially homogeneous environment. Similar to May \cite{may}, we also imagine that all species in each group have equal and negative self-interaction. However, it is possible to relax the simplifying constraints of uniform self-interaction and diffusion coefficients (see Section 9 in the Supplemental Material and Ref. \cite{barabas2017}). The probability that a particular pair of species interact with one another is $C$. This parameter is known as the `connectance' \cite{may}. The effect of a change in the abundance of species $j$ on species $i$, where $j$ belongs to trophic block $\beta$ and $i$ belongs to $\alpha$, is $A^{\alpha \beta}_{ij}$ [$\alpha, \beta \in \{u,v\}$].  The linearised reaction-diffusion equations thus take the following form 

\begin{align}
\frac{\partial u_i}{\partial t} &= D_u \frac{\partial^2 u_i}{\partial x^2}   - d_u u_i + \sum_{k=1}^{N_u} A_{ik}^{uu} u_k + \sum_{j=1}^{N_v} A_{ij}^{uv} v_j ,\nonumber \\
\frac{\partial v_j}{\partial t} &= D_v \frac{\partial^2 v_j}{\partial x^2}   - d_v v_j + \sum_{i=1}^{N_u} A_{ji}^{vu} u_i + \sum_{k=1}^{N_v} A_{jk}^{vv} v_k. \label{linearised}
\end{align}

If species $i$ and $j$ are non-interacting, $A^{\alpha \beta}_{ij}=A^{\beta \alpha}_{ji}= 0$. If the two species interact, then $A^{\alpha\beta}_{ij}$ and $A^{\beta\alpha}_{ji}$ are drawn from a joint Gaussian distribution with means $\overline{ A^{\alpha \beta}_{ij}} = \mu_{\alpha \beta}$ and $\overline{A^{\beta \alpha}_{ji}} = \mu_{\beta\alpha}$. The elements in the random matrix have variance $\sigma^2$ 

\begin{align}\label{var}
\overline{ ( A^{\alpha \beta}_{ij}- \mu_{\alpha \beta})^2} &= \sigma^2,  
\end{align}
and they are correlated according to

\begin{align}
\overline{(A^{\alpha \beta}_{ij}- \mu_{\alpha \beta}) (A^{\alpha' \beta'}_{ji}- \mu_{\alpha' \beta'})}&= \Gamma_{\alpha \beta \alpha' \beta'} \sigma^2  \label{rmstats}.
\end{align}
All other correlations are set to zero. Each of the model parameters can be interpreted ecologically. We define and interpret the complexity $c = CN\sigma^2$ in the main text. The interaction mean $\mu_{uu}$ indicates the degree to which different prey species cooperate (if $\mu_{uu}>0$) or compete (if $\mu_{uu}<0$). The coefficient $\mu_{vv}$ has a similar interpretation for predators. The means of the off-diagonal blocks $\mu_{uv}<0$ and $\mu_{vu}>0$ indicate the degree to which (on average) prey species suffer and predator species gain from predator-prey interactions. The parameters $\Gamma_{\alpha\beta\alpha'\beta'}$ describe the correlations between interaction coefficients. The only non-zero entries are taken to be $\Gamma_u \equiv \Gamma_{uuuu}$, $\Gamma_v \equiv \Gamma_{vvvv}$ and $\Gamma_{uv} \equiv \Gamma_{uvvu}$. That is, only elements which are diagonally opposite one another in $\mathbf{A}$ are correlated. A positive value of $\Gamma_{\alpha\beta\alpha'\beta'}$ indicates that if one species benefits more than average from an interaction, the other species involved does so as well. The opposite is true if $\Gamma_{\alpha\beta\alpha'\beta'}$ is negative. 

Taking the Fourier transform with respect to the spatial coordinate $x$ of Eq.~(\ref{linearised}), we arrive at dynamical equations (see Supplementary Information Section S1) for disturbances of wavenumber $q$ in the abundances of the various species (the wavenumber is related by $q = 2\pi/\lambda$ to the wavelength $\lambda$). We denote the combined vector of the Fourier transforms of species abundances by $\tilde{\mathbf{X}}_q = (\cdots, \tilde u_i(q,t), \cdots, \tilde v_j(q,t), \cdots)$, and arrive at the more compact matrix equation

\begin{align}
\dot{\tilde{\mathbf{X}}}_q = \mathbf{M}_q \tilde{\mathbf{X}}_q .\label{ftdynamics}
\end{align}
The matrix entry $\left(\mathbf{M}_q\right)_{ij}^{\alpha\beta} $ tells us what the effect of a disturbance of wavelength $q$ in species $j$ (belonging to trophic block $\beta$) is on species $i$ (belonging to trophic block $\alpha$). 

The vector $\mathbf{X}_q$ has dimension $N = N_u + N_v$ and is arranged such that the first $N_u$ elements are the Fourier-transformed abundances $\tilde u_i(q,t)$ and the last $N_v$ elements are the $\tilde v_j(q,t)$. The matrix $\mathbf{M}_q$ is depicted in Fig.~\ref{fig:mymatrix}. It is divided into blocks due to the trophic structure of the community. Its three contributions are a diagonal diffusion matrix, a diagonal self-interaction matrix and a random interaction matrix, whose variance and correlations are given in Eqs.~(\ref{var}) and (\ref{rmstats}). The indices $\alpha$ and $\beta$ correspond to the different blocks and $i$ and $j$ correspond to the position within the block. If the matrix $\mathbf{M}_q$ has eigenvalues with positive real parts for any value of $q \geq0$, the disturbances $u_i$ and $v_i$ will grow with time, indicating an unstable equilibrium. 

By setting $q =0$ in Eq.~(\ref{ftdynamics}), one recovers Eqs.~(\ref{linearised}) with the diffusion term removed. Focusing on $q=0$ in our model (or equivalently setting $D_u=D_v=0$) thus allows one to study the stability of a non-spatial model ecosystem with trophic structure. Further details can be found in Sections S1 and S5 in the Supplementary Material.

We note that a similar form for the matrix $\mathbf{M}_q$ can be found for a model of a meta-ecosystem with dispersal between discrete patches (similar to Ref. \cite{hanski, hanski1999}) instead of in continuous space (see Sections S1 C and S13 of the Supplementary Material).

The values of the model parameters used in the figures are given in full in Section S6 of the Supplementary Material.
\vspace{2em}
 
\noindent {\bf Calculation of the boundary surrounding the bulk of the eigenvalues.} The vast majority of the eigenvalues of the random matrices $\mathbf{M}_q$ reside within one or two `bulk' regions of the complex plane.  To determine stability we need to know if there are bulk eigenvalues with positive real parts. Identifying the boundaries of the regions containing the eigenvalues is sufficient for this purpose. Our calculation uses methods originally developed in statistical physics and follows lines similar to those of \cite{haake,sommers}. This approach converts the problem into the evaluation of a `potential' related to the eigenvalue density. This potential, in turn, can be expressed as a high-dimensional integral, which is carried out using the saddle-point method. A brief summary of the context of this approach in the wider literature is given Section S2 A of the Supplementary Material. Full details of the calculation are given in Sections S2 B and S2 C.

We write $\omega=\omega_x+i\omega_y$ for the eigenvalues of $\mathbf{M}_q$. The general expression for the boundary surrounding the bulk of the eigenvalues ($\omega_y$ as a function of $\omega_x$) is given by the simultaneous solution of the following equations (see Supplementary Material Eq.~(S81))

\begin{align}
\sum_\alpha \gamma_\alpha \left\vert \chi_\alpha \right\vert^2 - \frac{1}{c} &= 0, \nonumber \\
  - (\omega_x + i\omega_y + d_\alpha  +q^2 D_\alpha )\chi_\alpha + c\sum_\beta \Gamma_{\alpha \beta} \gamma_\beta \chi_\alpha\chi_\beta +1 &=0. \label{chis}
\end{align}

We note the free index $\alpha$ in the second of these equations ($\alpha\in\{u,v\}$). This is therefore a system of three coupled equations. One first eliminates the auxiliary variables $\chi_\alpha$, and then expresses $\omega_y$ in terms of $\omega_x$. This results in the red curves in Fig. \ref{fig:turinginstability}. 
 
The solution simplifies in several special cases, which we exploit to provide explicit stability criteria analogous to May's bound (Section S5 of the Supplementary Material).

\vspace{2em}

\noindent {\bf Calculation of the outlier eigenvalues.} In addition to the bulk eigenvalues the stability matrix can have isolated outlier eigenvalues. If any of these outliers have positive real part, the equilibrium is unstable. Their position in the complex plane is calculated along the lines of Ref. \cite{orourke}. Details can be found in the Supplementary Material Section S3. The outlier eigenvalues are given by the complex values $\omega$ satisfying the following equation (see Supplementary Material Eq.~(S82))

\begin{align}
\left[ \gamma_u NC \mu_{uu} - \frac{1}{\chi_u(\omega)}\right] &\left[\gamma_v NC \mu_{vv} - \frac{1}{\chi_v(\omega)}\right] - (NC)^2\gamma_u\gamma_v\mu_{uv}\mu_{vu} = 0,\label{simult1}
\end{align}

The auxiliary quantities $\chi_\alpha(\omega)$ in this relation satisfy

\begin{align}
- 1 &=  - (\omega + d_u + q^2 D_u) \chi_u +  c \Gamma_{u}\gamma_u \chi_u^2  +  c \Gamma_{uv}  \gamma_v\chi_v \chi_u , \nonumber \\
- 1 &= -  (\omega + d_v+ q^2 D_v) \chi_v  + c \Gamma_{uv}\gamma_u\chi_u \chi_v + c \Gamma_{v} \gamma_v\chi_v^2 , \label{simult2}
\end{align}

subject to the condition

\begin{align}
\sum_{\alpha} \gamma_\alpha \vert\chi_\alpha(\omega)\vert^2 < \frac{1}{c}.\label{cond}
\end{align}

Eqs.~(\ref{simult1}) and (\ref{simult2}) need to be solved simultaneously, subject to Eq.~(\ref{cond}). If there are no solutions then there are no outliers. In special cases the above expressions can be simplified, and explicit stability criteria can be found (Section S5 in the Supplementary Material).

\vspace{2em}

\noindent {\bf Finding the threshold for instability.} Instability can occur in one of several different ways:  (1) The bulk region of eigenvalues for $q=0$ can cross into the positive half-plane; (2) One of the outlier eigenvalues for $q=0$ can cross the imaginary axis; (3) An outlier eigenvalue for $q\neq 0$ can stray into the positive half-plane. We have not observed any circumstances under which the bulk crosses the imaginary axis for non-zero $q$ where it does not for $q=0$. The eigenvalues must have negative real parts for all $q$ in order for the equilibrium to be stable in the spatial system. This includes $q=0$. Cases (1) and (2) therefore indicate instabilities occurring both in the non-spatial and the spatial ecosystem. In case (3) the spatial system is unstable, but the non-spatial system remains stable. 
In each of these cases, the threshold for instability is found by identifying sets of parameters for which either the boundary for the bulk eigenvalues touches the imaginary axis (case 1) or where the outlier eigenvalues touch the imaginary axis (cases 2 and 3). 

This can be done using the analytical results for the spectrum of eigenvalues (Section S5 of the Supplementary Material), leading to the results shown in Figs. \ref{fig:phasediagram} and \ref{fig:turingvssigma}.

\vspace{2em}

\noindent{\bf Simulating the non-linear model.} Results in Figs.~\ref{fig:stabilitypreynonlinear} and \ref{fig:patterns} are from a numerical integration of the Levin-Segel-type model \cite{levinsegel},

\begin{align}
\frac{\partial u_i}{\partial t} &= D_u \frac{\partial^2 u_i}{\partial x^2} + u_i\left[ a - u_i + \sum_{k \in u} A_{ik}^{uu} u_k + \sum_{j \in v} A_{ij}^{uv} v_j\right] ,\nonumber \\
\frac{\partial v_j}{\partial t} &= D_v \frac{\partial^2 v_j}{\partial x^2} +v_j\left[  - v_j + \sum_{i \in x} A_{ji}^{vu} u_i + \sum_{k \in v} A_{jk}^{vv} v_j\right] . \label{levinsegel}
\end{align}
where $a>0$ is a constant. To integrate these equations numerically, the diffusion terms are discretised. The integration is then carried out using the Runge-Kutta (RK4) method \cite{sulimayers}. 

\vspace{2em}

\noindent{\bf Further variations on the model.} The flexibility of our analytical approach allows us to include additional features to our fairly austere model. For example, in Sections S7 and S8 of the Supplementary Material, we demonstrate the universality of our theoretical results. That is, we show that the matrix elements need not be drawn from a Gaussian distribution for our results to apply. Similar to Ref. \cite{barabas2017}, variation in the self-interaction and diffusions coefficients ($d_\alpha$ and $D_\alpha$ respectively) can also be taken into account (Section S9 in the Supplementary Material). In Section S13 of the Supplementary Material, we show that the dispersal-induced instability persists on a landscape of discrete patches (as opposed to diffusion in a continuous space) and when dispersal is non-local. These extensions highlight the robustness of the dispersal-induced instability in complex ecosystems and the versatility of the analytical formalism.

\noindent{\bf Inclusion of spatial heterogeneity in the interaction coefficients.} In analysing Eqs.~(\ref{linearised}) and the stability matrix in Eq.~(\ref{eq:mq}) we have assumed that the interaction coefficients $A_{ij}^{\alpha\beta}$ are the same at every point in space. In order to model spatial heterogeneity, we extend the setup along the lines of Ref. \cite{gravel} and imagine that dispersal takes place on a set of discrete patches indexed by their position $x$, similar to meta-population models \cite{hanski1999} (see Sections S10 in the Supplementary Material). The interaction matrix $A_{ij;x x'}^{\alpha\beta}$ then has two layers of block structure: one indicating the trophic structure as before and the second representing location in space. Multilayer networks have previously been used to encapsulate similar structures in ecological communities \cite{hutchinson2019, pilosof2017}. 
	
Adapting the prior calculation, the regions in the complex plane containing the bulk and outlier eigenvalues can also be computed for the model with spatial heterogeneity (Section S11 in the Supplementary Material). The model in the main text and that of Ref. \cite{gravel} are special cases of this setup. In particular, we can also recover the eigenvalue support and stability criteria of Ref. \cite{gravel}. In Section 14 of the Supplementary Material we show that the mechanism leading to dispersal-induced instability and the stabilising mechanism of Ref. \cite{gravel} can coexist in the same model. We also discuss the factors that determine whether dispersal acts to stabilise or destabilise equilibria of complex ecosystems.

\section*{Data Availability Statement}
The data in Figs. 2-6 is generated using the codes in the code availability statement. The data is also available upon reasonable request to the corresponding author. 
 
 \section*{Code Availability Statement}
Codes for Figs. 2-6 \cite{barongallacode} are available from the following link \url{https://doi.org/10.5281/zenodo.4068257}. They are written using Mathematica 12 and Python v3.8. Python packages matplotlib, numpy and scipy were used. The codes for producing the figures in Supplementary Material are available upon reasonable request to the corresponding author.

\bigskip

{\bf Acknowledgements.}  J.W.B. thanks the Engineering and Physical Sciences Research Council (EPSRC) for funding (PhD studentship, EP/N509565/1). J.W.B and T.G. are grateful for funding from the Spanish Ministry of Science, Innovation and Universities, the Agency AEI and FEDER (EU) under grant PACSS (RTI2018-093732-B-C22). T.G. acknowledges partial financial support from the Maria de Maeztu Program for Units of Excellence in R\&D (MDM-2017-0711).
\medskip

{\bf Author contributions.} J.W.B. designed the study, contributed to discussions guiding the work, carried out mathematical calculations, performed simulations and wrote the manuscript. T.G. designed the study, contributed to discussions guiding the work and wrote the manuscript. Both authors approved the final manuscript for submission.

\medskip

{\bf Competing interests.} The authors declare no competing interests.
\end{document}


\title{Dispersal-induced instability in complex ecosystems \\ \medskip \medskip{\Large--- Supplementary Information ---}}
\author{Joseph W. Baron}
\email{josephbaron@ifisc.uib-csic.es}
\affiliation{Department of Physics and Astronomy, School of Natural Sciences,
	The University of Manchester, Manchester M13 9PL, United Kingdom}
\affiliation{Instituto de F{\' i}sica Interdisciplinar y Sistemas Complejos IFISC (CSIC-UIB), 07122 Palma de Mallorca, Spain}
\author{Tobias Galla}
\email{tobias.galla@manchester.ac.uk}
\affiliation{Department of Physics and Astronomy, School of Natural Sciences,
	The University of Manchester, Manchester M13 9PL, United Kingdom}
\affiliation{Instituto de F{\' i}sica Interdisciplinar y Sistemas Complejos IFISC (CSIC-UIB), 07122 Palma de Mallorca, Spain}

\maketitle
\tableofcontents

\clearpage
\section*{Overview}

This Supplementary Material is structured as follows: \\

In Part I (Sections \ref{sec:model} -- \ref{sec:parameters}) we provide further details of the model setup and of the analysis. In particular we describe the calculation of the bulk region of the spectrum and of the outlier eigenvalues. We then discuss how this can be used to obtain criteria for stability. Most importantly, we show how one can use our results for the outlier eigenvalues to deduce parameter sets for which there is a dispersal-induced instability. We also provide the values of the model parameters used for the figures in the main text.
\\

In Part II (Sections \ref{sec:universality} -- \ref{sec:diagonal}) we test the robustness of our main result -- the possibility of dispersal-induced instability -- against a number of possible variations of the model. For example, we comment on universality (i.e., non-Gaussian interaction matrices), the effects of strictly enforced trophic interactions, and of heterogeneity in the diffusion or self-interaction coefficients.
\\

In Part III (Sections \ref{sec:heterogeneity} -- \ref{sec:interplay}) we connect with the work in Ref. \cite{gravel} who reported that dispersal, in combination with spatial heterogeneity, can stabilise equilibria of complex ecosystems. To this end, we include the effects of spatial heterogeneity in our model with trophic structure. This broader model encompasses the set-up in \cite{gravel} and that in our main paper as limiting cases. We show that both the stabilising and destabilising effects of dispersal may be observed, depending on the circumstances. Using our analytical approach to calculate bulk spectra and outlier eigenvalues, we identify the factors which can make dispersal a stabilising or a destabilising influence.

\clearpage
\part{Further details of model setup and analysis}\label{part:maintext}
This part of the Supplementary Material contains further details relating to the construction of the model (Section~\ref{sec:model}), followed by the calculation of the bulk eigenvalue spectrum in Section~\ref{section:support}, and of outlier eigenvalues in Section~\ref{sec:outliers}. In Section~\ref{section:scaling} we comment on further technical details used to carry out the random-matrix calculation, and on how a connectance parameter $C$ can be accounted for. In Section~\ref{section:stability} we show how these results can be used to obtain stability criteria for equilibria of the ecosystem. Section~\ref{sec:parameters} finally provides the detailed model parameters used for the figures in the main manuscript.

\section{Model construction}\label{sec:model}
\subsection{General setup: trophic structure, species interactions and diffusive dispersal}We suppose that the total community consists of a sub-community of predator species and a sub-community of prey species, which are distributed in space. We assume that there are $N_u = \gamma_u N$ different prey species and $N_v = \gamma_v N$ predator species, where $\gamma_u + \gamma_v=1$. $N$ is the total number of species such that $N=N_u+N_v$.

\medskip

We imagine that the system reaches an equilibrium and that the equations governing the dynamics of the species abundances, linearised about this equilibrium, take the following form
\begin{align}
\frac{\partial u_i}{\partial t} &= D_u \frac{\partial^2 u_i}{\partial x^2}  - d_u u_i + \sum_{k=1}^{N_u} A^{uu}_{ik} u_k+ \sum_{j=1}^{N_v} A^{uv}_{ij} v_j  ,\nonumber \\
\frac{\partial v_j}{\partial t} &= D_v \frac{\partial^2 v_j}{\partial x^2} - d_v v_j +  \sum_{i=1}^{N_u} A^{vu}_{ji} u_i+ \sum_{k =1}^{N_v} A^{vv}_{jk} v_k. \label{linearisedsupp}
\end{align}
 
The  $u_i(x,t)$, $i=1,\dots,N_u$, describe perturbations of the abundances of the prey species about the equilibrium at position $x$ and time $t$, and similarly $v_j(x,t)$, $j=1,\dots,N_v$, are perturbations of the predator abundances.

We presume for now that the interactions within a single-species population are competitive (such that $d_u>0$ and $d_v>0$) and that each species disperses via diffusion. The corresponding dispersal coefficients are $D_u$ for prey, and $D_v$ for predators. Crucially, the prey and predator species can have different dispersal coefficients $D_u \neq D_v$. In the model there are also inter-population interactions. These are described by the interaction coefficients $A_{ij}^{\alpha\beta}$, where $\alpha$ and $\beta$ can takes values $\alpha, \beta \in\{u,v\}$, where $u$ stands for prey and $v$ for predator. For example, $A_{ij}^{uu}$ describes interactions between different prey species, and $A_{ij}^{uv}$ describes the effect that predator species $j$ has on prey species $i$.

The interaction coefficients $A^{\alpha \beta}_{ij}$ are random numbers drawn from a Gaussian distribution (although this assumption can be relaxed -- see Section \ref{sec:universality}) with 
\begin{align}
\overline{A^{\alpha \beta}_{ij}} &=  \mu_{\alpha \beta} ,  \nonumber \\
\overline{(A^{\alpha \beta}_{ij}- \mu_{\alpha\beta})^2}  &= \sigma^2,   \nonumber \\
\overline{(A^{\alpha \beta}_{ij}- \mu_{\alpha \beta}) (A^{\alpha' \beta'}_{ji}- \mu_{\alpha' \beta'})} &= \Gamma_{\alpha \beta \alpha' \beta'} \sigma^2. \label{rmstats}
\end{align}
All other correlations are zero. We assume that the only non-zero $\Gamma_{\alpha \beta \alpha' \beta'}$ are $\Gamma_{uuuu} \equiv \Gamma_u$, $\Gamma_{uvvu} \equiv \Gamma_{uv}$ and $\Gamma_{vvvv} \equiv \Gamma_v$. With a probability $1-C$, species $i$ and $j$ are chosen \textit{not} to interact with one another and thus $A_{ij}^{\alpha\beta}$ and $A_{ji}^{\beta\alpha}$ are set to zero. The parameter $C$ therefore describes the `connectance' (the probability that any given pair of species interact with one another).

Each of the model parameters can be interpreted ecologically. We define and interpret the complexity $c = CN\sigma^2$ in the main text. The interaction mean $\mu_{uu}$ indicates the degree to which two different prey species cooperate on average (if $\mu_{uu}$ is positive) or compete (if $\mu_{uu}$ is negative). The coefficient $\mu_{vv}$ has a similar interpretation for predators. The means of the off-diagonal blocks $\mu_{uv}<0$ and $\mu_{vu}>0$ indicate the degree to which (on average) prey species suffer and predator species gain from predator-prey interactions. The coefficients $\Gamma_{\alpha\beta\alpha'\beta'}$ indicate statistical correlations between interaction coefficients. A positive value of $\Gamma_{\alpha\beta\alpha'\beta'}$ indicates a statistical tendency for one species to benefit more (than average) from an interaction if the other species involved does so as well. The opposite is true if $\Gamma_{\alpha\beta\alpha'\beta'}$ is negative. The only non-zero entries are taken to be $\Gamma_u \equiv \Gamma_{uuuu}$, $\Gamma_v \equiv \Gamma_{vvvv}$ and $\Gamma_{uv} \equiv \Gamma_{uvvu}$. That is, only elements which are diagonally opposite one another in $\mathbf{A}$ are correlated \cite{galla2006, bunin}.

Models of complex ecosystems with structured random matrices have been considered before. For example, the work by Grilli~et.~al.~\cite{grilli2017} involves matrices with similar block structure to those in our model, the blocks in \cite{grilli2017} are associated with plant and animal species. Ref. \cite{allesina2015} evaluates the eigenvalue spectra of random matrices with cascade structure. The elements above and below the diagonal respectively have different means and variances to the elements below the diagonal. However, this is different from the structure of the matrices in our model. We are not aware of any work combining both block structure and dispersal, and computing the eigenvalue spectra of such models.

\subsection{Fourier transform and stability matrix}
The perturbations $u_i(x,t)$ and $v_j(x,t)$ in Eq.~(\ref{linearisedsupp}) are functions of position (and time). Taking the Fourier transform with respect to the spatial coordinate $x$, we write
\be
\tilde u_i(q)=\frac{1}{\sqrt{2\pi}}\int dx ~e^{iqx} u_i(x), \label{ftdef}
\ee
and similar for $\tilde v_j(q)$. The $\tilde u_i(q), \tilde v_j(q)$ describe perturbations of wavenumber $q$, or equivalently, of wavelength $\lambda=2\pi/q$.

Using this in Eqs.~(\ref{linearisedsupp}) one obtains
\begin{align}
\dot{ \tilde u}_i = -(q^2 D_u  + d_u )\tilde u_i  + \sum_{k=1}^{N_u} A^{uu}_{ik}\tilde u_k+ \sum_{j =1}^{N_v} A^{uv}_{ij}\tilde v_j ,\nonumber \\
\dot{\tilde  v}_j = -(q^2 D_v + d_v)\tilde v_j + \sum_{i=1}^{N_u} A^{vu}_{ji}\tilde u_i+ \sum_{k=1}^{N_v} A^{vv}_{jk}\tilde v_k.\label{processq}
\end{align}
We note that by setting $q = 0$ in Eqs.~(\ref{processq}) one recovers Eqs.~(\ref{linearisedsupp}) with the diffusion term removed. From the definition of the Fourier transform Eq.~(\ref{ftdef}), we see these are then equations for the total $u_i$ and $v_j$ summed throughout space. This removes any sense of spatial distribution of species. Setting $q=0$ therefore recovers the na\"ive `non-spatial' model where species dispersal is not taken into account. Eqs.~(\ref{processq}) can be written in a more compact matrix form
\be\label{eq:mdef1}
\dot{\tilde X} = \mathbf{M}_q \tilde X ,
\ee
where the vector $X$ is of length $N$, and with
\be\label{eq:mdef2}
\mathbf{M}_q = -q^2\mathbf{D} - \mathbf{d} + \mathbf{A}.
\ee

\noindent The matrix $\mathbf{M}_q$ is of dimension $N\times N$. It is the sum of three terms: 
\medskip

\noindent (i) The first term is $-q^2 \mathbf{D}$, where the matrix $\mathbf{D}$ is of size $N\times N$, diagonal and of the form
\be
\mathbf{D}=\left(\begin{array}{cc} D_u \id_{N_u} & 0\\ 0 & D_v\id_{N_v}\end{array}\right),
\ee
where $\id_{N_u}$ and $\id_{N_v}$ are identity matrices of dimensions $N_u\times N_u$ and $N_v\times N_v$ respectively. 

\noindent (ii) The second term is of a similar form
\be
\mathbf{d}=\left(\begin{array}{cc} d_u \id_{N_u} & 0 \\ 0 & d_v\id_{N_v}\end{array}\right),
\ee
and describes intraspecific interaction.

\noindent (iii) The third term in $\mathbf{M}_q$ is a random matrix with the Gaussian statistics given in Eq.~(\ref{rmstats}). We note that the matrix $\mathbf{M}_q$ has a block structure; it is comprised of submatrices of dimensions $N_u \times N_u$, $N_u \times N_v$, $N_v \times N_u$ and $N_v \times N_v$.  This is illustrated in Fig. 1 of the main text.

\medskip

The eigenvalues of the matrix $\mathbf{M}_q$ tell us about the stability of the hypothetical fixed point about which we have linearised. More precisely, the fixed point is stable against perturbations of wavelength $\lambda=2\pi/q$ if and only if all eigenvalues of $\mathbf{M}_q$ have negative real part.  The fixed point is stable against perturbations of {\em any} wavelength only if this is the case for {\em all} $q$. That is to say, in order for the equilibrium to be stable, the matrix $\mathbf{M}_q$ must not have an eigenvalue with positive real part for any $q$.

We therefore wish to identify the eigenvalue spectrum of $\mathbf{M}_q$. We can then use this to deduce under what conditions the equilibrium is stable. The matrix $\mathbf{M}_q$ has $N_u+N_v$ eigenvalues in total. The eigenvalue spectrum consists of a `bulk' of eigenvalues with some possible outliers (see Fig. 2 in the main text). The bulk contains the vast majority of eigenvalues. They are confined to a compact area in the complex plane. Both the boundary of the bulk region and the location of the outliers can be calculated based only on the statistical properties of the matrix $\mathbf{M}_q$ (as opposed to the exact entries). Further, the boundary of the bulk region can be calculated independently of the outliers. We therefore split our calculation into two parts, and present the calculation of bulk eigenvalues in Section~\ref{section:support}, and that for the outliers in Section~\ref{sec:outliers}.

\subsection{Correspondence with a model of a meta-ecosystem with patches}\label{sec:patches1}
We now briefly consider a model in which species hop between discrete patches, as opposed to diffusion in a continuous space. A more complete treatment of a model with discrete patches (with the possibility of statistical difference between the patches) is given in Section \ref{sec:heterogeneity}. 

To do this we imagine each species to be located on one of a set of discrete patches, whose locations are given by $x$.  The continuous diffusion term in Eqs.~(\ref{linearisedsupp}) is then replaced by its discrete counterpart to yield
\begin{align}\label{eq:rdd0}
\frac{\partial u_ix)}{\partial t} &= \hat D_u  \sum_{x'} \left[\phi(x,x') u_i(x') - \phi(x',x) u_i(x)\right]  - d_u u_i(x)\nonumber \\
&+ \sum_{k=1}^{N_u} A_{ik}^{uu} u_k(x) + \sum_{j=1}^{N_v} A_{ij}^{uv} v_j(x) ,\nonumber \\
\frac{\partial v_j(x)}{\partial t} &= \hat D_v \sum_{x'} \left[\phi(x,x') v_j(x') - \phi(x',x) v_j(x) \right]   - d_v v_j( x)\nonumber \\
& + \sum_{i=1}^{N_u} A_{ji}^{vu} u_i(x) + \sum_{k=1}^{N_v} A_{jk}^{vv} v_j( x).
\end{align}
We have suppressed the dependence on time to keep the notation compact. The quantity $\phi(x,x')$ is a `dispersal kernel', describing the probability that an individual hops from patch $x'$ to patch $x$, given that a hop occurs. We assume that the patches form a linear chain, and we suppose that dispersal is only between neighbouring patches and occurs with equal probability to either side. We therefore choose 
\begin{align}
\phi(x,x') = \begin{cases}
\frac{1}{2} \,\,\,\,\,\, \mathrm{if} \,\,\,\,\, x = x' +\Delta x ,\\
\frac{1}{2} \,\,\,\,\,\, \mathrm{if} \,\,\,\,\, x = x' - \Delta x, \\
0 \,\,\,\,\,\, \mathrm{otherwise},
\end{cases}
\end{align}
where $\Delta x$ is the inter-patch separation. The constants $\hat D_u$ and $\hat D_v$ in Eqs.~(\ref{eq:rdd0}) describe the overall dispersal rates (for prey and predators respectively) to any of the two neighbouring patches. 

Next, we take the discrete Fourier transform with respect to the spatial variable in Eqs.~(\ref{eq:rdd0}) , 
\be
\tilde u_i(q)= \sum_x ~e^{iq x} u_i(x).
\ee
We obtain
\begin{align}\label{eq:rdd}
\frac{\partial \tilde u_i}{\partial t} &= \hat D_u [\tilde\phi(q)-1] \tilde u_i   - d_u \tilde u_i + \sum_{k=1}^{N_u} A_{ik}^{uu} \tilde u_k + \sum_{j=1}^{N_v} A_{ij}^{uv} \tilde v_j  \nonumber \\
\frac{\partial \tilde v_j}{\partial t} &= \hat D_v [\tilde\phi(q)-1] \tilde v_j  - d_v \tilde v_j + \sum_{i=1}^{N_u} A_{ji}^{vu} \tilde u_i + \sum_{k=1}^{N_v} A_{jk}^{vv} \tilde v_j ,
\end{align}
with
\be
 \tilde\phi(q) = \cos\left(q\Delta x\right).
 \ee
Provided the arrangement of patches is `dense', i.e., that there are many patches per unit distance, we can assume $\Delta x \ll 1$, and expand $\tilde \phi(x)= \cos\left(q\Delta x\right)$ in powers of $\Delta x$. To lowest non-trivial order we find
\be
\tilde\phi(x)\approx  1- \frac{1}{2}(q\Delta x)^2.
\ee
Eqs.~(\ref{eq:rdd}) then become
\begin{align}
\frac{\partial \tilde u_i}{\partial t} &= -(q^2 D_u  + d_u )\tilde u_i  + \sum_{k=1}^{N_u} A^{uu}_{ik}\tilde u_k+ \sum_{j =1}^{N_v} A^{uv}_{ij}\tilde v_j,\nonumber \\
\frac{\partial \tilde v_j}{\partial t} &=  -(q^2 D_v + d_v)\tilde v_j + \sum_{i=1}^{N_u} A^{vu}_{ji}\tilde u_i+ \sum_{k=1}^{N_v} A^{vv}_{jk}\tilde v_k ,
\end{align}
where we have identified the diffusion constants $D_\alpha$ in terms of the hopping rates $\hat D_\alpha$: $D_\alpha = \frac{\hat D_\alpha (\Delta x)^2}{2} $. We thus arrive at Eqs.~(\ref{linearisedsupp}). 

This shows that any stability criteria derived for the ecosystem with diffusion in continuous space also applies to a model with discrete patches and dispersal between adjacent patches. This is valid provided the distance between two patches in space is small compared to the characteristic wavelengths associated with the instability. It is also possible to analyse the model with discrete patches, and to find the eigenvalues as a function of $q$ without performing this approximation. One would still find the possibility for Turing instability (as demonstrated for a different system in \cite{barongallastochastic} and also in Section \ref{sec:all-to-all}).

\section{Boundary of the bulk of the eigenvalue distribution}\label{section:support}

\subsection{Overview: approach to calculating the eigenvalue spectral density}

Our approach to calculating the support of the eigenvalue spectra of random matrices is based on methods from statistical physics.  The study of random matrices has a long history in physics, going to back to Wigner who first established the semi-circle law for symmetric Gaussian random matrices in the context of nuclear physics \cite{Wigner1955,Wigner1958}. 

There is a wide range of other applications of random matrices in physics \cite{RMPhysicsBook}, in particular also in the theory of spin glasses and disordered systems \cite{mezard1987}. 

In this context, a number of different theoretical approaches have been developed to evaluate the spectra of random matrices (see e.g. \cite{mehta}). For example, one can recognise the resolvent of the random matrix [defined below in Eq.~(\ref{resolvent})] as the Stieltjes transform of the eigenvalue density in the limit of a large size $N$ of the matrix. This was done for Hermitian matrices in the seminal work by Pastur \cite{pastur1972}. This method has since been developed further -- the case of non-Hermitian matrices can be addressed through the use of quaternionic resolvents \cite{rogers2010, barabas2017}. Alternatively, one can evaluate the resolvent using a series expansion via the Dyson equation, dropping terms of sub-leading order in $N$ in the limit $N \to \infty$, as was done by Bray and Moore \cite{braymoore}. We use elements of this argument in Section \ref{sec:proofdiagonal}. Approaches involving series expansions of the resolvent have also been extended to take into account higher-order correlations \cite{aceitunorogersschomerus}. We note that the characterisation of spectra of random matrices is also a timely topic in applied mathematics. Recent work includes \cite{taovu2010,taovukrishnapur2010}, see also \cite{taobook} for a further overview.  

Our analytical approach to the calculation of bulk eigenvalue spectra in this section broadly proceeds along the lines of \cite{sommers}, who used the replica method to compute elliptic spectra of random matrices with correlations between elements, but with no particular sub-structure or compartments in the matrix. The approach of \cite{sommers} is based on the introduction of a suitable `potential' $\Phi$ (to be defined below), which is averaged over the ensemble of random matrices, and from which the resolvent can then be obtained. Essentially, the calculation converts the problem of finding the bulk region of eigenvalues into one of evaluating an integral over what would be called the (macroscopic) `order parameters' in statistical mechanics. In the limit of large matrices, this integral is then carried out using the saddle-point approximation (see e.g. \cite{altlandsimons, mezard1987,Coolen_MG} for modern accounts and applications to problems in disordered systems). The saddle-point method is a common approach in the theory of disordered system, with wide-ranging applications in spin-glass physics \cite{mezard1987}, neural networks \cite{gardner1988, Coolen_NN} and interacting-agent models and game theory \cite{Coolen_MG, Challet_MG}. Work using this approach to calculate spectra of random matrices includes \cite{haake}, see also \cite{mehta} for a more comprehensive overview. 

In Section \ref{sec:outliers} we then combine this calculation of the bulk region of the eigenvalue spectrum with the approach of Ref. \cite{orourke} to find the outlier eigenvalues. These arise from the introduction of a non-zero mean to the random matrix elements. The  resolvents necessary for the calculation of these outliers can readily be identified from the order parameters in the saddle-point calculation of the bulk spectrum.

The primary methodological contribution of our work is the calculation of the eigenvalue support for matrices with block structure. The block structure to which we refer is defined in more detail in the main paper and in \ref{sec:model} of this Supplement. In the analysis, the block structure is dealt with conveniently through the introduction of integration variables with an additional block index $\alpha$. In principle, this approach can be extended for matrices with more complicated structure and statistics (Parts II and III of this Supplement contain examples). The approach has further advantages, for instance, the origin of universality is quite transparent (see \ref{sec:universality}).

\medskip

\subsection{Preliminaries -- setting up the calculation}
\noindent For an ensemble of $N\times N$ random matrices $\mathbf{m}$, consider the so-called resolvent, defined by 
\begin{align}
G\left(\omega\right) = \frac{1}{N} \overline {\mathrm{Tr} \left[\left(\omega \id_N  - \mathbf{m}  \right)^{-1} \right]}  . \label{resolvent}
\end{align}
The overbar represents an average over the ensemble of matrices. The expression $\id_N$ denotes the identity matrix of size $N\times N$.

 We note the following identities from general linear algebra (for an $N\times N$ matrix $\mathbf{B}$):
\begin{enumerate}
	\item $\mathrm{Tr}\left[\mathbf{P}^{-1}\,\mathbf{B}\, \mathbf{P} \right] = \mathrm{Tr}\left[\mathbf{B}\right] $, for any invertible $N\times N$ matrix $\mathbf{P}$;
	\item If $\mathbf{B}$ is invertible, then the eigenvalues of $\mathbf{B}^{-1}$ are reciprocals of those of $\mathbf{B}$;
	\item The trace is the sum of the eigenvalues; 
	\item Adding the identity matrix to a matrix increases each eigenvalue of this matrix by $1$;
	\item Multiplying a matrix by a constant scales its eigenvalues by that same constant.
\end{enumerate}
Using these, one can show that
\begin{align}
G\left(\omega\right) =\frac{1}{N} \overline { \sum_\lambda \frac{1}{\omega - \lambda}}, \label{spectrum}
\end{align}
where the $\lambda$ are the eigenvalues of $\mathbf{m}$. In the limit $N \to \infty$, this sum can be written as an integral over the complex plane
\begin{align}
G\left(\omega\right) = \int d^2\lambda \frac{\rho(\lambda)}{\omega - \lambda} ,\label{spectint}
\end{align}
where $\rho(\lambda)$ is the density of eigenvalues. 

Bearing in mind that $G(\omega)$ is a complex quantity, we perform a contour integral around a closed path $\mathcal{C}$, which we presume contains none of the poles of $G$,
\begin{align}
\frac{1}{2\pi i}\int_{\mathcal{C}} d\omega~ G(\omega) &= \frac{1}{N} \overline{ \sum_\lambda\frac{1}{2\pi i}\int_{\mathcal{C}} d\omega \frac{1}{\omega - \lambda} } \nonumber \\
&= \frac{1}{N} \overline{ \sum_\lambda 1} = \int_{\mathcal{S}} d^2 \lambda~ \rho(\lambda) , \label{residue}
\end{align}
In this expression $\mathcal{S}$ denotes the region bounded by $\mathcal{C}$, and we have used the residue theorem to evaluate the contour integral. 

We next note the following complex version of Gauss' law (letting $\omega = x + i y$)
\begin{align}
\int_{\mathcal{S}} d^2\omega \left[  \frac{\partial G}{\partial x} + i \frac{\partial G}{\partial y} \right] = \frac{1}{i}\int_{\mathcal{C}} d\omega \,G(\omega) . \label{Gauss}
\end{align} 
Defining the real-valued potential $\Phi(\omega)$ via
\begin{align}
2\,\mathrm{Re}\, G = - \frac{\partial \Phi}{\partial x}, \,\,\,\, 2\,\mathrm{Im}\,G =  \frac{\partial \Phi}{\partial y}, \label{phiandg}
\end{align}
and using Eqs.~(\ref{residue}) and (\ref{Gauss}), along with the fact that the region $\mathcal{S}$ is arbitrary, we obtain Poisson's equation
\begin{align}
\nabla^2 \Phi = - 4 \pi \rho . \label{Poisson}
\end{align}
In analogy with electrostatics, the resolvent $G(\omega)$ can be thought of as being related to the `electric field' with components $E_x = 2 \,\mathrm{Re} ~G$ and $E_y = -2\, \mathrm{Im} ~G$ \cite{sommers}. Therefore, we can obtain the distribution of eigenvalues, if we can find the potential $\Phi$. 

The potential $\Phi$ can be related directly to the matrix $\mathbf{m}$ via
\begin{align}
\Phi(\omega) = -\frac{1}{N} \overline{\ln \det\left[ (\id_N\omega^\star - \mathbf{m}^T )( \id_N\omega - \mathbf{m} )\right]} . \label{potential}
\end{align}
This can be seen to agree with Eqs.~(\ref{phiandg}) and (\ref{spectrum}) once one realises that $\det\left(\id_N\omega - \mathbf{m} \right) = \prod_{\lambda} (\omega - \lambda)$. 

\noindent We now make the following observations about the resolvent and its relation to the potential $\Phi$ and the eigenvalue density $\rho$. Eq. (\ref{phiandg}) can be written more compactly as
\begin{align}
G(x,y) \equiv -\frac{1}{N}\overline{\mathrm{Tr} \left[ \frac{1}{\id_N\omega - \mathbf{m}} \right]} = \frac{\partial \Phi(\omega, \omega^\star)}{\partial \omega} .
\end{align}
Indeed, we also have
\begin{align}
G^\star(x,y) = \frac{\partial \Phi(\omega, \omega^\star)}{\partial \omega^\star} .
\end{align}
Combining these relations with Eq.~(\ref{Poisson}), one sees that the resolvent is related to the eigenvalue density by 
\begin{align}
\frac{\partial \mathrm{Re} [G]}{\partial x} - \frac{\partial \mathrm{Im} [G]}{\partial y} = \mathrm{Re}\left[\frac{\partial G(\omega, \omega^\star)}{\partial \omega^\star}\right]= 2 \pi \rho . \label{cauchyriemann}
\end{align}
The left-hand side of this equation is familiar from the Cauchy-Riemann equations for analytic functions. From this we deduce that if $G(x,y)$ is an analytic function in a particular region (i.e. we can write it in terms of $\omega$ only), the eigenvalue density must be zero in this region.

\subsection{Carrying out the average over the ensemble of random matrices}\label{sec:av}
\subsubsection*{Ensemble of random matrices}
 Adding a low-rank perturbation to a large random matrix is known not to change the bulk of the eigenvalue spectrum \cite{allesinatang1, orourke, edwardsjones}. This means that we can assume that the mean of the distribution from which the matrix elements are drawn is zero. The mean can have an effect on outlier eigenvalues, and we will therefore restore it for the calculation of the outliers in Section~\ref{sec:outliers}.

For the time being we consider a Gaussian random matrix with statistics as follows

\begin{align}
\overline{m^{\alpha \beta}_{ij}}&= 0,  \nonumber \\
\overline{ (m^{\alpha \beta}_{ij})^2}  &= \frac{\sigma^2}{N},   \nonumber \\
\overline{ m^{\alpha \beta}_{ij} m^{\alpha' \beta'}_{ji}}&= \frac{\Gamma_{\alpha \beta \alpha' \beta'} \sigma^2}{N} , \nonumber \\
m_{ii}^{\alpha \alpha} &= - d_\alpha,\label{eq:mstat}
\end{align}
The matrix $\mathbf{m}$ is similar to $\mathbf{M}_q$ (the matrix we focus on in the main paper), but it is different in a number of ways:

First, we note that we have re-scaled the matrix elements relative to the matrix $\mathbf{M}_q$ in the main paper -- the variance and co-variances are now proportional to $1/N$. This approach is standard in the theory of random matrices, its objective is of a technical nature. The re-scaling allows us to carry out the calculation in the limit $N\to\infty$. In practice the results derived in this way are often accurate also for finite $N$. The scaling is undone by replacing $\sigma^2 \to N \sigma^2$ in the result. For a more detailed discussion, see Section \ref{section:scaling}.

Further, we are assuming all-to-all interaction ($C=1$) for now between the species. We discuss how choosing a general connectance $0<C<1$ affects the result in Section~\ref{section:scaling}. 

Finally, the wavenumber $q$ is set to zero here. The result for non-zero $q$ can obtained by making the replacement $d_\alpha \to d_\alpha + q^2 D_\alpha$, see also Section~\ref{sec:dii}.

\subsubsection*{Replica method and Hubbard-Stratononich transformation}
In order to calculate the spectrum of the ensemble of random matrices we need to compute $\overline{\ln \det\left[ (\omega^*\id_N - \mathbf{m}^T )( \omega \id_N- \mathbf{m} )\right]} $. We recall that the overbar stands for the average over the randomness in $\mathbf{m}$. In order to calculate averages of logarithms, one can use the so-called replica method \cite{sommers, edwardsjones, sherrington, mezard1987}, based on the identity  $\ln~ y = \lim_{n\to 0} \frac{y^n -1}{n}$. This allows one to calculate $\overline{y^n}$ instead of the more intricate average $\overline{\ln~y}$. The object $y^n$ in turn represents an $n$-fold `replicated' system in statistical physics \cite{sommers, edwardsjones, sherrington, mezard1987}. In the context of the present problem one finds that the $n$ replicas decouple in the limit of large $N$, similar to what was observed in \cite{sommers, haake}. This means that
\be\label{eq:help}
\overline{\ln \det\left[ (\id_N\omega^\star - \mathbf{m}^T )( \id_N\omega - \mathbf{m} )\right]}= \ln \overline{\det\left[ (\id_N\omega^\star - \mathbf{m}^T )( \id_N\omega - \mathbf{m} )\right]}.
\ee
Using a Gaussian integral we can write the determinant in Eq.~(\ref{eq:help}) in the form
\begin{align}\
&\det\left[ (\id_N\omega^\star - \mathbf{m}^T )(\id_N\omega - \mathbf{m} )\right]^{-1} = \int \prod_{i, \alpha} \left( \frac{d^2z^\alpha_i}{\pi}\right) \nonumber \\
&\times \exp\left[ -\epsilon \sum_i \vert z^\alpha_i \vert^2 - \sum_{i,j,k, \alpha, \beta, \gamma}z^{\alpha \star}_i (\omega^\star\delta_{ik}\delta_{\alpha \gamma} - (m^T)_{ik}^{\alpha \gamma})(\omega\delta_{kj}\delta_{\gamma \beta} - m^{\gamma \beta}_{kj})z^\beta_j\right], \label{potentialdeterminant}
\end{align}
where $\epsilon$ is a positive infinitesimal quantity introduced so as to avoid singularities which occur when $\omega$ is equal to one of the eigenvalues of $\mathbf{m}$.
To carry out the average over the ensemble of random matrices we need to evaluate
\begin{align}
\int \prod_{ij\alpha\beta}\left(d m^{\alpha \beta}_{ij} \right) \sqrt{\frac{N}{2 \pi \sigma^2 (1-\Gamma_{\alpha\beta}^2)}} &\exp\left[-\frac{N}{ \sigma^2(1-\Gamma_{\alpha\beta}^2)}\left[(m^{\alpha \beta}_{ij})^2   - \Gamma_{\alpha \beta} m^{\alpha \beta}_{ij}m^{ \beta\alpha}_{ji} \right] \right] \nonumber \\
&\times\det\left[ (\id_N\omega^\star - \mathbf{m}^T )( \id_N\omega - \mathbf{m} )\right]^{-1} . \label{averagedeterminant}
\end{align}
We next perform a Hubbard--Stratonovich transformation \cite{hubbard} in order to linearise the terms involving $m^{\alpha \beta}_{ij}$. This yields 
\begin{align}
&\det\left[ (\id_N\omega^\star - \mathbf{m}^T )( \id_N\omega - \mathbf{m} )\right]^{-1} = \int \prod_{i\alpha} \left( \frac{d^2z^\alpha_i d^2y^\alpha_i}{2 \pi^2}\right) \exp\left[ - \sum_{i\alpha} y^{\alpha\star}_i y^\alpha_i + \epsilon z^{\alpha\star}_i z^\alpha_i \right]\nonumber \\
\times &\exp\left[ i \sum_{ij\alpha \beta} z^{\alpha\star}_i \left( m^{ \beta\alpha}_{ji} - \omega^\star\delta_{ij}\delta_{\alpha \beta} \right) y^\beta_j\right] \exp\left[ i \sum_{ij\alpha \beta} y^{\alpha\star}_i \left( m^{\alpha \beta}_{ij} - \omega \delta_{ij}\delta_{\alpha \beta}\right) z^\beta_j\right]. \label{HStransformed}
\end{align}
\subsubsection*{Average over the random matrix and introduction of order parameters}
Carrying out the average over the random matrix elements in Eq.~(\ref{HStransformed}) then gives [recalling the definition of the potential $\Phi(\omega)$ in Eq.~(\ref{potential})]
\begin{align}
\exp\left[ -N \Phi(\omega) \right] &= \int \prod_{i\alpha} \left( \frac{d^2z^\alpha_i d^2y^\alpha_i}{2 \pi^2}\right) \exp\left[ - \sum_{i\alpha} y^{\alpha\star}_i y^\alpha_i + \epsilon z^{\alpha\star}_i z^\alpha_i \right]\nonumber \\
\times & \exp\left[ -i \sum_{i \alpha} z^{\alpha\star}_i y^\alpha_i  (\omega^\star + d_\alpha) + z^\alpha_i y^{\alpha\star}_i ( \omega + d_\alpha)  \right]\nonumber \\
\times & \exp\left[ - \frac{\sigma^2}{2N}\sum_{i j \alpha \beta} (z_i^{\alpha\star} y^\beta_j + z^\alpha_i y_j^{\beta\star})^2 + \Gamma_{\alpha \beta} (z_i^{\alpha\star} y^\beta_j + z^\alpha_i y_j^{\beta\star})(z_j^{\beta\star} y^\alpha_i + z^\beta_j y_i^{\alpha\star}) \right] . \label{averaged}
\end{align}
Following \cite{haake, edwardsjones} we disregard contributions containing $N^{-1} \sum_i (y^{\alpha\star}_i)^2$, $N^{-1} \sum_i (z^{\alpha\star}_i)^2$, $N^{-1} \sum_i (y^\alpha_i)^2$, $N^{-1} \sum_i (z^\alpha_i)^2$, $N^{-1} \sum_i z^{\alpha\star}_i y^{\alpha\star}_i $ and $N^{-1} \sum_i z^\alpha_i y^\alpha_i $. These do not contribute to leading order in $N$.

We now introduce the order parameters 
\begin{align}
u &= \frac{1}{N} \sum_{i \alpha} z^{\alpha\star}_i z^\alpha_i, \,\,\,\,\, v = \frac{1}{N} \sum_{i \alpha} y^{\alpha\star}_i y^\alpha_i, \nonumber \\
w_\alpha &= \frac{1}{N_\alpha} \sum_i z^{\alpha\star}_i y^\alpha_i, \,\,\,\,\, w_\alpha^\star = \frac{1}{N_\alpha} \sum_i y^{\alpha\star}_i z^\alpha_i, \label{orderparameters}
\end{align}
which we impose in the integral Eq.~(\ref{averaged}) using Dirac delta functions in their complex exponential representation. For example
\begin{align}
\delta(u - \frac{1}{N} \sum_{i \alpha} z^{\alpha\star}_i z^\alpha_i ) &\propto \int d\hat u \exp\left[i \hat u (N u -  \sum_{i \alpha} z^{\alpha\star}_i z^\alpha_i)  \right], \nonumber \\
\delta(w_\alpha - \frac{1}{N_\alpha} \sum_i z^{\alpha\star}_i y^\alpha_i) &\propto \int d^2\hat w_\alpha \exp\left[i \hat w_\alpha^\star (N_\alpha w_\alpha -  \sum_i z^{\alpha\star}_i y^\alpha_i) + i \hat w_\alpha (N_\alpha w_\alpha^\star -  \sum_i z^{\alpha}_i y^{\alpha\star}_i)  \right] , \label{deltafunctions}
\end{align}
where we note that $w_\alpha$ is a complex quantity. We can thus rewrite Eq.~(\ref{averaged}) as 
\begin{align}
\exp\left[ -N \Phi(\omega) \right] &= \int \mathcal{D}\left[ \cdots \right]\exp\left[N(\Psi + \Theta + \Omega)\right] ,\label{saddlesetup}
\end{align}
where where $\mathcal{D}\left[ \cdots \right]$ denotes integration over all of the order parameters and their conjugate (`hatted') variables, and where 
\begin{align}
\Psi &= i  \hat u u  + i\hat v  v  + i \sum_{\alpha}\gamma_\alpha(\hat w_\alpha  w^\star_\alpha  +   \hat w^\star_\alpha  w_\alpha   ), \nonumber \\
\Theta &= -\epsilon  u - v - \sigma^2 u v + \sum_\alpha \gamma_\alpha\left[    -i w_\alpha (\omega^\star + d_\alpha)  - i w^\star_\alpha (\omega + d_\alpha)\right]  - \frac{1}{2} \sigma^2 \sum_{\alpha \beta}\gamma_\alpha \gamma_\beta\Gamma_{\alpha \beta}(w_\alpha w_\beta+ w_\alpha^\star w_\beta^\star) , \nonumber \\
\Omega &= \sum_\alpha \gamma_\alpha \ln  \left[\int   \left( \frac{d^2z^\alpha d^2y^\alpha}{2 \pi^2}\right)  \exp\left\{ -i  (\hat u z^{\alpha\star} z^\alpha  + \hat v y^{\alpha\star} y^\alpha  + \hat w_\alpha y^{\alpha\star} z^\alpha  + \hat w_\alpha^\star z^{\alpha\star} y^\alpha  ) \right\} \right].
\end{align}
We recall that $\alpha$ can take the values $\alpha=u$ (prey) and $\alpha=v$ (predators), and that $\gamma_u$ and $\gamma_v$ are the fraction of prey and predator species respectively. 

We note that the integrals over $y^\alpha_i$ and $z^\alpha_i$ are uncoupled for different values of $i$ as a result of introducing the order parameters in Eq.~(\ref{orderparameters}). Carrying out the integrals over the variables $y^\alpha$ and $z^\alpha$ in the expression for $\Omega$ one obtains
\begin{align}
\Omega &= -\sum_\alpha \gamma_\alpha\ln(\hat w_\alpha \hat w_\alpha^\star - \hat u \hat v).
\end{align}

\subsubsection*{Saddle-point integration and evalulation of order parameters}
We now take the limit $N\to\infty$, and carry out the integral in Eq.~(\ref{saddlesetup}) in the saddle-point approximation. To do this, we extremise the expression $\Psi + \Theta + \Omega$. Extremising with respect to the conjugate variables $\hat u, \hat v, \hat w_\alpha$ and $\hat w_\alpha^\star$, we find
\begin{align}
i u &= \sum_\alpha \frac{ -\gamma_\alpha \hat v}{\hat w_\alpha \hat w_\alpha^\star - \hat u \hat v } , \,\,\,\,\, iv = \sum_\alpha \frac{ -\gamma_\alpha \hat u}{\hat w_\alpha \hat w_\alpha^\star - \hat u \hat v} , \,\,\,\,\,  i\gamma_\alpha w_\alpha = \frac{\gamma_\alpha\hat w_\alpha}{\hat w_\alpha \hat w_\alpha^\star - \hat u \hat v } , \,\,\,\,\,  i\gamma_\alpha w_\alpha^\star = \frac{\gamma_\alpha \hat w_\alpha^\star}{\hat w_\alpha \hat w_\alpha^\star - \hat u \hat v } .\label{diffwrthatted}
\end{align}
One makes the following observation
\begin{align}
i\hat u u +i\hat v v +i \sum_\alpha \gamma_\alpha( \hat w_\alpha w_\alpha^\star +   \hat w_\alpha^\star w_\alpha )  =  2 .
\end{align}
Further, we find that
\begin{align}
- w_\alpha w^\star_\alpha = \frac{\hat w_\alpha \hat w_\alpha^\star}{(\hat w_\alpha \hat w_\alpha^\star-\hat u\hat v)^2} ,
\end{align}
which means that we can now express $\hat w_\alpha \hat w_\alpha^\star$ in terms of $w_\alpha w^\star_\alpha$ and $\hat u\hat v$. We also find that
\begin{align}
-uv = \sum_{\alpha \beta}\frac{\gamma_\alpha \gamma_\beta \hat u \hat v}{(\hat w_\alpha \hat w_\alpha^\star-\hat u\hat v)(\hat w_\beta \hat w_\beta^\star-\hat u\hat v)} ,
\end{align}
which means that we can in turn express $\hat u\hat v$ in terms of $w_\alpha w^\star_\alpha$ and $uv$.  Defining $r = uv$, we can write $\hat u\hat v$ and $\hat w_\alpha \hat w_\alpha^\star$ in terms of $w_\alpha w^\star_\alpha$ and $r$. This means that we can replace $\hat w_\alpha \hat w_\alpha^\star - \hat u \hat v$ in the expression for $\Omega$. 

So, letting $\hat q_\alpha = (\hat w_\alpha \hat w_\alpha^\star - \hat u \hat v)^{-1}$, we obtain an implicit equation for $\hat q_\alpha$ in terms of $r$ and $w_\alpha w^\star_\alpha$:
\begin{align}
- w_\alpha w^\star_\alpha &= \hat q_\alpha-\hat q_\alpha^2 \frac{r}{\left[\sum_\beta \gamma_\beta \hat q_\beta\right]^2} .
\end{align} 
We now extremise $\Psi + \Theta + \Omega$ in Eq.~(\ref{saddlesetup}) with respect to  $u, v, w_\alpha$ and $w_\alpha^\star$. We find
\begin{align}
i\hat u &= \epsilon + \sigma^2 v , \,\,\,\,\, i\hat v = 1 + \sigma^2 u , \nonumber \\
i\hat w_\alpha &=  i(\omega + d_\alpha) +  \sigma^2 \sum_{\beta}\Gamma_{\alpha \beta}\gamma_\beta w^\star_\beta , \,\,\,\,\, i\hat w_\alpha^\star = i (\omega^\star + d_\alpha) + \sigma^2 \sum_{\beta}\Gamma_{\alpha \beta}\gamma_\beta w_\beta . \label{diffwrtunhatted}
\end{align}
Combining these with Eqs.~(\ref{diffwrthatted}), one sees that 
\begin{align}
v &= (\epsilon + \sigma^2 v)\sum_\alpha \gamma_\alpha \hat q_\alpha , \nonumber \\
u &= (1 + \sigma^2 u )\sum_\alpha \gamma_\alpha \hat q_\alpha  , \nonumber \\
 w_\alpha &= -\left[  i(\omega + d_\alpha)+  \sigma^2 \sum_\beta \Gamma_{\alpha \beta} \gamma_\beta w^\star_\beta\right]\hat q_\alpha , \nonumber \\
 w^\star_\alpha &=  -\left[  i(\omega^\star + d_\alpha)+  \sigma^2 \sum_\beta \Gamma_{\alpha \beta} \gamma_\beta w_\beta\right]\hat q_\alpha . \label{unhatted}
\end{align}
The first equation in this set has two solutions for $\epsilon \to 0^+$. 
\medskip

\noindent\underline{First solution:}\\
 One solution is $v =0$, implying $ r = 0$, and $\hat q_\alpha =  - w_\alpha w^\star_\alpha$. The last two of the above equations yield
\begin{align}
1 &=    i(\omega + d_\alpha) w_\alpha^\star +  \sigma^2 \sum_\beta \Gamma_{\alpha \beta} \gamma_\beta w_\alpha^\star w^\star_\beta , \nonumber \\
1 &=   i (\omega^\star + d_\alpha) w_\alpha +  \sigma^2 \sum_\beta \Gamma_{\alpha \beta} \gamma_\beta w_\alpha w_\beta .
\end{align}
From this, one can solve for the resolvent $G(\omega) = \sum_\alpha  \gamma_\alpha w_\alpha^\star$, which we see is an analytic function of $\omega$. This means that the eigenvalue density vanishes in regions of the complex plane for which $r=0$ is the only valid solution (see Eq.~(\ref{cauchyriemann}).

\medskip

\noindent\underline{Second solution:} \\
The other solution to the first of Eq.~(\ref{unhatted}) is 
\begin{align}
\sum_\alpha \gamma_\alpha \hat q_\alpha = \frac{1}{\sigma^2} . \label{conditionforinside}
\end{align}
This can be solved simultaneously with the following equations in order to find $r$, $w_\alpha$ and $w_\alpha^\star$ as functions of $\omega = x + i y$
\begin{align}
- w_\alpha w^\star_\alpha &= \hat q_\alpha-\hat q_\alpha^2 \frac{r}{\left[\sum_\beta \gamma_\beta \hat q_\beta\right]^2} , \nonumber \\
w_\alpha &= - \left[ i (\omega + d_\alpha) +\sigma^2 \sum_\beta \Gamma_{\alpha \beta} \gamma_\beta w^\star_\beta\right]\hat q_\alpha , \nonumber \\
w^\star_\alpha &= - \left[ i (\omega^\star + d_\alpha)+  \sigma^2 \sum_\beta \Gamma_{\alpha \beta}\gamma_\beta w_\beta\right]\hat q_\alpha . \label{rneq0}
\end{align}

\subsubsection*{Significance of the order parameters}
It is important at this point to note that the order parameters defined in Eq.~(\ref{orderparameters}) have some broader significance. The quantities $w_\alpha$ and $w_\alpha^\star$, when evaluated at the saddle point, are related to the resolvent:
\begin{align}
G(\omega) &= \frac{\partial \Phi(\omega, \omega^\star)}{\partial \omega} = i\sum_\alpha  \gamma_\alpha w^\star_\alpha , \nonumber \\
G^\star(\omega) &= \frac{\partial \Phi(\omega, \omega^\star)}{\partial \omega^\star} =  i \sum_\alpha \gamma_\alpha w_\alpha . \label{resolventatsaddlepoint}
\end{align}
Imagine momentarily that instead of evaluating $- N \Phi(x,y) =  \overline{\ln \det\left[ (\id_N\omega^\star - \mathbf{m}^T )(\id_N\omega - \mathbf{m} )\right]}$, we had considered the `potential' of the matrix $m_{ij}^{\alpha\beta} - \omega_\alpha \delta_{\alpha\beta}\delta_{ij}$ [i.e. if we had redone the calculation up until this point but had replaced $\omega \id_N$ in the definition of the resolvent Eq.~(\ref{resolvent}) with a diagonal matrix $\left(\boldsymbol{\upomega}\right)_{ij}^{\alpha\beta}= \omega_\alpha \delta_{\alpha\beta}\delta_{ij}$]. Then, instead of arriving at Eq.~(\ref{resolventatsaddlepoint}), we would have been able to define 
\begin{align}
G_\alpha(\{\omega_\alpha, \omega_\alpha^\star\}) &\equiv \frac{1}{\gamma_\alpha}\frac{\partial \Phi(\{\omega_\alpha, \omega_\alpha^\star\})}{\partial \omega_\alpha} =  i w^\star_\alpha , \nonumber \\
G^\star_\alpha(\{\omega_\alpha, \omega_\alpha^\star\}) &\equiv \frac{1}{\gamma_\alpha}\frac{\partial \Phi(\{\omega_\alpha, \omega_\alpha^\star\})}{\partial \omega^\star_\alpha} = i  w_\alpha  .
\end{align}
The result of this line of reasoning is that, at the saddle point, we deduce $iw_\alpha = (iw^\star_\alpha)^\star$. Such a deduction is still valid once we set $\omega_\alpha = \omega$. This observation will prove important for ruling out invalid solutions when we solve the saddle-point equations. 

Although the fact that at the saddle-point $iw_\alpha = (iw^\star_\alpha)^\star$ may at first seem troubling, one should note the following. Let $w_\alpha = q_\alpha + i s_\alpha$ and $w^\star_\alpha = q_\alpha - i s_\alpha$. The (real) variables over which we integrate in Eq.~(\ref{saddlesetup}), in order to impose the delta function in Eq.~(\ref{deltafunctions}), are $q_\alpha$ and $s_\alpha$. Because the argument of the exponential in Eq.~(\ref{saddlesetup}) is complex, we have to use analytical continuation to deform the integrals, which are initially over the real axis, to contours in the complex plane. These contours include the saddle point. As such, the ostensibly `real' variables $q_\alpha$ and $s_\alpha$ take on complex values at the saddle point. Hence, $w_\alpha$ and $w_\alpha^\star$ are no longer necessarily complex conjugates when evaluated at the saddle point. 

\subsubsection*{Calculation of boundary of bulk spectrum}
Inside the support of the eigenvalue spectrum (where the eigenvalue density is non-zero), we have $r>0$ and the condition Eq.~(\ref{conditionforinside}). Assuming no discontinuities in the value of $r$ as a function of $\omega$, the boundary of the support of the eigenvalue spectrum is defined by points $\omega$ in the complex plane where $r$ approaches zero, but where Eq.~(\ref{conditionforinside}) is still satisfied. This yields the following set of equations which one can solve for the curve defining the boundary $\omega_y(\omega_x)$ (where $\omega = \omega_x + i \omega_y$): 
\begin{align}
\sum_\alpha \gamma_\alpha \hat q_\alpha &= \frac{1}{\sigma^2} , \nonumber \\
 -w_\alpha w^\star_\alpha &= \hat q_\alpha , \nonumber \\
 w_\alpha &= - \left[ i (\omega_x + i\omega_y+ d_\alpha) +\sigma^2 \sum_\beta \Gamma_{\alpha \beta}\gamma_\beta w^\star_\beta\right]\hat q_\alpha , \nonumber \\
 w^\star_\alpha &= - \left[ i (\omega_x - i\omega_y + d_\alpha)+  \sigma^2 \sum_\beta \Gamma_{\alpha \beta}\gamma_\beta w_\beta\right]\hat q_\alpha . \label{support}
\end{align}
That is, values of $\omega$ which lie on the boundary of the support of the eigenvalue spectrum must satisfy the above simultaneous equations. 

Since $iw_\alpha = (iw^\star_\alpha)^\star$ at the saddle point, the second relation in Eq.~(\ref{support}) dictates that $\hat q_\alpha$ is real and positive. In solving Eqs.~(\ref{support}), we therefore discard solutions which do not satisfy $\hat q_\alpha \in \mathbb{R}$ and $\hat q_\alpha>0$.
Alternatively, letting $i w_\alpha^\star = (i w_\alpha)^\star = \chi_\alpha$, we obtain
\begin{align}
\sum_\alpha \gamma_\alpha \left\vert \chi_\alpha \right\vert^2 &= \frac{1}{\sigma^2} , \nonumber \\
-1 &=  - (\omega_x + i\omega_y + d_\alpha)\chi_\alpha +\sigma^2 \sum_\beta \Gamma_{\alpha \beta} \gamma_\beta \chi_\alpha\chi_\beta. \label{chis}
\end{align}

 
\section{Outlier eigenvalues}\label{sec:outliers}
\subsection{Setup and general relations}
 We now wish to find the outliers that arise when we perturb the random matrix $\mathbf{m}$ by introducing a finite mean. That is, we are now interested in the matrix $\mathbf{m}'\equiv \mathbf{m}+ \frac{1}{N}\boldsymbol{\mathrm{\upmu}}$, where the $N\times $N matrix $\boldsymbol{\upmu}$ accounts for the non-zero mean. 
Specifically $\mathbf{m}'$ has elements 
\be\label{eq:mshift}
m'^{\alpha \beta}_{ij} =m^{\alpha \beta}_{ij} + \frac{\mu_{\alpha \beta}}{N},
\ee
in particular
\be
\left(\boldsymbol{\upmu}\right)^{\alpha\beta}_{ij}=\mu_{\alpha \beta},
\ee
independently of $i$ and $j$. The elements of $\mathbf{m}$ have zero mean as before. So, the elements within each of the $4$ blocks [$(\alpha,\beta)=(u,u), (u,v), (v,u), (v,v)$] of the matrix $\mathbf{m}'$ have the same mean, $\mu_{\alpha\beta}/N$. Following the conventions of statistical physics, the mean is chosen to be proportional to $1/N$ for technical convenience. In the main paper, the mean is independent of $N$. We explain the mapping between the two setups in Section \ref{section:scaling}, where we also discuss general values of connectance $0<C<1$.

The problem amounts to finding the eigenvalues of the matrix $\mathbf{m}'$ -- i.e. $\mathbf{m}$ with a rank-$2$ perturbation $\boldsymbol{\upmu}$. We follow a procedure similar to Ref. \cite{orourke}. The bulk of the eigenvalue spectrum is unchanged by the low-rank perturbation. Therefore we look for additional eigenvalues that lie outside of the bulk of eigenvalues of  $\mathbf{m}$. Such an eigenvalue $\lambda$ must satisfy
\begin{align}
\det\left[\lambda \id_N   - \mathbf{m} - \frac{1}{N} \boldsymbol{\upmu} \right] &= 0 . \label{orourkestart}
\end{align}

Any outlier $\lambda$, is by definition, not an eigenvalue of the unperturbed matrix $\mathbf{m}$. Therefore the matrix $\id_N \lambda - \mathbf{m}$ is invertible. Hence,
\begin{align}
\det\left[ \id_N   - \frac{1}{N}(\lambda\id_N  - \mathbf{m})^{-1} \boldsymbol{\upmu} \right] &= 0 . \label{determinantsetup}
\end{align}
\subsection{Simplification to an effective 2-species problem}
\noindent Now we define the resolvent matrix $\mathbf{G}= (\lambda\id_N  - \mathbf{m})^{-1}$. 

\medskip

\noindent\underline{Case $N_u=N_v$:}\\
We suppose for now that there are the same numbers of prey species as predator species, i.e. $N_u = N_v$. We will relax this assumption once we have outlined the proof for this simpler case. Suppose we rearrange the rows and columns of the matrices so that the first species is a predator species, the second a prey species, the third again a predator species, and so on. The matrix $\boldsymbol{\upmu}$ then consists of  $N_u \times N_u$  block of size $2\times 2$ of the form $\boldsymbol{\upmu}^{(2)}\equiv \left(\begin{array}{cc} \mu_{uu} & \mu_{uv} \\ \mu_{vu} & \mu_{vv}\end{array}\right)$. Then, the matrix $\boldsymbol{\upmu}$ can be written in the form
\be
\boldsymbol{\upmu}= \underline{u}\, \underline{v}^T, \label{decomposedmu}
\ee
where $\underline{u}$ and $\underline{v}$ are vectors of length $N_u$. Each entry of $\underline{v}$ is given by the $2\times 2$ matrix $\boldsymbol{\upmu}^{(2)}$,
\be
\underline{v}^T = \left( \boldsymbol{\upmu}^{(2)}, \boldsymbol{\upmu}^{(2)}, \cdots \right),
\ee
The vector $\underline{u}$ is of the form
\be
\underline{u} = (\id_2 , \id_2 , \id_2 , \cdots)^T.
\ee
We next use Sylvester's determinant identity
\begin{align}
\det\left[\id_m +\mathbf{A}\,\mathbf{B} \right] = \det\left[\id_k +\mathbf{B}\,\mathbf{A} \right] ,\label{eq:syl}
\end{align}
valid for combinations of $m\times k$ matrices $\mathbf{A}$ and $k\times m$ matrices $\mathbf{B}$. We note that the matrices on the left-hand side of Eq.~(\ref{eq:syl}) are of size $m\times m$, and those on the right-hand-side are of size $k\times k$.

Using this identity, we find that $\det(\id_2   - \frac{1}{N}\underline{v}^T (\id_N \lambda - \mathbf{m})^{-1} \underline{u} ) = 0$. Thus we obtain
\begin{align}
\det\left[ \id_2   -  \boldsymbol{\upmu}^{(2)} \frac{1}{N}\sum_{ij} \mathbf{G}^{(ij)}(\lambda) \right] &= 0 ,\label{reduceddimensiondet}
\end{align}
where $\left(\mathbf{G}^{(ij)} \right)_{\alpha\beta}= \left((\id_N \lambda - \mathbf{m})^{-1}\right)^{\alpha \beta}_{ij}$ . 

\vspace{2em}

\noindent\underline{General case ($N_u$ not necessarily equal to $N_v$):}\\
We now wish to show that Eq.~(\ref{reduceddimensiondet}) also applies in the case $N_u \neq N_v$. This is done using the following identity for block matrices:
If $A$, $B$, $C$ and $D$ are matrices of sizes $m \times m$, $m \times k$, $k \times m$ and $k \times k$ respectively, then we have
\begin{align}
\det
\begin{bmatrix}
A & B \\
0 & D
\end{bmatrix}
= \det(A)\det(D) = 
\begin{bmatrix}
A & 0 \\
C & D
\end{bmatrix} , \nonumber \\
\begin{bmatrix}
A & B \\
C & D
\end{bmatrix} \begin{bmatrix}
A' & B' \\
C' & D'
\end{bmatrix} = \begin{bmatrix}
AA' + BC' & AB' +BD'  \\
CA' + DC' &  CB' + DD'
\end{bmatrix} . \label{blockmatrixtheorems}
\end{align}
First we note that the determinant $\det\left[ \id_N   - \frac{1}{N} \mathbf{G}  \boldsymbol{\upmu} \right]$ in Eq.~(\ref{determinantsetup}) is unchanged by adding extra rows and columns to the matrices in the following way: 
We add rows and columns $\boldsymbol{\upmu}$ to produce a square matrix $\boldsymbol{\upmu}_{N^\star}$ of dimension $N^\star = N + \lvert N_u - N_v \rvert$ which can be decomposed as in Eq.~(\ref{decomposedmu}). We then augment $\mathbf{G}$ to obtain
\begin{align}
\mathbf{G}_{N^\star} = \begin{bmatrix}
\mathbf{G} & 0 \\
0 & 0 
\end{bmatrix}.
\end{align}
Then, using the theorems in Eq.~(\ref{blockmatrixtheorems}), one can show
\begin{align}
\det\left[ \id_{N^\star}   - \frac{1}{N} \mathbf{G}_{N^\star}  \boldsymbol{\upmu}_{N^\star} \right] = \det\left[ \id_N   - \frac{1}{N} \mathbf{G}  \boldsymbol{\upmu} \right] .
\end{align}
Hence, we can apply Sylvester's determinant identity as before and arrive at Eq.~(\ref{reduceddimensiondet}).

\subsection{Proof that the resolvent matrix is diagonal}\label{sec:proofdiagonal}

We have shown that Eq.~(\ref{reduceddimensiondet}) holds when the numbers of predator and prey species are unequal. It remains now to find the averaged Green's functions $\frac{1}{N}\sum_{ij} \mathbf{G}^{(ij)}(\lambda)$. For this, we follow \cite{braymoore} and use the so-called `locator expansion' to show that the resolvent matrix is diagonal in the limit $N \to \infty$, i.e. that $G^{\alpha \beta}_{ij} \propto \delta_{ij}\delta_{\alpha\beta}$.  Letting $(G^{-1})^{\alpha\beta}_{ij} = (f^{-1})^\alpha_{i}\delta_{\alpha\beta} \delta_{ij}- m^{\alpha\beta}_{ij}$ (where in our case $(f^{-1})^\alpha_{i} = \lambda$), it follows that $\mathbf{m} = \mathbf{G}^{-1}(\mathbf{G}- \mathbf{f})\mathbf{f}^{-1}$, from which we obtain a Dyson equation $\mathbf{G} = \mathbf{f} + \mathbf{G}\mathbf{m}\mathbf{f}$. This allows us to construct a series for the resolvent
\begin{align}
G_{ij}^{\alpha \beta} = f^\alpha_i + f^\alpha_i m^{\alpha\beta}_{ij} f^\beta_j + f^\alpha_i m^{\alpha\gamma}_{ik}f^\gamma_k m^{\gamma\beta}_{kj}f^\beta_j + f^\alpha_i m^{\alpha\gamma}_{ik}f^\gamma_k m^{\gamma \delta}_{kl}f^\delta_l m^{\delta\beta}_{lj}f^\beta_j + \dots \,\,,\label{expansion}
\end{align} 
where we have omitted sums over repeated indices for brevity. We find that we have to evaluate terms such as $\sum_{k, \gamma} m^{\alpha\gamma}_{ik}f^\gamma_k m^{\gamma\beta}_{kj} \approx \sum_{k \gamma} f^\gamma_k \overline{m^{\alpha\gamma}_{ik} m^{\gamma\beta}_{kj} }$. This involves averages over products of Gaussian variables. Such terms can be simplified using standard Gaussian combinatorics 
\begin{align}
\overline{ m^{\alpha_1\beta_1}_{i_1i'_1}m^{\alpha_2\beta_2}_{i_2i'_2}\cdots m^{\alpha_n\beta_n}_{i_ni'_n} } = \sum \prod \overline{m^{\alpha_a\beta_a}_{i_ai'_a}m^{\alpha_b\beta_b}_{i_bi'_b}} ,
\end{align}
which is valid for even $n$ and is zero otherwise, and where $\sum \prod$ stands for the sum of the products of all possible combinations of pairs $\overline{m^{\alpha_a\beta_a}_{i_ai'_a}m^{\alpha_b\beta_b}_{i_bi'_b}} $. Since we have
\begin{align}
\overline{m^{\alpha_a\beta_a}_{i_ai'_a}m^{\alpha_b\beta_b}_{i_bi'_b}} = \frac{\sigma^2}{N}\delta_{i_a i_b}\delta_{i'_a i'_b} \delta_{\alpha_a \alpha_b}\delta_{\beta_a \beta_b}+ \frac{\sigma^2\Gamma_{\alpha_a \beta_a}}{N}\delta_{i_a i'_b}\delta_{i'_a i_b}\delta_{\alpha_a \beta_b}\delta_{\beta_a \alpha_b},
\end{align}
one can see that (since the only free indices are $i$, $j$, $\alpha$ and $\beta$ in Eq.~(\ref{expansion})) that $G_{ij}^{\alpha\beta} \propto \delta_{ij}\delta_{\alpha \beta}$. We therefore find that 
\begin{align}
\sum_{ij} G^{\alpha\beta}_{ij}(\lambda) \approx  \delta_{\alpha\beta}\sum_i G^{\alpha\beta}_{ii}(\lambda),
\end{align}
for large $N$. 
\subsection{Final expression for outliers}
Given that the resolvent matrix is approximately diagonal for large $N$, we only need to calculate the sum of diagonal elements $\frac{1}{N}\sum_{i} G_{ii}^{\alpha\alpha}$ in Eq.~(\ref{reduceddimensiondet}). This can be found from the quantities involved in the bulk calculation as follows. Consider the potential $\Phi(\{\omega_\alpha, \omega_\alpha^\star\})$ associated with the generalised resolvent $G(\{\omega_\alpha\}) = \frac{1}{N} \sum_{i\alpha} \overline{(\boldsymbol{\upomega} -\mathbf{m} )^{-1}}$, where $(\boldsymbol{\upomega})_{ij}^{\alpha\beta}= \omega_\alpha \delta_{\alpha\beta}\delta_{ij}$. We know that at the saddle point
\begin{align}
\chi_\alpha \equiv i w^\star_\alpha = \frac{1}{\gamma_\alpha}\frac{\partial \Phi(\{ \omega_\alpha, \omega_\alpha^\star\})}{\partial \omega_\alpha}.
\end{align}
The potential is defined by $\Phi(\{ \omega_\alpha, \omega_\alpha^\star\}) = -\frac{1}{N} \overline{\ln \det \left[ (\boldsymbol{\upomega} -\mathbf{m} )^\dagger(\boldsymbol{\upomega} -\mathbf{m} )\right]}$. Now, differentiating with respect to $\omega_\alpha$ and using Jacobi's formula for the derivative of a determinant, we have
\begin{align}
\frac{\partial \Phi(\{ \omega_\alpha, \omega_\alpha^\star\})}{\partial \omega_\alpha} = \frac{1}{N}\mathrm{Tr}\left[ (\boldsymbol{\upomega} -\mathbf{m} )^{-1}\frac{\partial(\boldsymbol{\upomega} -\mathbf{m} )}{\partial \omega_\alpha}\right]  = \frac{1}{N}\sum_{i}G^{\alpha\alpha}_{ii}(\{ \omega_\alpha, \omega_\alpha^\star\}) .
\end{align}
Now setting $\omega_\alpha = \omega$, we see from Eq.~(\ref{reduceddimensiondet}) that the outliers $\lambda$ obey
\begin{align}
\det\left[ \id_2   -  \boldsymbol{\upmu}^{(2)}\,\boldsymbol{\upgamma}\, \boldsymbol{\upchi}(\lambda)\right] &= 0 , \nonumber \\
\boldsymbol{\upgamma}(\lambda) = \begin{bmatrix}
\gamma_u & 0 \\
0 & \gamma_v
\end{bmatrix}, \,\,\,\,\,
\boldsymbol{\upchi}(\lambda) =   \begin{bmatrix}
\chi_u & 0 \\
0 & \chi_v
\end{bmatrix} .
\end{align}
This can be written as
\begin{align}
\left[ \gamma_u\mu_{uu} - \frac{1}{\chi_u(\lambda)}\right] \left[\gamma_v \mu_{vv} - \frac{1}{\chi_v(\lambda)}\right] - \gamma_u\gamma_v\mu_{uv}\mu_{vu} = 0.\label{simult1}
\end{align}
The quantities $\chi_\alpha(\lambda)$ satisfy [c.f. Eq.~(\ref{chis})]
\begin{align}
- 1 &=  - (\lambda + d_u) \chi_u +  \sigma^2 \Gamma_{u}\gamma_u \chi_u^2  +  \sigma^2 \Gamma_{uv}  \gamma_v\chi_v \chi_u , \nonumber \\
- 1 &= -  (\lambda + d_v) \chi_v  + \sigma^2 \Gamma_{uv}\gamma_u\chi_u \chi_v + \sigma^2 \Gamma_{v} \gamma_v\chi_v^2 . \label{simult2}
\end{align}
Note that not all solutions of this set of equations are valid. At the saddle point, the order parameter $u = \frac{1}{N}\sum_{i \alpha} z_i^{\alpha\star} z_i^\alpha$ must satisfy $u\geq 0$, and we have $u = (1 + \sigma^2 u) \sum_{\alpha}\gamma_\alpha \vert\chi_\alpha\vert^2$. This means that we must have
\begin{align}
\sum_{\alpha} \gamma_\alpha \vert\chi_\alpha(\lambda)\vert^2 < \frac{1}{\sigma^2}\label{cond}
\end{align}
 in order for the set $(\lambda, \{\chi_\alpha(\lambda)\})$ to be a valid solution to the combination of Equations (\ref{simult1}) and (\ref{simult2}).
  
Eqs.~(\ref{simult1}) and (\ref{simult2}) can be solved numerically for $\lambda$, subject to the constraint in (\ref{cond}). Note that when the left-hand side is equal to the right-hand side in the inequality (\ref{cond})  the point $\lambda$ in question is inside the bulk region. 

\section{Scaling with $N$ and connectance $C$}\label{section:scaling}
There are two key differences between the random matrices used in Sections \ref{section:support} and \ref{sec:outliers}  [c.f. Eq.~(\ref{eq:mstat})] and those in the main text [see Methods section]. These are: (a) The model parameters are rescaled by factors of $N$;  (b) Not all species interact with all others in the main text (i.e. the connectance can take values smaller than one, $0<C<1$). We now describe how to relate the results of the previous sections to the setup in the main text.
\subsection{Rescaling with $N$}
In this Supplement, we used the statistics given in Eqs.~(\ref{eq:mstat}) and (\ref{eq:mshift}) for the entries of the random matrix $\mathbf{m}$. These are
 \begin{align}
\overline{ m^{\alpha \beta}_{ij}} &=  \frac{\mu_{\alpha \beta}}{N} ,  \nonumber \\
\overline{( m^{\alpha \beta}_{ij})^2}- \left(\mu_{\alpha \beta}\right)^2 &= \frac{\sigma^2}{N},   \nonumber \\
\overline{ (m^{\alpha \beta}_{ij}- \mu_{\alpha \beta}) (m^{\alpha' \beta'}_{ji}- \mu_{\alpha' \beta'})} &= \frac{\Gamma_{\alpha \beta \alpha' \beta'} \sigma^2}{N}.
\end{align}
 
The statistics used in Eq. (4) and (5) of the the main paper (and also given in Eq.~(\ref{rmstats}) of this Supplement) are different. They can be obtained from the above by making the replacements $\mu_{\alpha\beta} \to N \mu_{\alpha\beta}$ and $\sigma^2 \to N \sigma^2$. The results in the main paper have been obtained from those in Sections \ref{section:support}, \ref{sec:outliers} and \ref{section:stability} by making this replacement. 

The theoretical predictions for stability or instability are formally derived in the limit $N\to\infty$, but in practice they are accurate for finite values of $N$, provided $N$ is not too small. Sizes of $N\approx 50$ are often sufficient to obtain satisfactory agreement (see Fig. \ref{fig:pairofoutliers} below for an example). The spectra in Fig. 2 of the main paper are obtained numerically from matrices of size $750\times 750$, and can be seen to agree very well with the results of the mathematical theory.

\subsection{Taking into account connectance $C$}
A connectance $C<1$ between species can be accounted for by making a similar replacement of $\sigma^2$ and $\mu_{\alpha\beta}$ to the one described above. We now describe how to see this.

In order to find the potential $\Phi(\omega)$ (which contains all the information about the eigenvalue spectrum and could in principle also be used to find the outliers \cite{edwardsjones}) defined in Eq.~(\ref{potential}), one needs to carry out an average over the ensemble of random matrices, see e.g. the step from Eq.~(\ref{HStransformed}) to Eq.~(\ref{averaged}). We now wish to take into account that any one pair of species only interacts with probability $C$. We then obtain [c.f. Eq.~(\ref{averagedeterminant})]
 
\begin{align}
&\int \prod_{ij\alpha\beta}\left(d m^{\alpha \beta}_{ij} \right) P(\{m_{ij}^{\alpha\beta}\})\det\left[ (\id_N\omega^\star - \mathbf{m}^T )( \id_N\omega - \mathbf{m} )\right]^{-1}\nonumber \\
 =& \int \prod_{ij\alpha\beta}\left(d m^{\alpha \beta}_{ij} \right)\det\left[ (\id_N\omega^\star - \mathbf{m}^T )( \id_N\omega - \mathbf{m} )\right]^{-1} \nonumber \\
& \Bigg\{(1-C)\delta(m_{ij}^{\alpha\beta})\delta(m_{ji}^{\beta\alpha}) \nonumber \\
&+ C\sqrt{\frac{N}{2 \pi \sigma^2 (1-\Gamma_{\alpha\beta}^2)}} \exp\left[-\frac{N}{ \sigma^2(1-\Gamma_{\alpha\beta}^2)}\left[(m^{\alpha \beta}_{ij} - \mu_{\alpha\beta})^2   - \Gamma_{\alpha \beta} (m^{\alpha \beta}_{ij}-\mu_{\alpha\beta})(m^{ \beta\alpha}_{ji}-\mu_{\beta\alpha}) \right] \right] \Bigg\} .
\end{align}
Upon evaluating the integral, one obtains
\begin{align}
&\exp\left[ -N \Phi(\omega) \right] = \int \prod_{i\alpha} \left( \frac{d^2z^\alpha_i d^2y^\alpha_i}{2 \pi^2}\right) \exp\left[ - \sum_{i\alpha} y^{\alpha\star}_i y^\alpha_i + \epsilon z^{\alpha\star}_i z^\alpha_i \right]\nonumber \\
\times & \exp\left[ -i \sum_{i \alpha} z^{\alpha\star}_i y^\alpha_i  (\omega^\star + d_\alpha) + z^\alpha_i y^{\alpha\star}_i ( \omega + d_\alpha)  \right]\nonumber \\
\times & \prod_{i,j,\alpha,\beta}\Bigg\{(1-C)+ C\exp\Bigg[\frac{i}{N} \mu_{\alpha\beta} \left(z_i^{\beta\star}y_j^\alpha  + z_i^{\beta}y_j^{\alpha\star}  \right)\nonumber \\
&- \frac{\sigma^2}{2N} \left[(z_i^{\alpha\star} y^\beta_j + z^\alpha_i y_j^{\beta\star})^2 + \Gamma_{\alpha \beta} (z_i^{\alpha\star} y^\beta_j + z^\alpha_i y_j^{\beta\star})(z_j^{\beta\star} y^\alpha_i + z^\beta_j y_i^{\alpha\star})\right] \Bigg]\Bigg\} . \label{phiwithconnectance}
\end{align}
If we set $C=1$, we recover the term we would normally obtain with all species connected
\begin{align}
&\exp\left[ -N \Phi(\omega) \right] = \int \prod_{i\alpha} \left( \frac{d^2z^\alpha_i d^2y^\alpha_i}{2 \pi^2}\right) \exp\left[ - \sum_{i\alpha} y^{\alpha\star}_i y^\alpha_i + \epsilon z^{\alpha\star}_i z^\alpha_i \right]\nonumber \\
\times & \exp\left[ -i \sum_{i \alpha} z^{\alpha\star}_i y^\alpha_i  (\omega^\star + d_\alpha) + z^\alpha_i y^{\alpha\star}_i ( \omega + d_\alpha)  \right]\nonumber \\
\times & \exp\Bigg[\frac{i}{N} \sum_{i,j,\alpha,\beta} \mu_{\alpha\beta} \left(z_i^{\beta\star}y_j^\alpha  + z_i^{\beta}y_j^{\alpha\star}  \right)\nonumber \\
&- \frac{\sigma^2}{2N} \sum_{i,j,\alpha,\beta} (z_i^{\alpha\star} y^\beta_j + z^\alpha_i y_j^{\beta\star})^2 + \Gamma_{\alpha \beta} (z_i^{\alpha\star} y^\beta_j + z^\alpha_i y_j^{\beta\star})(z_j^{\beta\star} y^\alpha_i + z^\beta_j y_i^{\alpha\star}) \Bigg] . \label{phiwithoutconnectance}
\end{align}
For individual sets of $i$, $j$, $\alpha$ and $\beta$, and assuming $N\gg 1$, the expression in the curly brackets in Eq.~(\ref{phiwithconnectance}) can be expanded in orders of $1/N$ to obtain
\begin{align}
&\Bigg\{(1-C)+ C\exp\Bigg[\frac{i}{N} \mu_{\alpha\beta} \left(z_i^{\beta\star}y_j^\alpha  + z_i^{\beta}y_j^{\alpha\star}  \right)\nonumber \\
&- \frac{\sigma^2}{2N} (z_i^{\alpha\star} y^\beta_j + z^\alpha_i y_j^{\beta\star})^2 + \Gamma_{\alpha \beta} (z_i^{\alpha\star} y^\beta_j + z^\alpha_i y_j^{\beta\star})(z_j^{\beta\star} y^\alpha_i + z^\beta_j y_i^{\alpha\star}) \Bigg]\Bigg\} \nonumber \\
&\approx 1 + \frac{i}{N} C\mu_{\alpha\beta} \left(z_i^{\beta\star}y_j^\alpha  + z_i^{\beta}y_j^{\alpha\star}  \right)\nonumber \\
&- \frac{C\sigma^2}{2N} (z_i^{\alpha\star} y^\beta_j + z^\alpha_i y_j^{\beta\star})^2 + \Gamma_{\alpha \beta} (z_i^{\alpha\star} y^\beta_j + z^\alpha_i y_j^{\beta\star})(z_j^{\beta\star} y^\alpha_i + z^\beta_j y_i^{\alpha\star}) \nonumber \\
&\approx\exp\Bigg[\frac{i}{N} C\mu_{\alpha\beta} \left(z_i^{\beta\star}y_j^\alpha  + z_i^{\beta}y_j^{\alpha\star}  \right)\nonumber \\
&- \frac{C\sigma^2}{2N} \left[(z_i^{\alpha\star} y^\beta_j + z^\alpha_i y_j^{\beta\star})^2 + \Gamma_{\alpha \beta} (z_i^{\alpha\star} y^\beta_j + z^\alpha_i y_j^{\beta\star})(z_j^{\beta\star} y^\alpha_i + z^\beta_j y_i^{\alpha\star})\right] \Bigg] . \label{expandedexponentialconnectance}
\end{align}
Comparing Eq.~(\ref{phiwithconnectance}) [with Eq.~(\ref{expandedexponentialconnectance}) substituted in] to Eq.~(\ref{phiwithoutconnectance}), we see that the consequence of introducing a connectance $0<C<1$ is to scale the system parameters in the following way 
\begin{align}
\mu_{\alpha\beta} &\to C \mu_{\alpha\beta} , \nonumber \\
\sigma^2 &\to C \sigma^2 . 
\end{align}
\subsection{Combined effect}
In summary, the scaling that one needs to perform in order to obtain the results in the main text from those in Sections \ref{section:support} and \ref{sec:outliers} is given by 
\begin{align}
\mu_{\alpha\beta} &\to NC \mu_{\alpha\beta} , \nonumber \\
\sigma^2 &\to NC \sigma^2 . 
\end{align}
All simulation results in the main text were performed using the model as defined in Section \ref{sec:model} (which is the model defined in the main text). We note that Eqs.~(\ref{chis}) and (\ref{simult1}-\ref{cond}) become respectively
\begin{align}
\sum_\alpha \gamma_\alpha \left\vert \chi_\alpha \right\vert^2 - \frac{1}{c} &= 0, \nonumber \\
  - (\omega_x + i\omega_y + d_\alpha  +q^2 D_\alpha )\chi_\alpha + c\sum_\beta \Gamma_{\alpha \beta} \gamma_\beta \chi_\alpha\chi_\beta +1 &=0,
\end{align}
and 
\begin{align}
\left[ \gamma_u NC \mu_{uu} - \frac{1}{\chi_u(\omega)}\right] \left[\gamma_v NC \mu_{vv} - \frac{1}{\chi_v(\omega)}\right] - (NC)^2\gamma_u\gamma_v\mu_{uv}\mu_{vu} &= 0,\nonumber \\
  - (\omega + d_u + q^2 D_u) \chi_u +  c \Gamma_{u}\gamma_u \chi_u^2  +  c \Gamma_{uv}  \gamma_v\chi_v \chi_u  &=- 1, \nonumber \\
 -  (\omega + d_v+ q^2 D_v) \chi_v  + c \Gamma_{uv}\gamma_u\chi_u \chi_v + c \Gamma_{v} \gamma_v\chi_v^2 & = - 1,  \nonumber \\
\sum_{\alpha} \gamma_\alpha \vert\chi_\alpha(\omega)\vert^2 &< \frac{1}{c},
\end{align}
where $c = NC\sigma^2$.
\vspace{2em}

\section{Stability criteria}\label{section:stability}
\noindent We class the instabilities in our system into two types:

\begin{itemize}
\item[(1)] The bulk of the eigenvalue distribution may cross the imaginary axis. This is the type of instability May identified \cite{may}.
\item[(2)] One of the outlier eigenvalues may cross the imaginary axis, with the bulk staying in the negative half-plane.  Instabilities of this type are discussed for non-spatial systems for example in \cite{allesinatang2}.
 \end{itemize}
Below, we provide analytical criteria for each of these instabilities. Crucially, we note that an instability of the second kind can be introduced to an otherwise stable system by the inclusion of diffusion. That is, for a finite band of wavenumbers, the spatial system is unstable where the non-spatial system would be stable. 

First, we provide criteria for the stability of the non-spatial system, and then we characterise the spatial instability. In the following the model parameters $\sigma^2$ and $\mu_{\alpha\beta}$ are scaled in the same way as in the main text, and we allow for general connectance $0<C<1$.
\subsection{Instabilities of the non-spatial system}
\subsubsection*{Instability due to bulk eigenvalues, and relation to May's bound}
 
\noindent\underline{Mathematical condition for instability:}\\
The transition occurs when the boundary of the bulk eigenvalue spectrum first hits the imaginary axis. We generally observe that the bulk of the eigenvalue spectrum is a convex set. Since the matrix we are dealing with is real, we also know that the spectrum is symmetric with respect to the real axis. As a consequence we find that point of first contact is at $\omega = 0$. We note that, from the correspondence with the resolvent shown in the previous section, this means that the variables $\chi_\alpha(0)$ must be real. In order for the system to be stable, $\lambda = 0$ must necessarily be outside the bulk (otherwise there are positive real eigenvalues). Hence, a necessary condition for stability is 
\begin{align}
\sum_\alpha \gamma_\alpha  \chi_\alpha^2 &< \frac{1}{c} , \label{bulkstabilitycondition}
\end{align}
We have used Eq.~(\ref{cond}) with the rescaling $ \mu_{\alpha\beta} \to C \mu_{\alpha\beta}$, and $\sigma^2 \to C \sigma^2$, and  we recall that the complexity parameter is given by $c=NC\sigma^2$. The quantities $\chi_\alpha$ satisfy
\begin{align}\label{eq:bulk2}
-1 &=  - d_\alpha \chi_\alpha +c \sum_\beta \Gamma_{\alpha \beta} \gamma_\beta \chi_\alpha\chi_\beta.
\end{align}
The equations above cannot be solved analytically in general, but a numerical solution can be obtained. 
\medskip

\noindent\underline{Special cases:}\\
We can make the connection with May's original criterion \cite{may} in several special cases. First, consider the case where $\gamma_v=0$ -- that is, we do not distinguish between predator and prey species, and consider a general community of $N=N_u$ species (as May did). In this case, Eqs.~(\ref{bulkstabilitycondition}) and (\ref{eq:bulk2}) yield
\begin{align}
\sqrt{c} (1 + \Gamma_{uu}) < d_u .\label{bulkstability1}
\end{align}
We recover May's criterion for $\Gamma_{uu} = 0$ and $d_u = 1$. Secondly, the criterion for stability also simplifies greatly when the prey and predator sub-communities have uncorrelated internal interactions so that $\Gamma_{u} = \Gamma_v = 0$, the same number of species (i.e., $\gamma_u = \gamma_v = \frac{1}{2}$), and equal death rates $d_u = d_v = d$. In this case, Eqs.~(\ref{bulkstabilitycondition}) and (\ref{eq:bulk2}) give
\begin{align}
\sqrt{2 c}(1 + \frac{1}{4}\Gamma_{uv}) < d .\label{bulkstability2}
\end{align}
Similarly, if $\Gamma_{uv} = 0$ and $\Gamma_u = \Gamma_v = \Gamma$, we obtain 
\begin{align}
\sqrt{2 c} (1 + \frac{1}{4}\Gamma)< d .\label{bulkstability3}
\end{align}
A further special case is a situation in which all interactions are uncorrelated ($\Gamma_{\alpha\beta} = 0$ for all $\alpha$ and $\beta$). Here we find 
\begin{align}
c \sum_{\alpha} \frac{\gamma_\alpha}{d_\alpha^2} < 1 .\label{bulkstability4}
\end{align}
From these examples, we get a general idea of how each of the model parameters affect the onset of instability. The intra-species interaction $I_\alpha = -d_\alpha$ is required to be negative in order to have stability. The more negative $I_\alpha$ is, the more stable the equilibrium is. The equilibrium can also be stabilised by asymmetry in the interaction coefficients. That is, the lower the parameters $-1\leq \Gamma_{\alpha\beta}\leq 1$ are, the more stable the equilibrium. Finally and crucially, as per May's conclusion, a lower complexity $c= NC\sigma^2$ contributes to making the equilibrium more stable in all the special cases outlined above. 

\subsubsection*{Instability due to outlier eigenvalues}
\noindent\underline{Mathematical condition for instability:}\\
If we suppose that the bulk eigenvalues remain to the left of the imaginary axis, an instability may still occur if one or more of the outlier eigenvalues take on a positive real part. This may occur in two possible ways: (I) A single outlier with no imaginary part may cross at $\lambda = 0$; or (II) A pair of complex conjugate outliers may cross the imaginary axis. The point at which an instability of the first kind occurs can be obtained by substituting $\lambda = 0$ in Eqs.~(\ref{simult1}) and (\ref{simult2}). Re-scaling $\sigma^2$ and $\mu_{\alpha\beta}$ to match the conventions in the main paper, we obtain
\begin{align}
\left[ NC\gamma_u\mu_{uu} - \frac{1}{\chi_u}\right] \left[NC\gamma_v \mu_{vv} - \frac{1}{\chi_v}\right] - (NC)^2\gamma_u\gamma_v\mu_{uv}\mu_{vu} &= 0, \nonumber \\
  c \Gamma_{u}\gamma_u \chi_u^2  +  c\Gamma_{uv}  \gamma_v\chi_v \chi_u  -  d_u \chi_u +1 &= 0  , \nonumber \\
 c \Gamma_{uv}\gamma_u\chi_u \chi_v + c\Gamma_{vv} \gamma_v\chi_v^2  - d_v \chi_v  +1 &= 0. \label{criticaloutlierreal}
\end{align}

The model parameters at which an instability of type (II) above occurs are found as follows: We evaluate the outlier eigenvalues by solving Eqs.~(\ref{simult1}) and (\ref{simult2}) subject to Eq.~(\ref{cond}), and we carry out the re-scaling  $\mu_{\alpha\beta} \to NC \mu_{\alpha\beta}$, $\sigma^2 \to NC \sigma^2$. We do this for varying model parameters until one of the outliers gains a positive real part. This way, we ensure that we spot any complex conjugate pairs crossing the imaginary axis. The sloped lines in Fig. 3 in the main text, for example, were generated in this way. 
\medskip

\begin{figure}[H]
	\centering 
	\includegraphics[scale = 0.2]{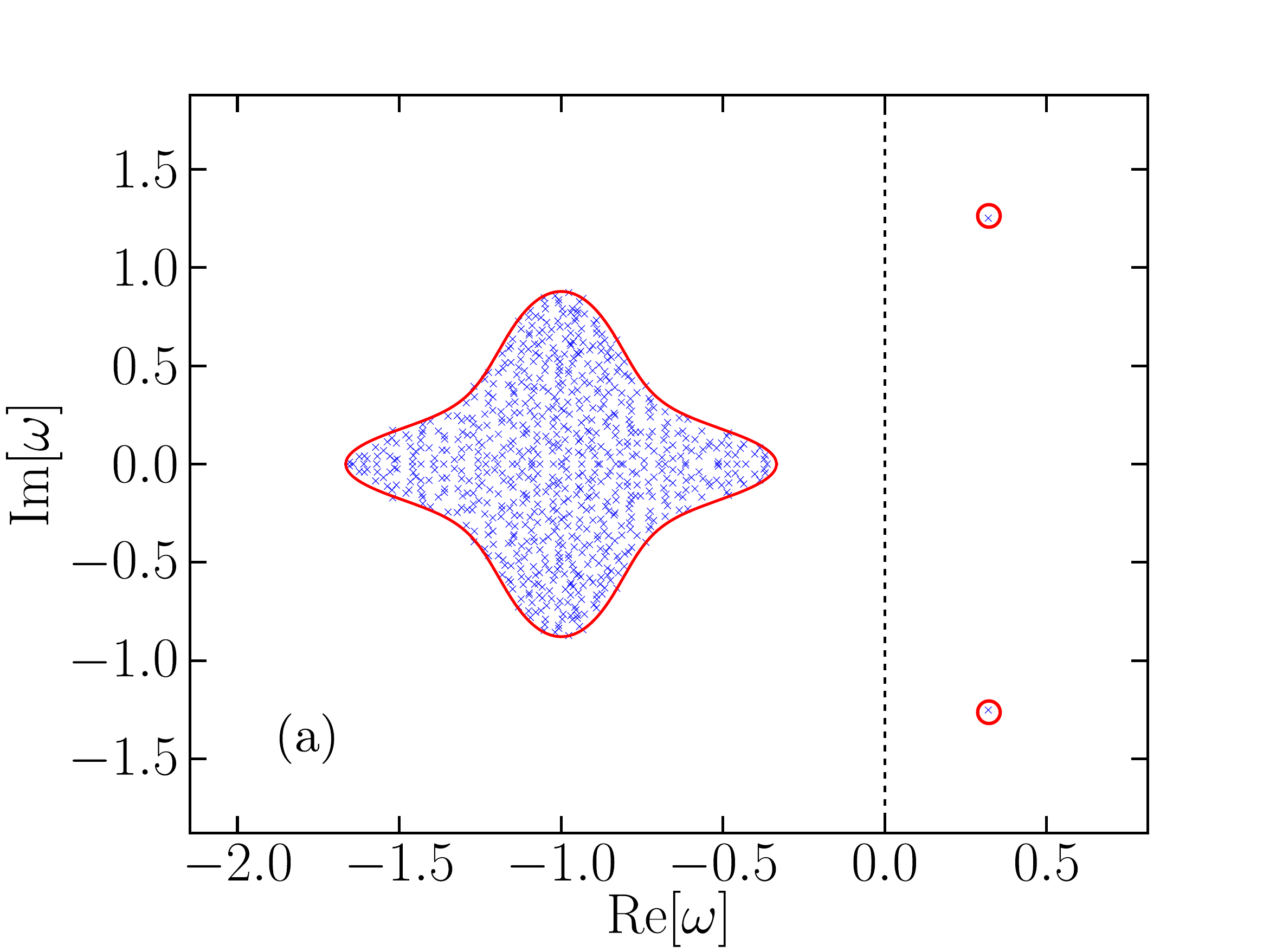}
    \includegraphics[scale = 0.2]{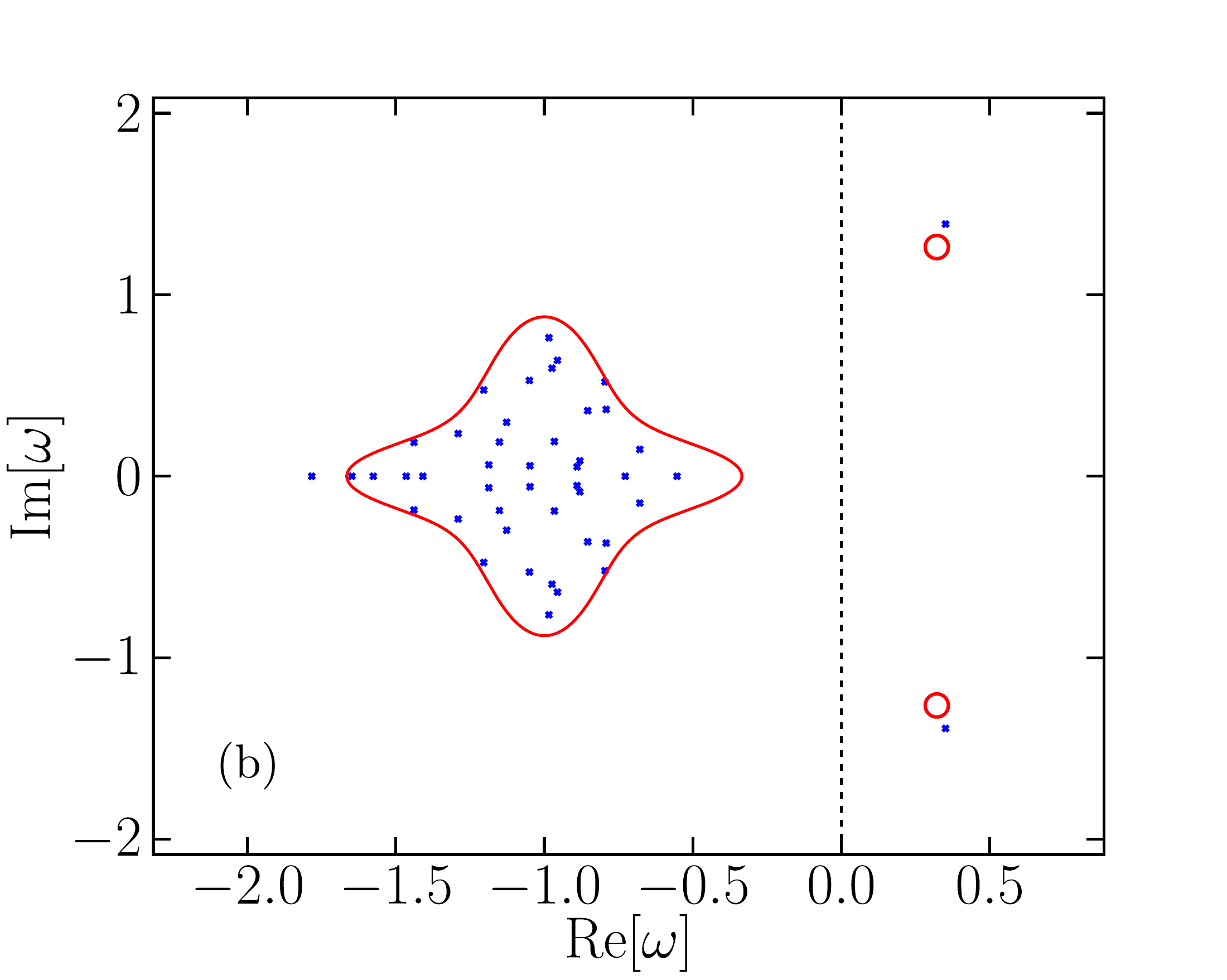}
	\caption{ {\bf Further example of instability due to outlier eigenvalues crossing the imaginary axis.} Eigenvalues obtained numerically from computer-generated random matrices are shown as blue crosses. Red lines are the theoretical predictions for the boundary for the bulk region, and red hollow circles the predictions for the locations of any outliers. The model parameters are the same as in Fig. 2 (a) (with $q=0$) of the main text, with the exception of the means $\mu_{\alpha\beta}$ which are now given by $CN\mu_{uu} = 4$, $CN\mu_{uv} = -4$, $CN\mu_{vu}= 4$, $CN\mu_{vv} = -2$, where the scaling of the system parameters is the same as in the main text. In the left-hand panel we use $N = 750$, on the right $N = 45$, keeping the products $CN\mu_{\alpha\beta}$ constant.  }
	\label{fig:pairofoutliers}
\end{figure}

\noindent \underline{Example:}\\
\noindent Fig. 2 in the main text shows an example in which a single outlier crosses the imaginary axis. In Fig. \ref{fig:pairofoutliers} we show a case in which a complex-conjugate pair of outliers has crossed the axis. The model parameters are the same as in Fig. 2 of the main text, with the exception of the $\mu_{\alpha\beta}$. This shows that altering the means of the matrix elements can turn an instability of type (I) above into one of type (II). We note that the bulk of the eigenvalue spectrum is the same in Fig. 2a and in Fig.~\ref{fig:pairofoutliers}. This confirms that the bulk spectrum is not affected by changes in the means $\mu_{\alpha\beta}$.

\noindent\underline{Special cases:}\\
We can make analytical progress in the same set of special cases as in the previous section. First, consider the case $\gamma_u= 0$ (we could similarly use $\gamma_v= 0$). Then, the outlier eigenvalue [from Eq.~(\ref{simult2})] is real and given by 
\begin{align}
\lambda = NC\mu_{vv} + \frac{\Gamma_{v} c}{NC\mu_{vv}} .
\end{align}
For $C=1$ this reduces further to results mentioned in \cite{orourke, edwardsjones}. A necessary condition for stability is then 
\begin{align}\label{eq:sc1}
NC\mu_{vv} + \frac{\Gamma_{v} c}{NC\mu_{vv}} <0.
\end{align}

In the case where the prey and predator communities have uncorrelated internal interactions so that $\Gamma_{u} = \Gamma_v = 0$, the same numbers of species $\gamma_u = \gamma_v = \frac{1}{2}$, and the same death rates $d_u = d_v = d$ and $\mu_{uu }= -\mu_{vv}$,  the outlier eigenvalues are given by
\begin{align}
\lambda = -d \pm \frac{{(NC)^2(\mu_{uu}^2 + \mu_{uv}\mu_{vu}) + 2\Gamma_{uv}c}}{2NC\sqrt{\mu_{uu}^2 + \mu_{uv}\mu_{vu}}} .
\end{align}
If $\mu_{uu}^2 + \mu_{uv}\mu_{vu}<0$, there is no real solution for $\lambda$ and there is guaranteed to be no instability, so long as $d>0$. We note that complex conjugate pairs of outliers are made possible only by the random matrix having trophic (block) structure. 

Similarly, in the case  $\Gamma_{uv} = 0$ and $\Gamma_u = \Gamma_v = \Gamma$, the eigenvalues are given by 
\begin{align}\
\lambda = -d \pm \frac{{(NC)^2(\mu_{uu}^2 + \mu_{uv}\mu_{vu}) + 2\Gamma c}}{2 NC \sqrt{\mu_{uu}^2 + \mu_{uv}\mu_{vu}}} . \label{eigenvaluenoxy}
\end{align}
In both special cases, the more negative the quantity $\mu_{uv}\mu_{vu}$ is, the more stable the equilibrium. We note that $-\mu_{uv}\mu_{vu}$ is a measure for the `strength' of predation (taking into account both the effects on prey and predators). So, predation is a stabilising influence. Further, asymmetry in the interaction coefficients has a stabilising effect. 

A further special case is $\Gamma_{\alpha\beta} =0$ for all $\alpha$ and $\beta$. Here, we obtain
\begin{align}\label{eq:sc3}
\left[ NC\gamma_u\mu_{uu} - d_u\right] \left[NC\gamma_v \mu_{vv} - d_v\right] - (NC)^2 \gamma_u\gamma_v\mu_{uv}\mu_{vu} &> 0 .
\end{align}
Once again, the more negative the quantity $\mu_{uv}\mu_{vu}$ is, the more stable the equilibrium.

\subsection{Dispersal-induced instability}\label{sec:dii}
So far we have considered only a non-spatial ecosystem in this Supplement, i.e. we have set $q=0$ in Eq.~(\ref{processq}), or equivalently $D_u = D_v =0$. We now address instabilities induced by dispersal. The additional terms arising from a non-zero dispersal term can be absorbed into a re-definition of $d_\alpha$,
\begin{align}\label{eq:dqD}
d_\alpha \to d_\alpha + q^2 D_\alpha .
\end{align}
This can be seen directly from Eq.~(\ref{eq:mdef2}). With the replacement in (\ref{eq:dqD}), we can then use the stability criteria derived in Section~\ref{section:stability} to obtain criteria for dispersal-induced instability.

\subsubsection*{The instability due to bulk eigenvectors is unaffected by dispersal}
With the inclusion of dispersal Eqs. (\ref{bulkstabilitycondition}) and (\ref{eq:bulk2}) become
 \begin{align}
\sum_\alpha \gamma_\alpha  \chi_\alpha^2 &< \frac{1}{c} , \nonumber \\
-1 &=  -( d_\alpha+ q^2 D_\alpha) \chi_\alpha +c\sum_\beta \Gamma_{\alpha \beta} \gamma_\beta \chi_\alpha\chi_\beta .
\end{align} 
This is the condition for the stability of the Fourier mode with wavenumber $q$.

Although it is difficult to prove analytically, we find in general that the inclusion of dispersal serves only to shift the bulk of the eigenvalues in the negative real direction, never in the positive. This is seen to be the case in Fig. 2 in the main text. This can also be seen to be true in the special cases we have discussed already -- increasing the magnitude of $d_\alpha$ (which is effective what one does by introducing non-zero $q$) in Eqs. (\ref{bulkstability1})-(\ref{bulkstability4}) serves only to expand the range of parameters for which the equilibrium is stable.

\subsubsection*{Turing-type instability due to outlier eigenvalues}
The Turing instability is a `stationary' instability \cite{yadavhorsthemke} which exhibits no oscillations. This means that the eigenvalue crossing the imaginary axis has no imaginary part (unlike the example given in Fig.~\ref{fig:pairofoutliers}). The point in parameter space at which an instability of this type occurs in the non-spatial case ($q=0$) is given by the solution of Eq.~(\ref{criticaloutlierreal}). Using this, and the replacement $d_\alpha \to d_\alpha + q^2 D_\alpha $ to account for $q\neq 0$, we arrive at the condition for stability with respect to disturbances of non-zero wavenumber $q$:
\begin{align}
P(q^2) \equiv \left[ NC\gamma_u\mu_{uu} - \frac{1}{\chi_u}\right] \left[NC\gamma_v \mu_{vv} - \frac{1}{\chi_v}\right] - (NC)^2\gamma_u\gamma_v\mu_{uv}\mu_{vu} &>0, \nonumber \\
c \Gamma_{u}\gamma_u \chi_u^2  +  c\Gamma_{uv}  \gamma_v\chi_v \chi_u -  (d_u + D_u q^2) \chi_u +1 &= 0  , \nonumber \\
c \Gamma_{uv}\gamma_u\chi_u \chi_v + c \Gamma_{vv} \gamma_v\chi_v^2  - (d_v + D_v q^2) \chi_v  +1 &= 0. \label{turinginst}
\end{align}
In order for the ecosystem to show a Turing instability this condition must be satisfied for $q=0$ (i.e., the system is stable with respect to perturbations with $q=0$), but violated for some non-zero $q$ (there are unstable modes with $q\neq 0$). That this can indeed be possible is exemplified by the special case where $\Gamma_{\alpha\beta} = 0$ for all $\alpha$ and $\beta$. In this case, the stability criterion for the Fourier mode $q$ becomes
\begin{align}
P_0(q^2) \equiv \left[ NC\gamma_u\mu_{uu} -  (d_u + D_u q^2)\right] \left[NC\gamma_v \mu_{vv} -(d_v + D_v q^2)\right] - (NC)^2\gamma_u\gamma_v\mu_{uv}\mu_{vu} &> 0 .
\end{align}
The expression for $P_0(q^2)$ is quadratic in $q^2$. It has a minimum at 
\begin{align}
q^2_{\mathrm{min}} = \frac{(NC\gamma_u\mu_{uu}-d_u)D_v+ (NC\gamma_v \mu_{vv} - d_v)D_u  }{2 D_u D_v}. 
\end{align}
If $P_0(0)>0$ and $P_0(q^2_{\mathrm{min}})<0$, then there is a Turing instability. This occurs when 
\begin{align}
\frac{D_v}{D_u} > \gamma_u\gamma_v\left[\frac{1}{\gamma_u \mu'_{uu}}\left( \sqrt{\mu'_{uu}\mu'_{vv}  -\mu_{uv} \mu_{vu}} + \sqrt{-\mu_{uv} \mu_{vu} } \right) \right]^2, \label{criticalratio}
\end{align}
where $NC\gamma_u\mu'_{uu}  = NC\gamma_u\mu_{uu} -d_u$ and $NC\gamma_v\mu'_{vv}  = NC\gamma_v\mu_{vv} -d_v$. This is very reminiscent of the criterion for Turing instability in more conventional reaction-diffusion systems \cite{crosshohenberg}. 

In the more general case, we need to find the minimum possible value of the expression on the left-hand side of the first relation in Eq.~(\ref{turinginst}). If the minimum value of $P(q^2)$ is negative (i.e. the condition in the first line of Eq.~(\ref{turinginst}) is violated), but the value of $P(0)$ is positive, then we say that there is a Turing (dispersal-induced) instability for this parameter set. 

Again, we first find the minimum value of $P(q^2)$ for a given set of model parameters. To do this, we differentiate $P(q^2)$ with respect to $q^2$, noting that the three relations in Eq.~(\ref{turinginst}) are coupled. This leads to the following set of simultaneous equations to find $P(q^2_{\mathrm{min}})$

\begin{align}
P'(q^2_{\mathrm{min}})  &=  NC\gamma_u\mu_{uu} \chi_u' \left[NC\gamma_v \chi_v\mu_{vv} - 1\right] +  NC\gamma_v \mu_{vv} \chi_v'\left[ NC\gamma_u\chi_u\mu_{uu} - 1\right] \nonumber \\
 &\,\,\,\,- (NC)^2\gamma_u\gamma_v\mu_{uv}\mu_{vu}\left[ \chi_u \chi_v' +  \chi_v \chi_u'\right] = 0 , \nonumber \\
0 &= 2c \Gamma_{u}\gamma_u \chi_u \chi_u'  +  c \Gamma_{uv}  \gamma_v\left(\chi_v \chi_u' + \chi_v' \chi_u\right)  -  (d_u + D_u q^2_{\mathrm{min}}) \chi_u' - D_u \chi_u   , \nonumber \\
0 &= c \Gamma_{uv}\gamma_u\left(\chi_v \chi_u' + \chi_v' \chi_u\right) + 2c \Gamma_{vv} \gamma_v\chi_v\chi_v'  - (d_v + D_v q^2_{\mathrm{min}}) \chi_v' - D_v\chi_v , \nonumber \\
0 &= c \Gamma_{u}\gamma_u \chi_u^2  +  c \Gamma_{uv}  \gamma_v\chi_v \chi_u  -  (d_u+ D_u q^2_{\mathrm{min}}) \chi_u +1  , \nonumber \\
0 &= c\Gamma_{uv}\gamma_u\chi_u \chi_v + c \Gamma_{vv} \gamma_v\chi_v^2  - (d_v+ D_v q^2_{\mathrm{min}}) \chi_v  +1 ,\label{turingsimulteqs}
\end{align}
where $\chi_\alpha' = \frac{\partial \chi_\alpha}{\partial q^2}$. One solves the above equations simultaneously to find $\chi_u$, $\chi_u'$, $\chi_v$, $\chi_v'$ and $q^2_{\mathrm{min}}$. One then evaluates 
\begin{align}
P(q^2_{\mathrm{min}}) &= \left[ NC\gamma_u\chi_u\mu_{uu} - 1\right] \left[NC\gamma_v \chi_v\mu_{vv} - 1\right] - (NC)^2\gamma_u\gamma_v\mu_{uv}\mu_{vu}\chi_u\chi_v.
\end{align}
So, one can find the minimum value of $P(q^2)$ for a particular parameter set and thus deduce whether or not there is dispersal-induced instability (i.e. whether the condition in Eq.~(\ref{turinginst}) is satisfied). For the purpose of constructing stability diagrams, one can instead set $P(q^2_{\mathrm{min}}) = 0$, equivalent to assuming that the outlier eigenvalue is just about to cross the imaginary axis ($\lambda=0$). One then solves for the critical model parameters instead which give $P(q^2_{\mathrm{min}}) = 0$. This can be used to produce solid sloped lines in Figs. 3b and 4 in the main text .

Although the above set of equations appears cumbersome, we note that they are at most linear in $\chi_u'$ and $\chi_v'$, so these can be solved for relatively easily. In practice, Eqs.~(\ref{turingsimulteqs}) reduce to three simultaneous equations for $q^2_{\mathrm{min}}$, $\chi_u$ and $\chi_v$.

\subsection{Dispersal alone cannot make an unstable equilibrium stable}\label{sec:dispersalalone}
In the model discussed thus far, there are no instances in which an unstable equilibrium of the non-spatial ecosystem becomes stable when dispersal is introduced. This is because a stable equilibrium in the spatial model must be stable with respect to disturbances for all values of $q$, including $q=0$ (which is equivalent to setting $D_\alpha=0$ and corresponds to the non-spatial case). So, if the non-spatial system is unstable, the $q=0$ mode in the spatial system must also be unstable.

With the introduction of spatial heterogeneity in the interaction coefficients however, this reasoning no longer applies. This is because spatial heterogeneity in the interaction coefficients introduces a non-trivial interdependence between the eigenvalue spectra for different Fourier modes $q$. That is, the spectra for different values of $q$ can no longer be calculated independently and, more specifically, the bulk of the $q=0$ spectrum may `shrink' as a result of this coupling. Further details can be found in Section \ref{sec:heterogeneity}. 
\section{Parameters used for figures in the main text}\label{sec:parameters}

\noindent {\bf Figure 2:}\\
\noindent $N = 750$, $C = 0.5$, $\gamma_u = 2/3$, $\gamma_v = 1/3$, $c = C N \sigma^2 = 0.375$, $d_u = 1$, $d_v = 1$, $\Gamma_u = 0.5$, $\Gamma_{uv} = -0.9$, $\Gamma_v = -0.5$, $CN\mu_{uu} = 3$, $CN\mu_{uv} = -3$, $CN\mu_{vu}= 3$, $CN\mu_{vv} = -1.5$, $D_u = 1$, $D_v = 5$.

\bigskip

\noindent {\bf Figure 3:}\\
\noindent  $\Gamma_{u} = 0.5$, $\Gamma_v = -0.5$, $\Gamma_{uv} = -0.9$, $D_u = 1$, $D_v = 5$, $d_u = d_v = 1$, $CN\mu_{uu} = 3$, $CN\mu_{vv} = -1.5$, $\gamma_u = 2/3$, $\gamma_v = 1/3$, $\mu_{uv} = -\mu_{vu}$. 

\bigskip

\noindent {\bf Figure 4:}\\
\noindent $\Gamma_{u} = 1$, $\Gamma_v = -1$, $\Gamma_{uv} = 0$, $D_u = 1$,  $d_u = d_v = 1$, $CN\mu_{uu} = CN\mu_{vu}= 3$, $CN\mu_{u} = -3$, $CN\mu_{vv} = -1.5$, $\gamma_u = 2/3$, $\gamma_v = 1/3$. 

\bigskip

\noindent {\bf Figure 5:}\\
\noindent $N = 100$, $\Gamma_{u} = 0.5$, $\Gamma_v = -0.5$, $\Gamma_{uv} = 0$, $D_u = 0.2$, $D_v = 20$, $\sqrt{CN}\sigma = 0.1$, $d_u = d_v = 0.1$, $CN\mu_{uu} = 1.2$, $CN\mu_{vu} = 2$, $CN\mu_{uv} = -2$, $CN\mu_{vv} = -0.8$, $a = 0.5$, $\gamma_u = 1/2$, $\gamma_v = 1/2$. In (a), $D_u, D_v=0$ and in (b) $D_u=0.2, D_v=20$

\clearpage
\part{Robustness against variations of the model}
In this part of the Supplement we consider several variations on the model presented previously. In Section~\ref{sec:universality}, we comment on the possibility of universality and show that our main results carry over to non-Gaussian random interaction matrices under relatively mild conditions. Further, while the model in the main text describes trophic structure in a statistical sense, we show in Section~\ref{sec:trophicenforce} that the formalism we develop can also be used when trophic interactions are enforced more strictly. Finally, in Section~\ref{sec:diagonal} we address the case in which the diagonal elements of the random interaction matrix are not uniform within blocks. To this end we extend the analytical calculation of the eigenvalue spectra in Part I. We use this to show that heterogeneity in the self-interaction or diffusion coefficients tends to broaden the eigenvalue spectrum, and hence promotes instability. In particular, the possibility of dispersal-induced instability persists in the face of such heterogeneity.

\section{Universality of the theory -- non-Gaussian random matrices}\label{sec:universality}
\subsection{Analytical demonstration of universality}
We have so far considered the interaction coefficients (the elements of the matrix $m_{ij}^{\alpha\beta}$) to be drawn from a Gaussian ensemble. 

We now demonstrate that our results apply to other distributions as well provided some fairly liberal conditions are fulfilled.

Suppose that the random matrix elements $m_{ij}^{\alpha\beta}$, which we take for now to have zero mean, are drawn from a non-Gaussian distribution. We imagine however that the statistics of this non-Gaussian distribution are the same as the previously specified Gaussian distribution up to the second moment and that higher order moments decay sufficiently quickly with $N$. What is meant by `sufficiently quickly' will emerge in the calculation that follows.

The first and second moments of the random matrix are 
\begin{align}
\overline{m^{\alpha \beta}_{ij}}&= 0,  \nonumber \\
\overline{ (m^{\alpha \beta}_{ij})^2}  &= \frac{\sigma^2}{N},   \nonumber \\
\overline{ m^{\alpha \beta}_{ij} m^{\alpha' \beta'}_{ji}}&= \frac{\Gamma_{\alpha \beta \alpha' \beta'} \sigma^2}{N} , \nonumber \\
m_{ii}^{\alpha \alpha} &= - d_\alpha. \label{rmstatsagain}
\end{align}
We begin once again with Eq.~(\ref{HStransformed}). The object of primary interest when performing the average over the ensemble of random interaction matrices is
\begin{align}
U_{ij}^{\alpha\beta} = \overline{\exp\left[ i  (z^{\alpha\star}_i y^\beta_j + z_i^\alpha y_j^{\beta\star}) m_{ij}^{\alpha\beta} +  i  (z^{\beta\star}_j y^\alpha_i + z_j^\beta y_i^{\alpha\star}) m_{ji}^{\beta\alpha} \right]}.
\end{align}
We now expand the exponential as a series to obtain
\begin{align}
U_{ij}^{\alpha\beta} = \sum_{r = 0}^\infty \sum_{s = 0}^\infty  \frac{i^{r+s}}{r!s!} (z^{\alpha\star}_i y^\beta_j + z_i^\alpha y_j^{\beta\star})^r  (z^{\beta\star}_j y^\alpha_i + z_j^\beta y_i^{\alpha\star})^s \overline{(m_{ij}^{\alpha\beta})^r (m_{ji}^{\beta\alpha})^s } . \label{expseries}
\end{align}
We now assume that the higher-order moments (such that $r+s>2$) are ${\cal O}(N^{-(1+\epsilon)})$ where $\epsilon>0$, as would be the case with the original Gaussian distribution. This means that all moments $r+s>0$ fall off with $N$ faster than $1/N$. Since we take into account only the leading-order contribution in $N^{-1}$, this allows us to truncate the series in Eq.~(\ref{expseries}), carry out the average over the disorder and re-exponentiate as follows
\begin{align}
&U_{ij}^{\alpha\beta} \approx 1 - \frac{\sigma^2}{2 N} \left[ (z^{\alpha\star}_i y^\beta_j + z_i^\alpha y_j^{\beta\star})^2 +(z^{\beta\star}_j y^\alpha_i + z_j^\beta y_i^{\alpha\star})^2 + 2 \Gamma_{\alpha\beta} (z^{\alpha\star}_i y^\beta_j + z_i^\alpha y_j^{\beta\star})  (z^{\beta\star}_j y^\alpha_i + z_j^\beta y_i^{\alpha\star})  \right]  \nonumber \\
&\approx \exp\left\{ - \frac{\sigma^2}{2 N} \left[ (z^{\alpha\star}_i y^\beta_j + z_i^\alpha y_j^{\beta\star})^2 +(z^{\beta\star}_j y^\alpha_i + z_j^\beta y_i^{\alpha\star})^2 + 2 \Gamma_{\alpha\beta} (z^{\alpha\star}_i y^\beta_j + z_i^\alpha y_j^{\beta\star})  (z^{\beta\star}_j y^\alpha_i + z_j^\beta y_i^{\alpha\star})  \right]\right\}.
\end{align}
Once we sum over all combinations of $(\alpha, \beta;i,j)$ in the exponent, we find the same result as in Eq.~(\ref{averaged})  and the calculation continues as before. This reasoning implies that the results in Eqs.~(\ref{support}) are valid for a wide class of matrices, whose elements may be drawn from non-Gaussian distributions. We emphasize that all that is required for Eq.~(\ref{averaged}) to be valid is that Eq.~(\ref{rmstatsagain}) be satisfied and that any higher moments than the variance decrease sufficiently quickly with $N$. Similar arguments can also be used when the first moments of the random matrix elements are non-zero and so we should expect Eqs.~(\ref{simult1}-\ref{cond}) to be generally applicable in the same fashion.

\subsection{Verification with Bernoulli, Laplace and uniformly distributed random numbers}
\begin{figure}[H]
	\centering 
	\includegraphics[scale = 0.4]{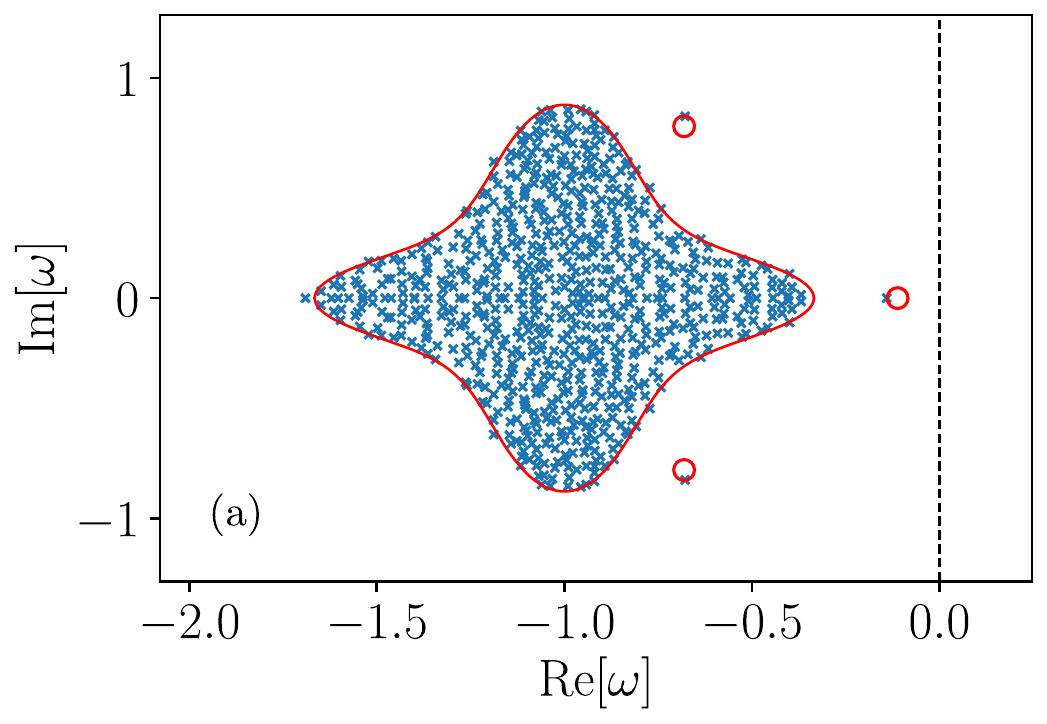}
	\includegraphics[scale = 0.4]{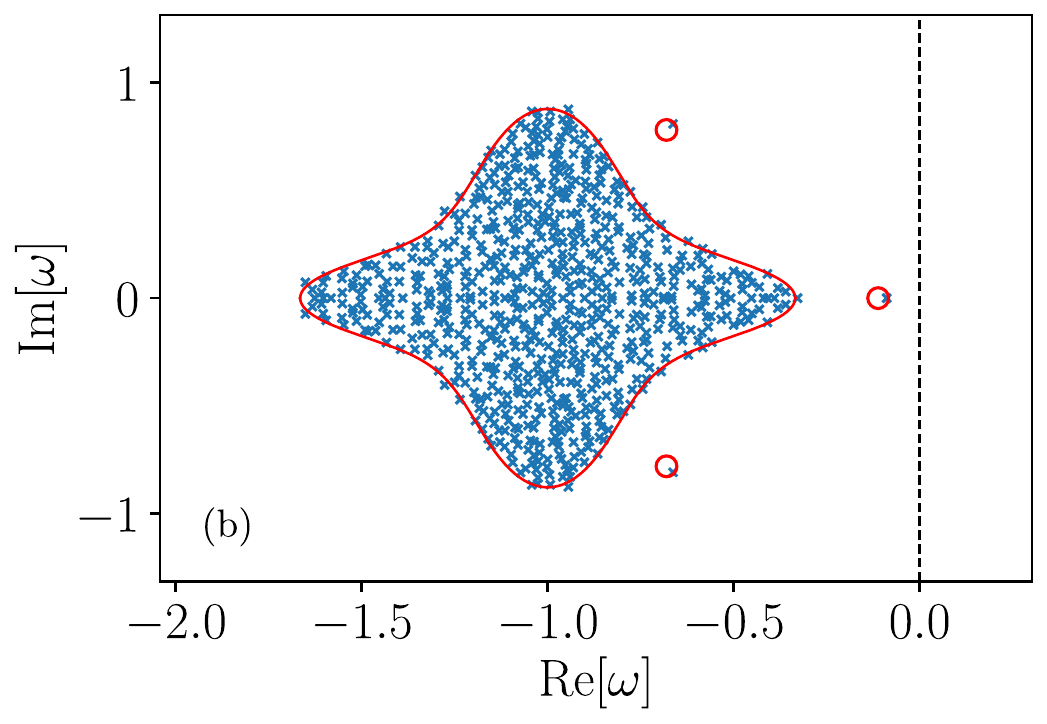}
	\includegraphics[scale = 0.4]{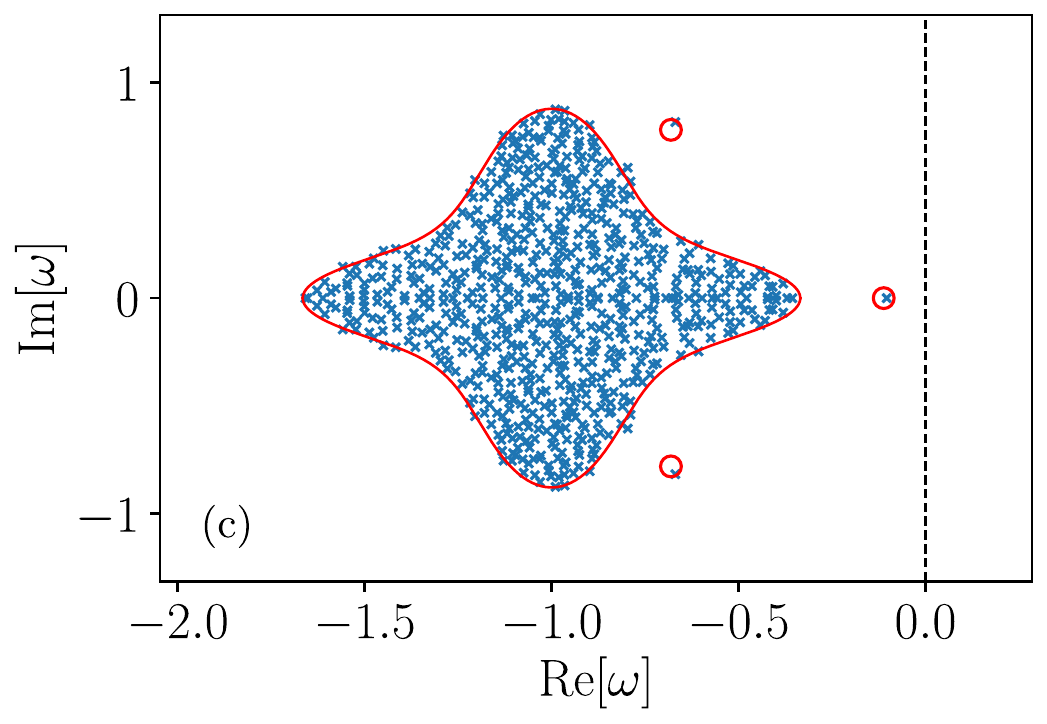}
	\caption{ {\bf Universality: eigenvalue spectra only depend on first two moments of interaction matrix.} The eigenvalues of a random matrix $A_{ij}^{\alpha\beta}$ with non-Gaussian statistics. Panel (a): Bernoulli distribution [Eq.~(\ref{bernoulli_distribution})], panel (b): Laplace distribution [Eq.~(\ref{laplace_distribution})], panel (c): uniform distribution [Eq.~(\ref{uniform_distribution})]. All distributions fulfil Eq.~(\ref{rmstatsagain}) but also have non-zero $ \overline{m_{ij}^{\alpha\beta}} = \mu_{\alpha\beta}/N$. Model parameters are the same as in Fig.~2a in the main text with $N = 750$. Blue markers are from computer-generated matrices, the red lines and circles are the predictions of the analytical theory.} \label{fig:universal_verify}
\end{figure}

To demonstrate the universality of our theory, we compare the eigenvalue spectra for random matrices whose elements are drawn from different distributions. In Fig. \ref{fig:universal_verify} we show that the locations of the eigenvalues are predicted accurately by our theory. 

We first verify the universality of our theoretical approach using Bernoulli random numbers. Here, the joint probability distribution for the random matrix entries is 
\begin{align}
P(m_{ij}^{\alpha\beta}, m_{ji}^{\beta\alpha} ) &= \frac{(1+\Gamma_{\alpha\beta})}{4 } \left[\delta(m_{ij}^{\alpha\beta} - \sigma/\sqrt{N})\delta(m_{ji}^{\beta\alpha} - \sigma/\sqrt{N}) + \delta(m_{ij}^{\alpha\beta} + \sigma/\sqrt{N})\delta(m_{ji}^{\beta\alpha} + \sigma/\sqrt{N}) \right] \nonumber \\
&+ \frac{(1-\Gamma_{\alpha\beta})}{4 } \left[\delta(m_{ij}^{\alpha\beta} - \sigma/\sqrt{N})\delta(m_{ji}^{\beta\alpha} + \sigma/\sqrt{N}) + \delta(m_{ij}^{\alpha\beta} + \sigma/\sqrt{N})\delta(m_{ji}^{\beta\alpha} -\sigma/\sqrt{N}) \right]. \label{bernoulli_distribution}
\end{align}
We note that this is for the case in which the mean vanishes, $\mu_{\alpha\beta} = 0$. A non-zero mean can be added subsequently.

\noindent We also verify our universality claim with random numbers drawn from a Laplace distribution,
\begin{align}
P(m_{ij}^{\alpha\beta}, m_{ji}^{\beta\alpha} ) &= \sqrt{\frac{N}{ 2\sigma^2}}\exp\left(-\frac{ \sqrt{2N}\left\vert m_{ij}^{\alpha\beta} \right\vert }{\sigma}\right)\nonumber \\
&\times \left[\frac{(1+\Gamma_{\alpha\beta})}{2 } \delta(m_{ij}^{\alpha\beta} - m_{ji}^{\beta\alpha})  + \frac{(1-\Gamma_{\alpha\beta})}{2 } \delta(m_{ij}^{\alpha\beta} + m_{ji}^{\beta\alpha})\right], \label{laplace_distribution}
\end{align}
and a uniform distribution,
\begin{align}
P(m_{ij}^{\alpha\beta}, m_{ji}^{\beta\alpha} ) &= \frac{\sqrt{N}}{2\sqrt{3}\sigma}\theta\left(m_{ij}^{\alpha\beta} - \frac{\sqrt{3}\sigma}{\sqrt{N}}\right)\theta\left(m_{ij}^{\alpha\beta} + \frac{\sqrt{3}\sigma}{\sqrt{N}}\right)\nonumber \\
&\times \left[\frac{(1+\Gamma_{\alpha\beta})}{2 } \delta(m_{ij}^{\alpha\beta} - m_{ji}^{\beta\alpha})  + \frac{(1-\Gamma_{\alpha\beta})}{2 } \delta(m_{ij}^{\alpha\beta} + m_{ji}^{\beta\alpha})\right]. \label{uniform_distribution}
\end{align}
In Section~\ref{sec:trophicenforce}, we also use a Gamma distribution. The probability distributions in Eqs.~(\ref{bernoulli_distribution}--\ref{uniform_distribution}) are constructed such that the elements of the random matrices obey the relations in Eq.~(\ref{rmstatsagain}). 

Examples of spectra of matrices with these statistics are shown in Fig.~\ref{fig:universal_verify}. As can be seen from the data, the limiting spectra are independent of the details of the distribution and are well-described by the analytical theory [given in Eqs.~(\ref{chis}), (\ref{simult1}) and (\ref{simult2})].

\section{Strict enforcement of trophic interactions}\label{sec:trophicenforce}
In the first instance, we employed Gaussian distributions as the prototype for the distribution from which the elements of our random matrices were drawn. This does not necessarily guarantee that all interactions between, say, predator species and prey species will be assigned negative coefficients for the prey-like species and positive coefficients for the predator-like species. Due to the choice of the first moments there is a statistical tendency for this to be the case but it is not strictly enforced for each and every pair due to the unbounded support of the bivariate Gaussian distribution. Similarly, these distributions do not preclude the possibility of predator-prey-like interactions between two prey species or two predator species (intraguild predation). The blocks in the matrix therefore do not describe strict trophic levels. Instead, the block structure divides the community into species which are more \textit{likely} to act as prey or predators respectively. That is, there is a statistical trophic structure to the ecosystem represented by these matrices. 

Starting from a given Gaussian ensemble, we now use the universality discussed in the previous section, to find matrices which enforce a given set of sign relations \textit{and} produce the same eigenvalue spectrum as the Gaussian ensemble. We impose a $(-,+)$ relation for pairs in which one species belongs to the prey block and the other to the predator block. Further, we eliminate intraguild predation by constraining the interaction coefficients between two prey species (or two predator species) all to have the same sign (positive for prey-prey interactions and negative for predator-predator in the example below). Here, for the sake of the analysis, we also constrain $\mu_{uv} = - \mu_{vu}$, $\Gamma_{uv}<0$, $\Gamma_{uu}>0$ and $\Gamma_{vv}>0$. A similar procedure could in principle be repeated for other set-ups [for example where prey-prey interactions were not constrained always to be $(+,+)$ but could also be $(-,-)$].

A natural starting point for the construction of such a matrix is a distribution with a support only on the positive (or negative) half of the real axis, guaranteeing the sign of the random variable. One such distribution is the gamma distribution. Consider a pair of independent gamma-distributed random variables $x_{ij}^{\alpha\beta}>0$ and $y_{ij}^{\alpha\beta}>0$, each distributed according to
\begin{align}\label{eq:gamma_aux}
p(\xi) = \frac{1}{\Gamma(k_{\alpha\beta}) \theta_{\alpha\beta}^{k_{\alpha\beta}}} \xi^{k_{\alpha\beta}-1} e^{-\xi/\theta_{\alpha\beta}}
\end{align}
for non-negative $\xi$, where $\Gamma(k)$ here indicates the standard gamma function (not to be confused with the correlation coefficients $\Gamma_{uu}$ etc.). The mean and variance of this distribution are
\begin{align}
\langle \xi \rangle &= k_{\alpha\beta} \theta_{\alpha\beta}  , \nonumber \\
\langle (\xi - \langle \xi \rangle)^2 \rangle &= k_{\alpha\beta}  \theta_{\alpha\beta} ^2 . 
\end{align}
Now focus on a given pair of species $i$ and $j$, where $i$ is in block $\alpha\in\{u,v\}$, and $j$ in block $\beta\in\{u,v\}$. The diagonally opposing entries in the random matrix describing the interactions between these two species are then constructed as linear combinations of $x_{ij}^{\alpha\beta}$ and $y_{ij}^{\alpha\beta}$,
\begin{align}
m_{ij}^{\alpha\beta} &= a^{\alpha\beta}_{xx} x_{ij}^{\alpha\beta} + a^{\alpha\beta}_{xy} y_{ij}^{\alpha\beta} , \nonumber \\
m_{ji}^{\beta \alpha} &= a^{\alpha\beta}_{yx} x_{ij}^{\alpha\beta} + a^{\alpha\beta}_{yy} y_{ij}^{\alpha\beta}.\label{eq:lincomb}
\end{align}

It can then be verified that the following choice for the coefficients in the linear transformation produces the correct signs, moments and correlations for the matrix elements $m_{ij}^{\alpha\beta}$, 
\begin{align}\label{tophicmapping}
a^{\alpha\beta}_{xx} &=  (-1)^{\delta_{\beta v}}\left[ \sqrt{1+\Gamma_{\alpha\beta}} + \sqrt{1- \Gamma_{\alpha\beta}} \right] , \nonumber \\
a^{\alpha\beta}_{xy} &=   (-1)^{\delta_{\beta v}} (-1)^{1-\delta_{\alpha\beta}} \left[ \sqrt{1+\Gamma_{\alpha\beta} }- \sqrt{1- \Gamma_{\alpha\beta}} \right] , \nonumber \\
a^{\alpha\beta}_{yx} &=   (-1)^{\delta_{\alpha v}}(-1)^{1-\delta_{\alpha\beta}} \left[ \sqrt{1+\Gamma_{\alpha\beta} }- \sqrt{1- \Gamma_{\alpha\beta}} \right] , \nonumber \\
a^{\alpha\beta}_{yy} &=  (-1)^{\delta_{\alpha v}}\left[ \sqrt{1+\Gamma_{\alpha\beta}} + \sqrt{1- \Gamma_{\alpha\beta}} \right] ,
\end{align}
if we choose $k_{\alpha\beta} = \frac{(\mu'_{\alpha\beta})^2}{N\sigma^2}, ~~\theta_{\alpha\beta} = \frac{\sigma^2}{\mu'_{\alpha\beta}}$, where $
\mu'_{\alpha\beta}=\frac{\vert \mu_{\alpha\beta} \vert}{\sqrt{1+\Gamma_{\alpha\beta}}}$. The exponents on the right-hand side of Eq.~(\ref{tophicmapping}) indicate Kronecker deltas, for example $\delta_{\beta v}=1$ if $\beta=v$ (species $j$ is in the predator block), and $\delta_{\beta v}=0$ if $\beta=u$ (species $j$ is in the prey block). \\

\begin{figure}[H]
	\centering 
	\includegraphics[scale = 0.31]{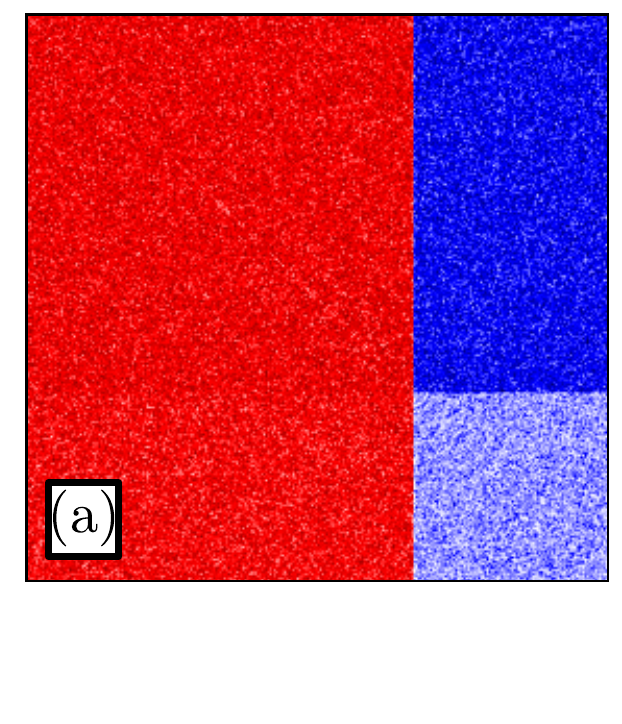}
	\includegraphics[scale = 0.31]{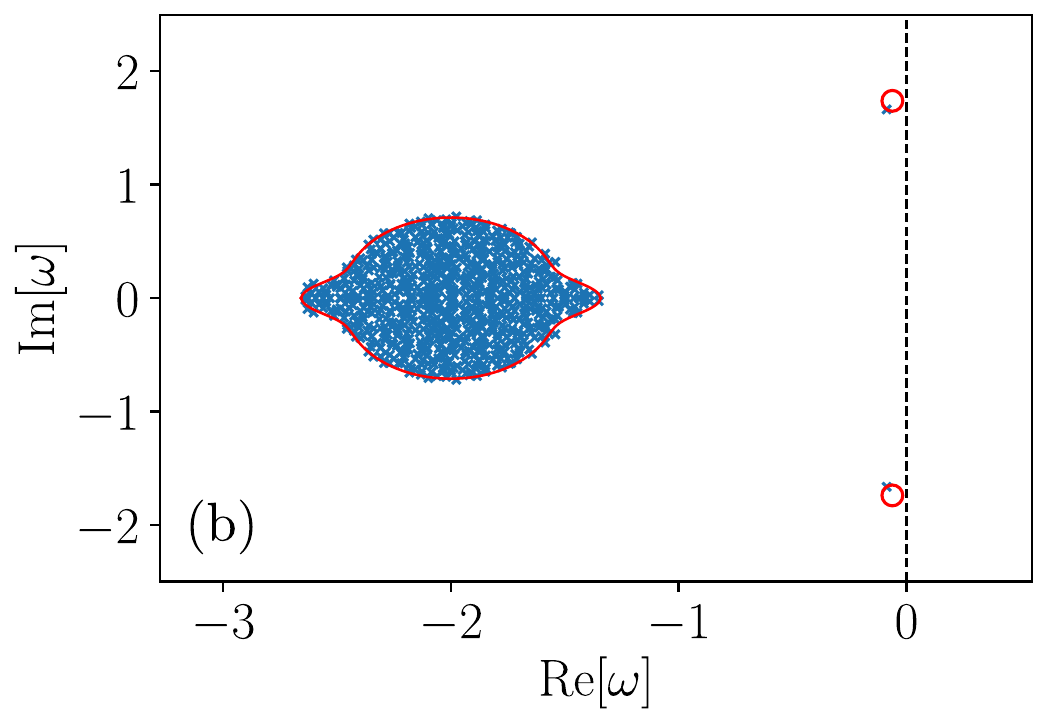}
	\includegraphics[scale = 0.31]{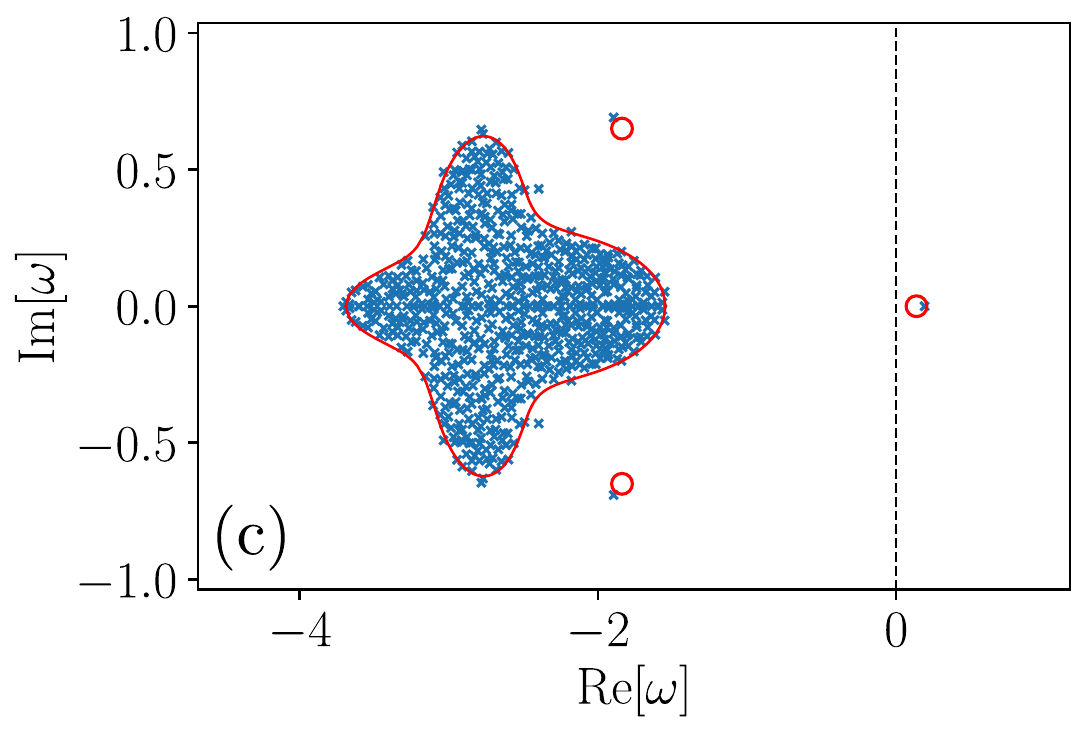}
	\includegraphics[scale = 0.31]{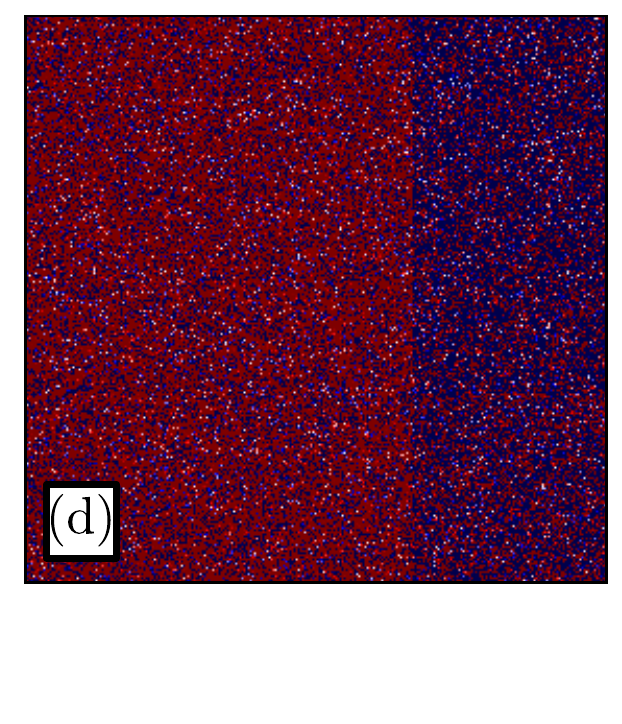}
	\includegraphics[scale = 0.31]{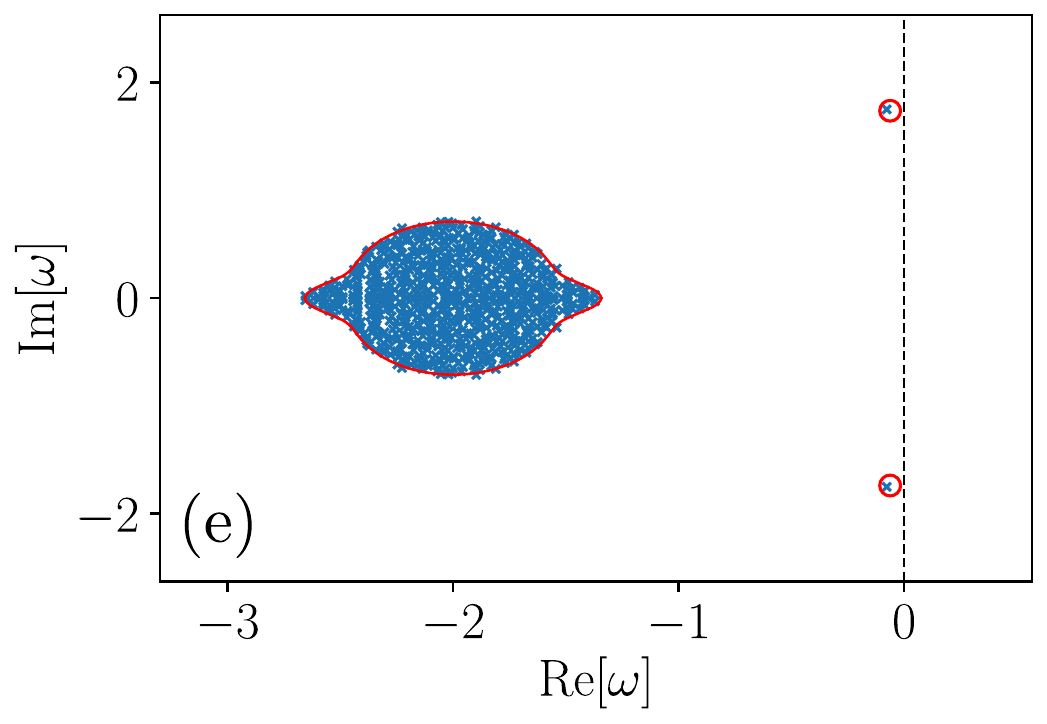}
	\includegraphics[scale = 0.31]{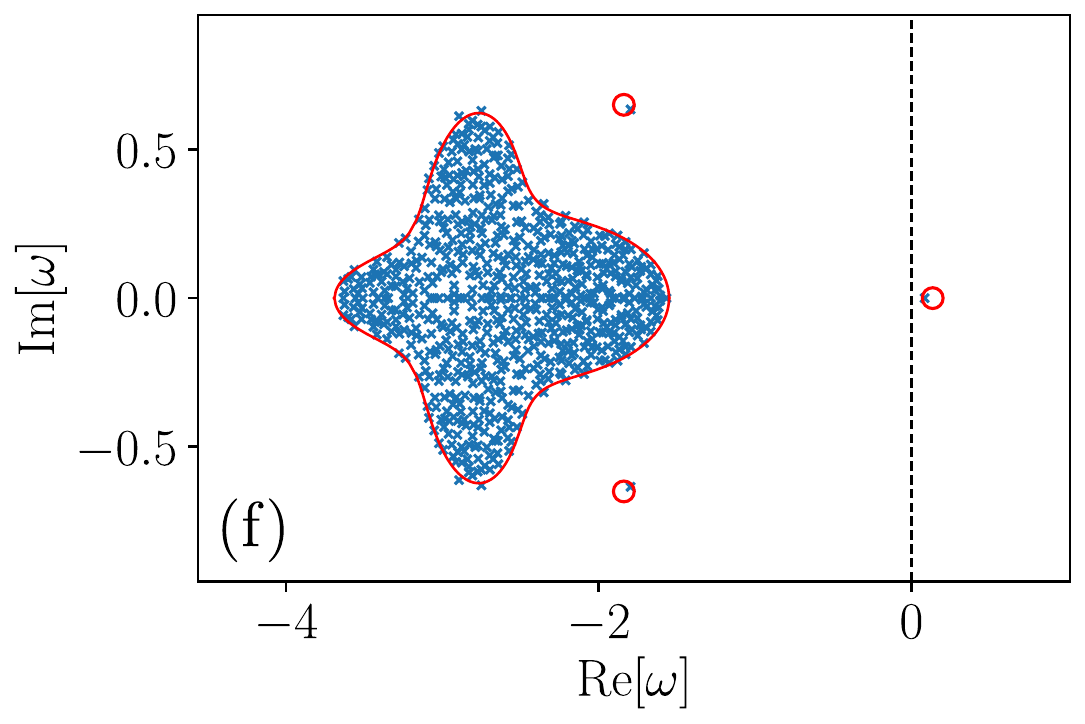}
	\caption{{\bf Universality allows the application of the Gaussian theory to a model with strict trophic interactions.} Comparison between the eigenvalue spectra of random matrices with entries drawn from a gamma distribution and a Gaussian distribution. Panels (a) and (d) show the entries of the matrix $m_{ij}^{\alpha\beta}$ with red indicating a positive entry and blue a negative one. In panels (a), (b) and (c), random matrix entries are constrained to have signs befitting the trophic level to which they belong. In panels (d), (e) and (f), random matrix elements are drawn from a Gaussian distribution and so do not necessarily obey the sign relations, although they have a statistical tendency to do so. The random matrix entries in (a) and (d) are drawn from distribution which have the same moments, so the spectra in panels (b)/(c) and (e)/(f), respectively, are statistically equivalent. The only difference between (b) and (c) [(e) and (f)] is the value of $q$, affecting only the diagonal elements of $\mathbf{M}_q$. The figure demonstrates that the dispersal-induced instability remains possible for a model with strict trophic levels. The system parameters (after rescaling, see Sec.~\ref{section:scaling}) here are $D_u = 1$, $D_v = 5$, $NC\mu_{uu} = NC\mu_{vu} = -NC\mu_{uv} = 8$, $NC\mu_{vv} = -4$, $\gamma_u = 2/3$, $\gamma_v = 1/3$, $c = NC\sigma^2 = 0.375$, $d_u = d_v = 2$, $N = 750$, $C = 1$, $\Gamma_{u} = 0.5$, $\Gamma_{uv}= -0.9$, $\Gamma_{v} = 0.6$. Panels (b) and (e) $q = 0$, panels (c) and (f) $q =0.5$. }\label{fig:enforce_trophic}
\end{figure}

That the correct signs of the elements $m_{ij}^{\alpha\beta}$ are upheld by Eq.~(\ref{tophicmapping}) can be seen case-by-case as follows. We note that in $\sqrt{1+\Gamma_{\alpha\beta}} + \sqrt{1- \Gamma_{\alpha\beta}}>0$ trivially in all cases. Take for example the case $\alpha = \beta  = u$. Then we require all $a_{\gamma\delta}^{uu}$ to be positive. Since we constrain $\Gamma_{uu}>0$, $\sqrt{1+\Gamma_{uu} }> \sqrt{1- \Gamma_{uu}}$. All the powers of the factors of $-1$ in the prefactors are zero in this case and hence all the coefficients $a_{\gamma\delta}^{uu}$ are positive. In the case, $\alpha= \beta = v$, the coefficients must all be negative. Again, since $\Gamma_{vv}>0$, we have $\sqrt{1+\Gamma_{vv} }> \sqrt{1- \Gamma_{vv}}$ so the signs of the coeffcients $a_{\gamma\delta}^{\alpha\beta}$ are determined by the prefactors, which one can see each yield $-1$ in this case. In the case $\alpha = u$ and $\beta = v$, we require $a_{xx}^{uv}<0$, $a_{xy}^{uv}<0$, $a_{yx}^{uv}>0$ and $a_{yy}^{uv}>0$. Now, since $\Gamma_{uv}<0$, we have $\sqrt{1+\Gamma_{\alpha\beta} }< \sqrt{1- \Gamma_{\alpha\beta}}$. The prefactor of $a_{xx}^{uv}$ is $-1$ and that of $a_{xy}^{uv}$ is $1$ -- both coefficients are negative. In contrast, the prefactor of $a_{yx}^{uv}$ is $-1$ and that of $a_{yy}^{uv}$ is $1$ and both coefficients are positive. A similar outcome is true for the case $\alpha = v$ and $\beta = u$. This scheme is verified in Fig. \ref{fig:enforce_trophic}(a).

We can thus produce random matrices whose elements are certain to have particular signs. This sign constraint still permits us to vary the statistics of the random matrix elements. As is shown in Fig. \ref{fig:enforce_trophic}, one can reproduce the eigenvalue spectrum of a Gaussian random matrix with such a sign-constrained random matrix. This speaks to the validity of the block structure of the random matrices as a representation of trophic structure in an ecosystem. Conversely, we note that given a model with strictly enforced signs in each block of the matrix, one can find a Gaussian ensemble with the same first and second moments. One can then use the theory to compute the eigenvalue support and stability.

\section{Variation in diagonal elements}\label{sec:diagonal}
In our original model, following May \cite{may}, we took the diagonal elements of the interaction matrix to be constant within the groups of predator and prey species for the sake of mathematical simplicity. We now relax this assumption and allow these diagonal elements to vary. We extend our analytical results for the boundary of the bulk region of the eigenvalue spectrum and the locations of any outlier eigenvalues, taking this variation in diagonal entries into account. We then discuss the effect on stability when the self-regulation rates or the dispersal coefficients are allowed to vary.

\subsection{Boundary of the bulk of the eigenvalue spectrum}
We begin by returning to the model used in the main text [see also Eq.~(\ref{averaged})], making the alteration $d_\alpha \to d_i^\alpha$. That is, the diagonal elements may now vary from species to species. This formulation also allows us to study the effect of variation in the dispersal coefficients $D_\alpha$ on stability, as we discuss in Section \ref{sec:variationdispersal}.

The diagonal elements $d_i^\alpha$ are now drawn from an as-yet-unspecified distribution. Eq.~(\ref{averaged}) becomes
\begin{align}
\exp\left[ -N \Phi(\omega) \right] &= \int \prod_{i\alpha}\left[ \left( \frac{d^2z^\alpha_i d^2y^\alpha_i}{2 \pi^2}\right) \right] \exp\left[ - \sum_{i\alpha} y^{\alpha\star}_i y^\alpha_i + \epsilon z^{\alpha\star}_i z^\alpha_i \right]\nonumber \\
\times & \exp\left[ -i \sum_{i \alpha} z^{\alpha\star}_i y^\alpha_i  (\omega^\star + d_i^\alpha) + z^\alpha_i y^{\alpha\star}_i ( \omega + d_i^\alpha)  \right]\nonumber \\
\times & \exp\left[ - \frac{\sigma^2}{2N}\sum_{i j \alpha \beta} (z_i^{\alpha\star} y^\beta_j + z^\alpha_i y_j^{\beta\star})^2 + \Gamma_{\alpha \beta} (z_i^{\alpha\star} y^\beta_j + z^\alpha_i y_j^{\beta\star})(z_j^{\beta\star} y^\alpha_i + z^\beta_j y_i^{\alpha\star}) \right] . 
\end{align}
We once again introduce the order parameters 
\begin{align}
u &= \frac{1}{N} \sum_{i \alpha} z^{\alpha\star}_i z^\alpha_i, \,\,\,\,\, v = \frac{1}{N} \sum_{i \alpha} y^{\alpha\star}_i y^\alpha_i, \nonumber \\
w_\alpha &= \frac{1}{N_\alpha} \sum_i z^{\alpha\star}_i y^\alpha_i, \,\,\,\,\, w_\alpha^\star = \frac{1}{N_\alpha} \sum_i y^{\alpha\star}_i z^\alpha_i, 
\end{align}
and we arrive at 
\begin{align}
\exp\left[ -N \Phi(\omega) \right] &= \int \mathcal{D}\left[ \cdots \right]\exp\left[N(\Psi + \Theta + \Omega)\right] ,
\end{align}
where $\mathcal{D}\left[ \cdots \right]$ denotes integration over all of the order parameters and their conjugate (`hatted') variables, and where 
\begin{align}
\Psi &= i  \hat u u  + i\hat v  v  + i \sum_{\alpha}\gamma_\alpha(\hat w_\alpha  w^\star_\alpha  +   \hat w^\star_\alpha  w_\alpha   ), \nonumber \\
\Theta &= -\epsilon  u - v - \sigma^2 u v + \sum_\alpha \gamma_\alpha\left[    -i w_\alpha \omega^\star   - i w^\star_\alpha \omega \right]  - \frac{1}{2} \sigma^2 \sum_{\alpha \beta}\gamma_\alpha \gamma_\beta\Gamma_{\alpha \beta}(w_\alpha w_\beta+ w_\alpha^\star w_\beta^\star) , \nonumber \\
\Omega &=\frac{1}{N} \sum_{\alpha, i} \gamma_\alpha \ln  \left[\int \left[ \left( \frac{d^2z^\alpha_i d^2y^\alpha_i}{2 \pi^2}\right) \right]  \exp\left\{ -i  [\hat u z_i^{\alpha\star} z_i^\alpha  + \hat v y_i^{\alpha\star} y_i^\alpha  + (\hat w_\alpha + d_i^\alpha) y_i^{\alpha\star} z_i^\alpha  + (\hat w_\alpha^\star+ d_i^\alpha) z_i^{\alpha\star} y_i^\alpha  ] \right\} \right] \nonumber \\
\approx &  \sum_{\alpha} \gamma_\alpha\int d(d^\alpha) P(d^\alpha) \ln  \Bigg\{\int \left[ \left( \frac{d^2z^\alpha d^2y^\alpha}{2 \pi^2}\right) \right]  \nonumber \\
&\;\;\;\;\;\times\exp\left\{ -i  [\hat u z^{\alpha\star} z^\alpha  + \hat v y^{\alpha\star} y^\alpha  + (\hat w_\alpha + d^\alpha) y^{\alpha\star} z^\alpha  + (\hat w_\alpha^\star+ d^\alpha) z^{\alpha\star} y^\alpha  ] \right\} \Bigg\}. \label{varydiagonalexponent}
\end{align}
Once again the integrals over $y_i^\alpha$ and $z_i^\alpha$ can be carried out, yielding
\begin{align}
\Omega &= -\sum_{\alpha} \gamma_\alpha \int d(d^\alpha) P_\alpha(d^\alpha)\ln \left\{ (\hat w_\alpha + d^\alpha)(\hat w^\star_\alpha + d^\alpha) - \hat u \hat v \right\} \nonumber \\
&\equiv -\sum_\alpha \gamma_\alpha\left\langle \ln \left\{ (\hat w_\alpha + d^\alpha)(\hat w^\star_\alpha + d^\alpha) - \hat u \hat v \right\}\right\rangle_d .
\end{align}
The notation $\avg{\cdots}_d$ indicates an average over the distribution $P_\alpha(d_\alpha)$ of diagonal elements. At the saddle point, we therefore obtain by differentiating Eq.~(\ref{varydiagonalexponent}) with respect to the `hatted' quantities [c.f. Eqs.~(\ref{diffwrthatted})]
\begin{align}
iu = -\sum_\alpha \gamma_\alpha\left\langle \frac{\hat v}{ (\hat w_\alpha + d_\alpha)(\hat w_\alpha^\star + d_\alpha) - \hat u \hat v } \right\rangle_d  ,& \,\,\,\,\, iv =- \sum_\alpha \gamma_\alpha \left\langle \frac{\hat u}{ (\hat w_\alpha + d_\alpha)(\hat w_\alpha^\star + d_\alpha) - \hat u \hat v } \right\rangle_d, \nonumber \\
i w_\alpha^\star = \left\langle \frac{\hat w_\alpha^\star + d_\alpha}{ (\hat w_\alpha + d_\alpha)(\hat w_\alpha^\star + d_\alpha) - \hat u \hat v } \right\rangle_d,& \,\,\,\,\, iw_\alpha = \left\langle \frac{\hat w_\alpha + d_\alpha}{ (\hat w_\alpha + d_\alpha)(\hat w^\star_\alpha + d_\alpha) - \hat u \hat v } \right\rangle_d. \label{diffwrthatted3}
\end{align}
We now define the following
\begin{align}
\hat f = \sum_\alpha \gamma_\alpha\left\langle \frac{1}{ (\hat w_\alpha + d_\alpha)(\hat w^\star_\alpha + d_\alpha) - \hat u \hat v } \right\rangle_d , \,\,\,\,\, \hat d_\alpha = \frac{1}{\hat f}\left\langle \frac{d_\alpha}{ (\hat w_\alpha + d_\alpha)(\hat w^\star_\alpha + d_\alpha) - \hat u \hat v } \right\rangle_d .
\end{align}
Thus, 
\begin{align}
iu &=  -\hat v \hat f, \,\,\,\,\, iv = -\hat u \hat f, \nonumber \\
i w^\star_\alpha &= (\hat w^\star_\alpha +\hat d_\alpha) \hat f , \,\,\,\,\, iw_\alpha =(\hat w_\alpha+ \hat d_\alpha) \hat f . \label{diagonalunhatted}
\end{align}
We also obtain the complementary set of saddle-point relations by differentiating  Eq.~(\ref{varydiagonalexponent}) with respect to the `unhatted' quantities [c.f. Eqs.~(\ref{diffwrtunhatted})]
\begin{align}
i\hat u &= \epsilon + \sigma^2 v , \,\,\,\,\, i\hat v = 1 + \sigma^2 u , \nonumber \\
i\hat w_\alpha &=  i \omega +  \sigma^2 \sum_{\beta} \gamma_\beta \Gamma_{\alpha\beta} w^\star_\beta , \nonumber \\
i\hat w^\star_\alpha &=  i \omega^\star +  \sigma^2 \sum_{\beta} \gamma_\beta \Gamma_{\alpha\beta} w_\beta  .
\end{align}
We now employ a slightly different solution strategy to previous calculations, opting instead to leave the solution in terms of the hatted order parameters. For $\epsilon \to 0^+$ we see that 
\begin{align}
\hat u = \sigma^2 \hat u \hat f .
\end{align}
This has two solutions -- either $\hat u = 0$, in which case the resolvent $iw^\star$ is analytic and the eigenvalue density vanishes, or $\hat f = \frac{1}{\sigma^2}$. For the boundary of the support of the eigenvalue spectrum, we solve for the values of $\omega$ which satisfy both conditions simultaneously. That is, we solve
\begin{align}
\hat f &= \sum_\alpha \gamma_\alpha\left\langle \frac{1}{ (\hat w_\alpha + d_\alpha)(\hat w^\star_\alpha + d_\alpha) } \right\rangle_d = \frac{1}{\sigma^2} , \nonumber \\ \hat d_\alpha &= \sigma^2 \left\langle \frac{d_\alpha}{ (\hat w_\alpha + d_\alpha)(\hat w^\star_\alpha + d_\alpha)  } \right\rangle_d , \nonumber \\
\hat w_\alpha &=   \omega -   \sum_{\beta}\gamma_\alpha \gamma_\beta \Gamma_{\alpha\beta} (\hat w_\beta^\star + \hat d_\beta) , \nonumber \\
\hat w^\star_\alpha &=   \omega^\star - \sum_{\beta}\gamma_\alpha \gamma_\beta \Gamma_{\alpha\beta} (\hat w_\beta + \hat d_\beta). \label{diagonalboundary}
\end{align}
We note that the boundary of the eigenvalue spectrum is dependent on the shape of the probability distribution from which we draw the diagonal elements. This is in contrast to the universality property for the off-diagonal random matrix elements discussed in Section \ref{sec:universality}.

\subsubsection*{Recovering the results of Ref. \cite{barabas2017}}
Eqs.~(\ref{diagonalboundary}) simplify in the case $\gamma_u = 1$ to results previously obtained in Ref. \cite{barabas2017}, which were obtained using a different method (that of quaternionic resolvents). Letting $i w^\star_{u} = \alpha(-\omega)$ and suppressing the subscript $u$ for brevity, one deduces from the first and third of Eqs.~(\ref{diagonalboundary}) and the third of Eqs.~(\ref{diagonalunhatted})
\begin{align}
\left\langle \left[ \vert d -\omega - \Gamma \sigma^2 \alpha(\omega)  \vert^2 \right]^{-1} \right\rangle_d = \frac{1}{\sigma^2} . \label{barabas1}
\end{align}
Now using the fourth of Eqs.~(\ref{diagonalboundary}) in combination with the third of Eqs.~(\ref{diagonalunhatted}) one obtains 
\begin{align}
\sigma^2 \alpha(\omega) - \hat d = -\omega^\star - \Gamma \sigma^2 \alpha^\star(\omega).
\end{align}
Letting $\alpha(\omega )=\alpha_{\mathrm{re}} + i\alpha_{\mathrm{im}}$ and $\omega = \omega_x + i \omega_y$, the imaginary part of this equation obeys
\begin{align}
\alpha_{\mathrm{im}} = \frac{\omega_y}{\sigma^2 (1-\Gamma)},\label{bbrackets1}
\end{align}
and the real part obeys 
\begin{align}
 \sigma^2 \alpha_{\mathrm{re}} = \left\langle \left[d - \sigma^2 \omega_x - \Gamma \sigma^4 \alpha_{\mathrm{re}}   \right]\left[ \vert d -\omega - \Gamma \sigma^2 \alpha(\omega)  \vert^2 \right]^{-1} \right\rangle_d,\label{bbrackets2}
\end{align}
where we have used Eq.~(\ref{barabas1}) and the second of Eqs.~(\ref{diagonalboundary}). Writing the averages over $d$ explicitly in Eqs.~(\ref{bbrackets2}) and (\ref{barabas1}), one obtains 
\begin{align}
\frac{1}{\sigma^2} &=\int d(d) P(d) \frac{1}{\left[d - \omega_x - \Gamma \sigma^2 \alpha_{\mathrm{re}} \right]^2 + \omega_y^2/(1-\Gamma)^2}  , \nonumber \\
\sigma^2 \alpha_{\mathrm{re}} &= \int d(d) P(d) \frac{d - \sigma^2 \omega_x - \Gamma \sigma^4 \alpha_{\mathrm{re}}}{\left[d - \omega_x - \Gamma \sigma^2 \alpha_{\mathrm{re}} \right]^2 + \omega_y^2/(1-\Gamma)^2} .
\end{align}
These are exactly Eqs.~(4.42) and (4.43) in the Supplement of Ref. \cite{barabas2017} (once one sets $\sigma^2 = 1$ in the above equations and takes the limit $\vert \beta \vert \to 0$ in those of Ref. \cite{barabas2017} so as yield the equations for the boundary of the support).

\subsection{Diagonal variation stretches the bulk support along the real axis}
In order provide and example and to compare to previous work \cite{barabas2017}, we take the case of a matrix without trophic structure. 
When the diagonal elements are uniformly distributed between $d_0 -\Delta_0/2$ and $d_0 + \Delta_0/2$ one finds from the first two of Eqs.~(\ref{diagonalboundary}) (noting that $\hat w$ and $\hat w^\star$ are complex conjugates and defining $\hat w = \hat w_x + i \hat w_y$)
\begin{align}
\hat f &= \frac{1}{\hat w_y \Delta_0}\left[\arctan\left( \frac{\hat w_x + d_0 + \Delta_0/2}{\hat w_y}\right) -\arctan\left( \frac{\hat w_x + d_0 - \Delta_0/2}{\hat w_y}\right) \right], \nonumber \\
\frac{\Delta_0 \hat d}{\sigma^2} &= \frac{1}{2} \ln \left[\frac{(\hat w_x + d_0 +\Delta_0)^2+\hat w_y^2}{(\hat w_x + d_0-\Delta_0)^2+\hat w_y^2} \right] - \frac{\hat w_x \Delta_0 }{\sigma^2} . \label{uniformdiagonal}
\end{align}
In the case where $\Gamma  = 0$, from last two of Eqs.~(\ref{diagonalboundary}) we have $\hat w = \omega \equiv \omega_x + i \omega_y$ and we obtain the following closed form solution for the boundary of the support
\begin{align}
\frac{1}{\omega_y \Delta_0}\left[\arctan\left( \frac{\omega_x + d_0 + \Delta_0/2}{\omega_y}\right) -\arctan\left( \frac{\omega_x + d_0 - \Delta_0/2}{\omega_y}\right) \right] = \frac{1}{\sigma^2}, 
\end{align}
which can be written alternatively
\begin{align}
\frac{1}{\omega_y \Delta_0}\arctan\left[ \frac{\Delta_0 \omega_y}{(\omega_x + d_0)^2 + \omega_y^2 - \frac{\Delta_0^2}{4}}\right] = \frac{1}{\sigma^2}. 
\end{align}
One notes that for small $\Delta_0$ this reduces to the circular law. We can solve this exactly when $\omega_y \to 0$. In this case, one finds
\begin{align}\label{eq:auxaux}
\omega_x = -d_0 \pm \sqrt{\sigma^2 + \frac{\Delta_0^2}{4}} .
\end{align}
This agrees with the result for the abscissa derived in Ref. \cite{barabas2017} (see Eq. (5.13) in the Supplementary Material of Ref. \cite{barabas2017}). 
 
\begin{figure}[H]
	\centering 
	\includegraphics[scale = 0.4]{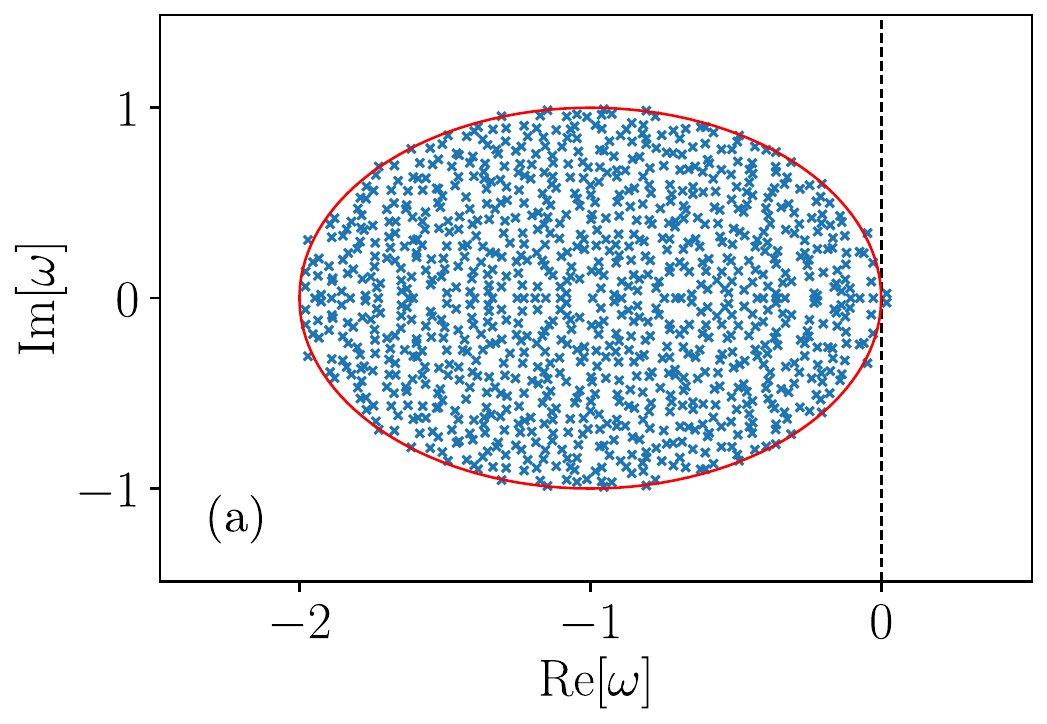}
	\includegraphics[scale = 0.4]{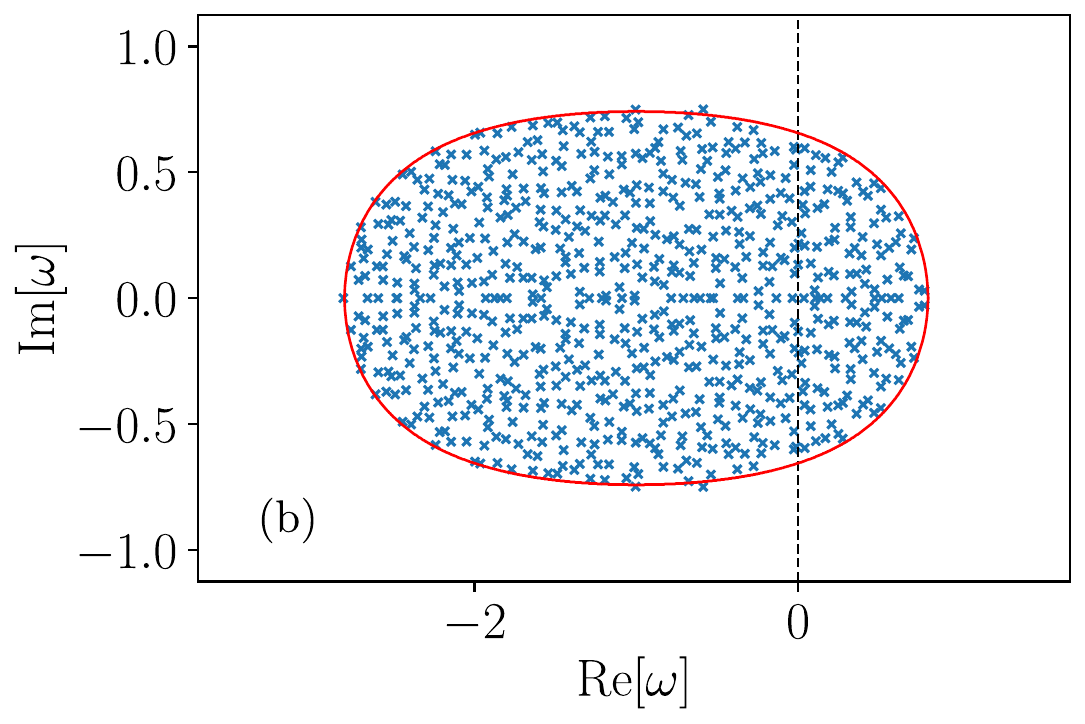}
	\caption{{\bf Destabilising effect of heterogeneity in diagonal matrix elements.} Eigenvalue spectrum for $NC\mu_{uu} = 0$, $NC\mu_{uv} = 0$, $NC\mu_{vu} = 0$, $NC\mu_{vv} = 0$, $c = NC\sigma^2 = 1$,  $\Gamma_{uv} = 0$, $\Gamma_{uu} = 0$, $\Gamma_{vv} = 0$, $\gamma_u = 1$, $\gamma_v = 0$, $C= 1$ and $N = 750$.  The diagonal entries are uniformly distributed with parameters $d_0 =1$ and (a) $\Delta_0 = 0$, (b) $\Delta_0 = 3$.  }\label{fig:diagonalvarying_uniform}
\end{figure}

Eq.~(\ref{eq:auxaux}) indicates that the variation in diagonal elements acts to elongate the bulk region of the eigenvalue spectrum. This is demonstrated in Fig. \ref{fig:diagonalvarying_uniform}.

\subsection{Calculation of outlier eigenvalues}
The saddle-point approximation to the resolvent $iw_\alpha^\star$ can be extracted from Eqs.~(\ref{diagonalboundary}) using  Eq.~(\ref{diagonalunhatted}). From this, the outlier eigenvalues which arise from the inclusion of non-zero random matrix entry means $\mu_{\alpha\beta}$ can be deduced as usual. The outlier eigenvalues $\lambda$ must satisfy, as in Section \ref{sec:outliers},
\begin{align}
\left[ \gamma_u\mu_{uu} - \frac{1}{\chi_u(\lambda)}\right] \left[\gamma_v \mu_{vv} - \frac{1}{\chi_v(\lambda)}\right] - \gamma_u\gamma_v\mu_{uv}\mu_{vu} = 0.\label{simultdiag1}
\end{align}
However, with the inclusion of variation in the diagonal elements, the quantities $\chi_\alpha(\lambda)$ now satisfy 
\begin{align}
\chi_\alpha &= (\hat w_\alpha^\star + \hat d_\alpha)\hat f, \nonumber \\
\hat f &= \sum_\alpha \gamma_\alpha\left\langle \frac{1}{ (\hat w_\alpha + d_\alpha)(\hat w^\star_\alpha + d_\alpha)  } \right\rangle_d , \nonumber \\
\hat d_\alpha &= \frac{1}{\hat f}\left\langle \frac{d_\alpha}{ (\hat w_\alpha + d_\alpha)(\hat w^\star_\alpha + d_\alpha) } \right\rangle_d , \nonumber \\
\hat w_\alpha &=   \lambda -   \sum_{\beta} \gamma_\beta \Gamma_{\alpha\beta} (\hat w_\beta^\star + \hat d_\beta) , \nonumber \\
\hat w^\star_\alpha &=   \lambda^\star - \sum_{\beta} \gamma_\beta \Gamma_{\alpha\beta} (\hat w_\beta + \hat d_\beta)  , \label{simultdiag2}
\end{align}
subject to the condition
\begin{align}
\sum_{\alpha} \gamma_\alpha \vert\chi_\alpha(\lambda)\vert^2 < \frac{1}{\sigma^2} . \label{conddiag}
\end{align}
\subsubsection*{Example: Random matrix without trophic structure}
\begin{figure}[H]
	\centering 
	\includegraphics[scale = 0.4]{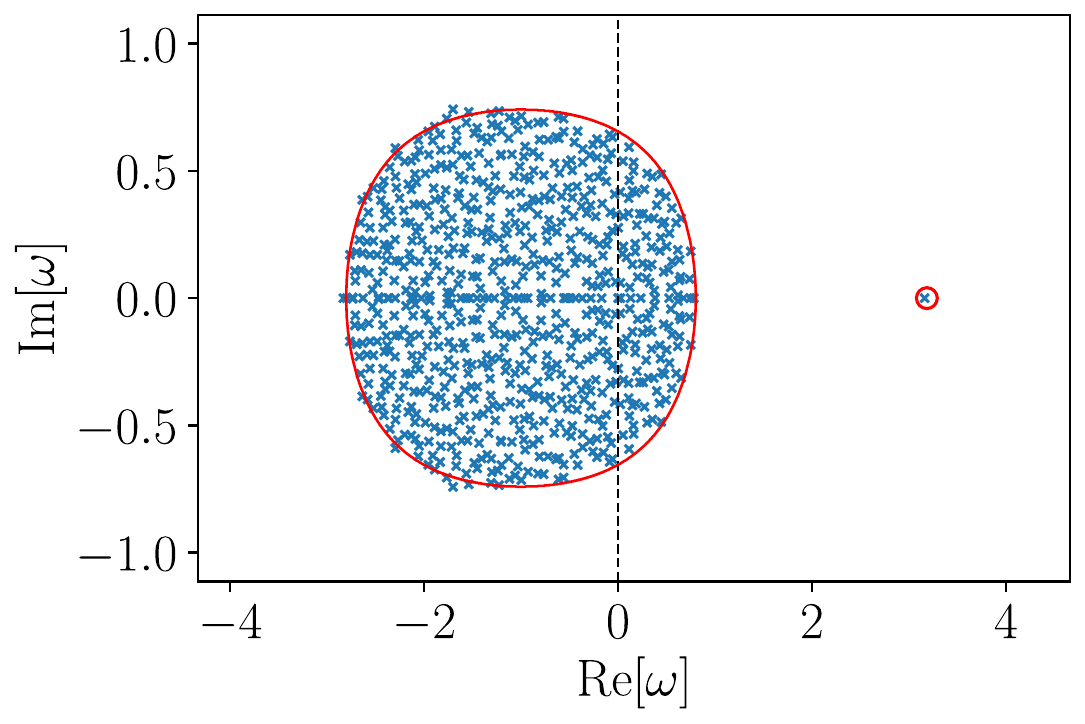}
	\caption{{\bf Verification of theoretical predictions for outlier eigenvalues of matrices with heterogeneous diagonal elements.} Eigenvalue spectrum for $NC\mu_{uu} = 4$, $NC\mu_{uv} = 4$, $NC\mu_{vu} = 4$, $NC\mu_{vv} = 4$, $c = NC\sigma^2 = 1.0$,  $\Gamma_{uv} = 0$, $\Gamma_{uu} = 0$, $\Gamma_{vv} = 0$, $\gamma_u = 1$, $\gamma_v = 0$, $C = 1$ and $N = 750$.  The diagonal entries are uniformly distributed with parameters $d_0 =1$ and $\Delta_0 = 3$.  }\label{fig:diagonalvarying_uniform_outlier}
\end{figure}

\noindent We once again take a simple example to verify this approach. The equations for the outlier simplify greatly in the case where $\gamma_v = 0$ and $\Gamma_{\alpha\beta} = 0$ (i.e. where there is no trophic structure or correlations). We find
\begin{align}
\lambda^\star &=-\hat d+ \frac{1}{\mu \hat f},\nonumber \\
\hat f &= \left\langle \frac{1}{ (\lambda + d)(\lambda^\star + d)  } \right\rangle_d , \nonumber \\
\hat d &= \frac{1}{\hat f}\left\langle \frac{d}{ (\lambda + d)(\lambda^\star + d)  } \right\rangle_d . \label{simplediagoutlier}
\end{align}
Taking the example once again of the uniform distribution for the diagonal elements centred on $d_0$ with width $\Delta_0$, one obtains for the last two of Eqs.~(\ref{simplediagoutlier})
\begin{align}
\hat f &= \frac{1}{\lambda_y \Delta_0}\arctan\left[ \frac{\Delta_0 \lambda_y}{(\lambda_x + d_0)^2 + \lambda_y^2 - \frac{\Delta_0^2}{4}}\right], \nonumber \\
\Delta_0 \hat f\hat d &= \frac{1}{2} \ln \left[\frac{(\lambda_x + d_0 +\Delta_0/2)^2+\lambda_y^2}{(\lambda_x + d_0-\Delta_0/2)^2+\lambda_y^2} \right] - \lambda_x \Delta_0 \hat f .
\end{align}
Presuming that the outlier is real, we find
\begin{align}
\hat f &= \frac{1}{(\lambda_x + d_0)^2 - \Delta_0^2/4} , \nonumber \\
\hat d &=  \frac{1}{\Delta_0 \hat f } \ln \left[\frac{(\lambda_x + d_0 +\Delta_0/2)}{(\lambda_x + d_0-\Delta_0/2)} \right] - \lambda_x  . 
\end{align}
Hence, combining this with the first of Eqs.~(\ref{simplediagoutlier})
\begin{align}
\lambda_x = - d_0 + \frac{\Delta_0}{2} \frac{ \exp\left[ \Delta_0/\mu \right] + 1}{\exp\left[ \Delta_0/\mu \right] - 1}  . 
\end{align}
This expression is verified in Fig. \ref{fig:diagonalvarying_uniform_outlier}.

When $\Delta_0 \to 0$, we recover $\lambda_x = -d_0 + \mu$. As $\Delta_0 \to \infty$, $\lambda_x \approx -d_0 + \Delta_0/2$ for $\mu>0$ and $\lambda_x \approx -d_0 - \Delta_0/2$ for $\mu<0$. So variation in the diagonal entries can act to change the outlier eigenvalues considerably, depending on the degree of variation.
\subsection{Dispersal-induced instability remains possible in models with trophic structure when self-interaction is heterogeneous}
\begin{figure}[H]
	\centering 
	\includegraphics[scale = 0.4]{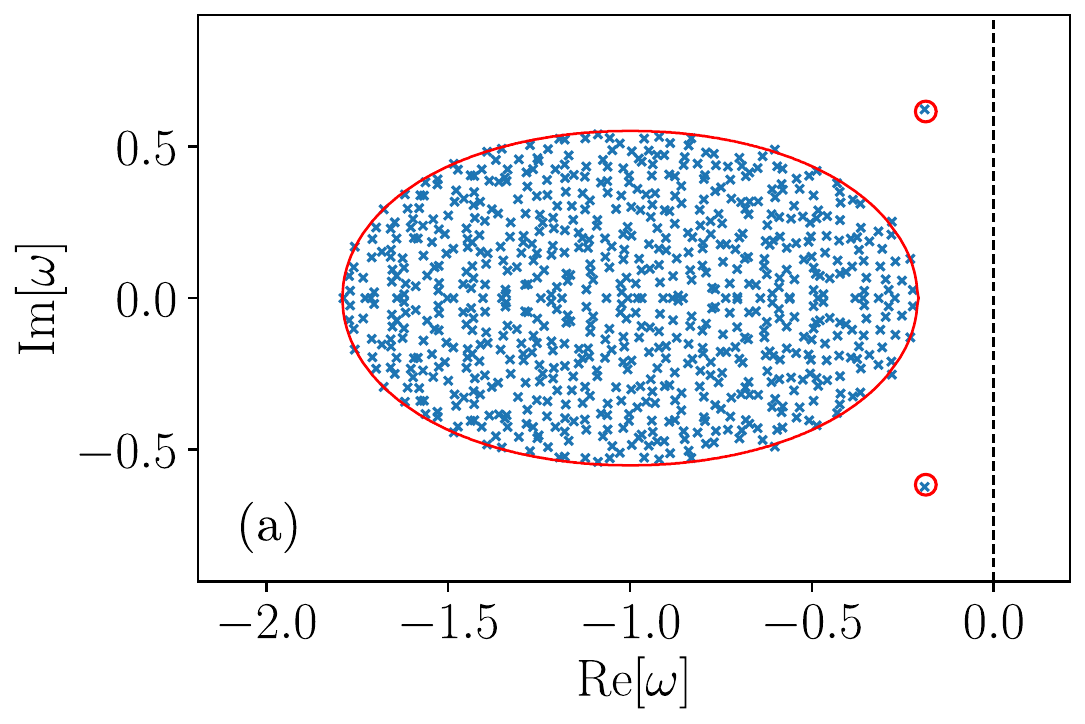}
	\includegraphics[scale = 0.4]{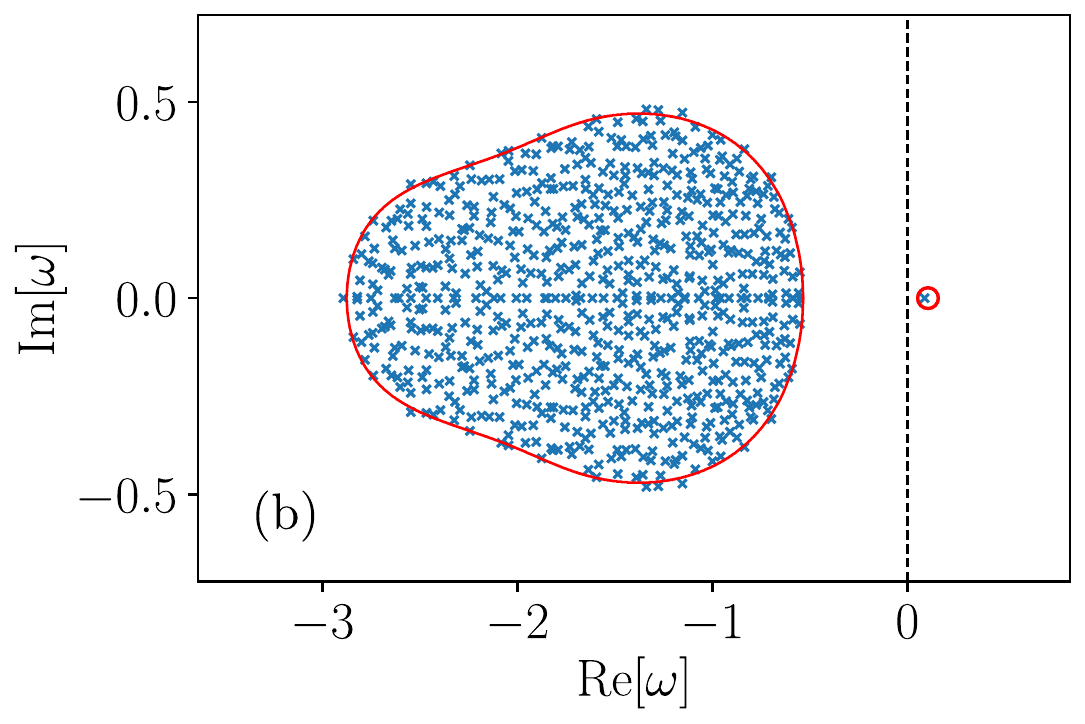}
	\caption{{\bf Example of dispersal-induced instability when self-interaction coefficients are heterogeneous.} Eigenvalue spectra for random matrices with trophic structure and statistical variation in the diagonal elements. The system parameters (after rescaling, see Sec.~\ref{section:scaling}) here are $D_u = 1$, $D_v = 5$, $NC\mu_{uu} = NC\mu_{vu} = -NC\mu_{uv} = 3$, $NC\mu_{vv} = -1.5$, $\gamma_u = 2/3$, $\gamma_v = 1/3$, $c = NC\sigma^2 = 0.375$, $d_0^u = d_0^v = 1$, $\Delta_0^u = \Delta_0^v = 1$. Panel (a) $q = 0$, panel (b) $q =1$.     }\label{fig:diagonalvariationTuring}
\end{figure}

We now turn our attention to whether or not variation in the diagonal elements disrupts the possibility of a Turing instability. To do this, we return to the model with trophic structure. We once again take the simple case of $\Gamma_{\alpha\beta } = 0$. In this case, from Eq.~(\ref{diagonalboundary}) we find the following closed-form solution for the boundary of the bulk region of the eigenvalue spectrum 
\begin{align}
\sum_\alpha \frac{\gamma_\alpha}{\omega_y \Delta^\alpha_0}\arctan\left[ \frac{\Delta^\alpha_0 \omega_y}{(\omega_x + d^\alpha_0)^2 + \omega_y^2 - \frac{(\Delta^\alpha_0)^2}{4}}\right] = \frac{1}{\sigma^2} .
\end{align}
From Eqs.~(\ref{simultdiag2}), we see that the outlier eigenvalues obey
\begin{align}
0 &=\left[ \gamma_u\mu_{uu} - \frac{1}{\chi_u(\lambda)}\right] \left[\gamma_v \mu_{vv} - \frac{1}{\chi_v(\lambda)}\right] - \gamma_u\gamma_v\mu_{uv}\mu_{vu} , \nonumber \\
\chi_\alpha(\lambda_x, \lambda_y) &= (\lambda_x + \hat d_\alpha)\hat f - i \lambda_y \hat f, \nonumber \\
\hat f &=\sum_\alpha \frac{\gamma_\alpha}{\lambda_y \Delta^\alpha_0}\arctan\left[ \frac{\Delta^\alpha_0 \lambda_y}{(\lambda_x + d^\alpha_0)^2 + \lambda_y^2 - \frac{(\Delta^\alpha_0)^2}{4}}\right], \nonumber \\
\hat d_\alpha &= \frac{1}{2 \Delta^\alpha_0 \hat f} \ln \left[\frac{(\lambda_x + d^\alpha_0 +\Delta^\alpha_0/2)^2+\lambda_y^2}{(\lambda_x + d^\alpha_0-\Delta^\alpha_0/2)^2+\lambda_y^2} \right] -  \lambda_x,
\end{align}
where $\lambda = \lambda_x + i \lambda_y$. These equations can be evaluated numerically. We find that despite considerable diagonal variation, the Turing instability remains possible, as is demonstrated in Fig. \ref{fig:diagonalvariationTuring}.

\subsection{Variation in dispersal coefficients can promote dispersal-induced instability}\label{sec:variationdispersal}
\begin{figure}[t!]
	\centering 
	\includegraphics[scale = 0.4]{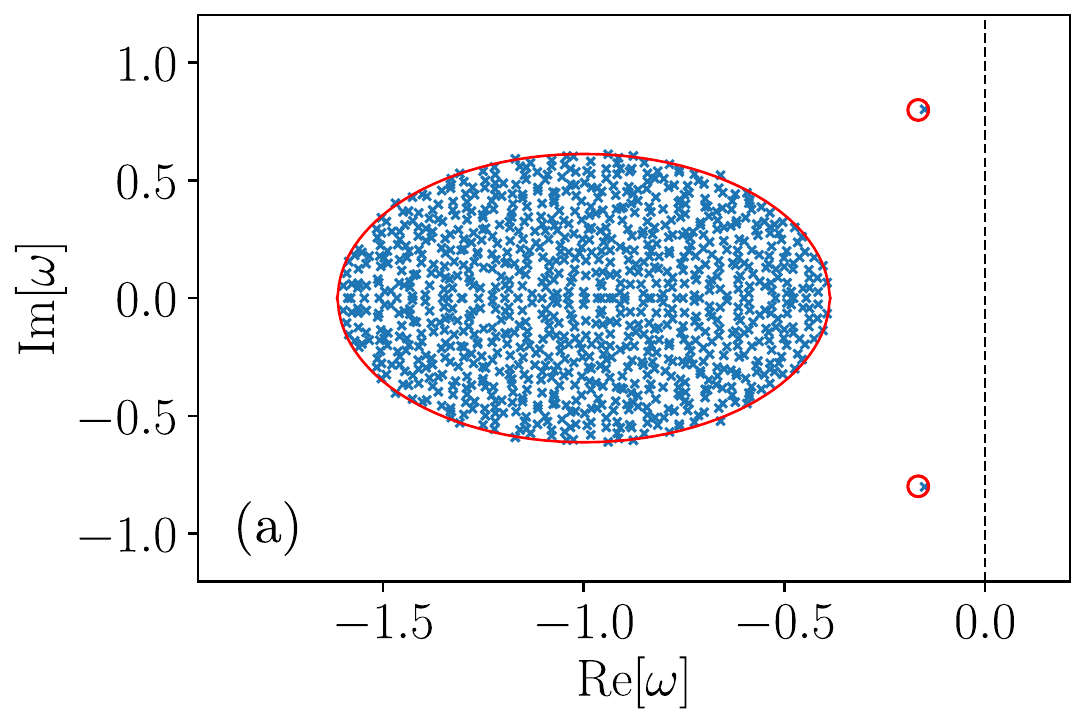}
	\includegraphics[scale = 0.4]{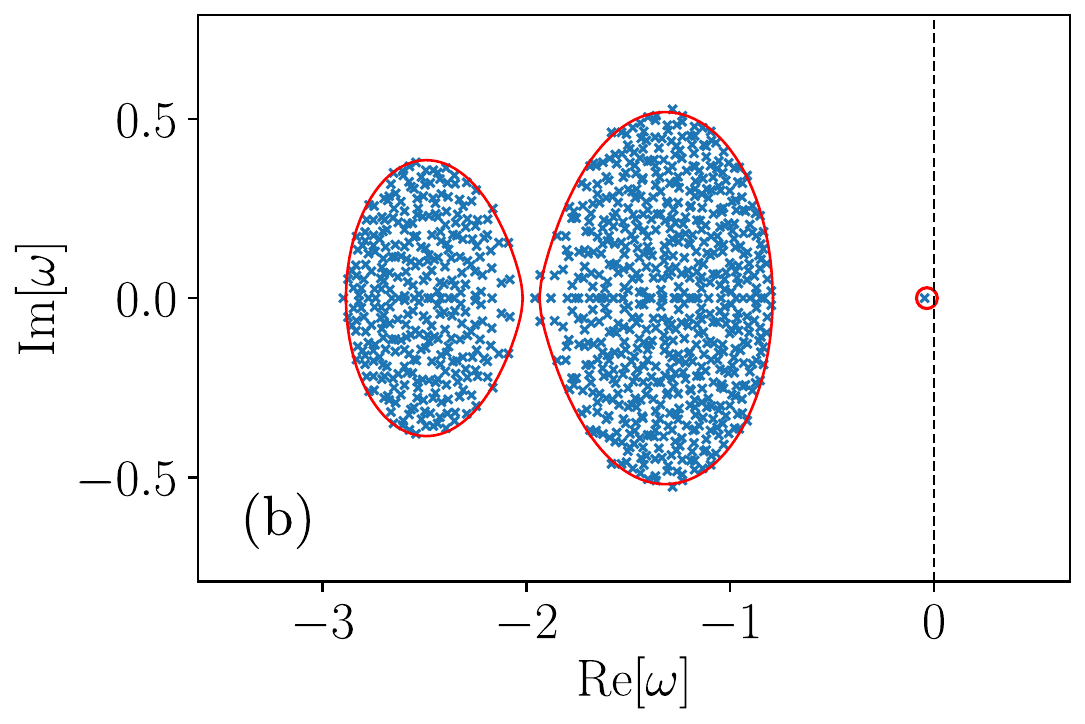}
	\includegraphics[scale = 0.4]{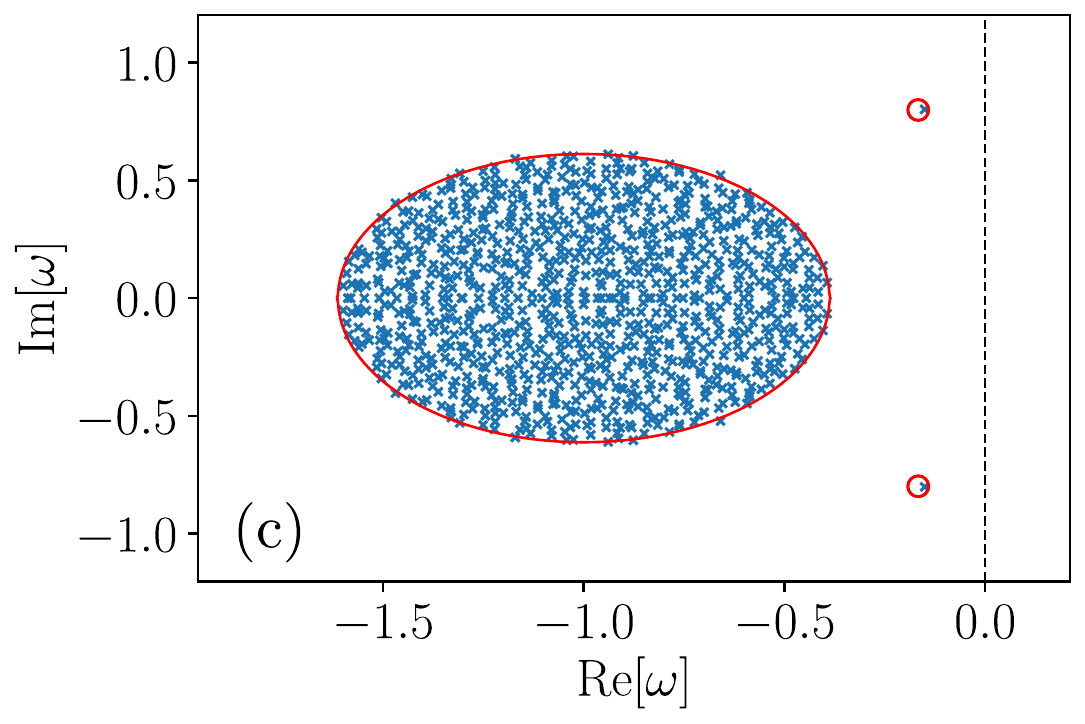}
	\includegraphics[scale = 0.4]{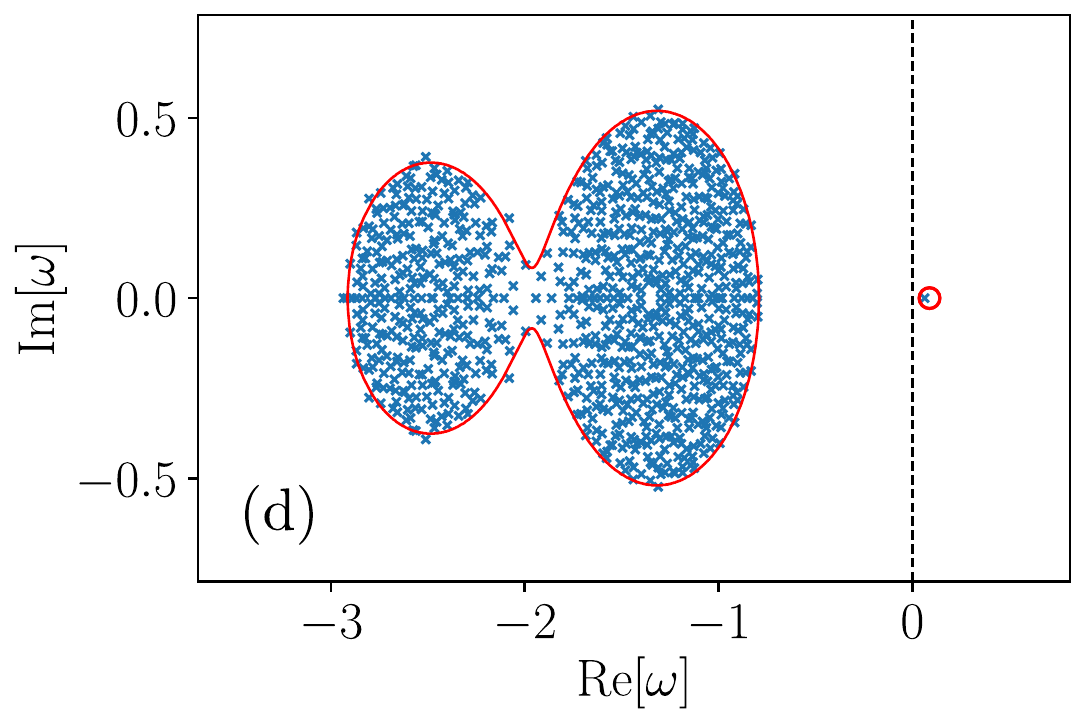}
	\caption{{\bf Example of dispersal-induced instability when diffusion coefficients are heterogeneous.} Eigenvalue spectra for random matrices with trophic structure and statistical variation in the dispersal coefficients. (a) Variation in dispersal coefficients has no effect on the $q=0$ mode. In (b) there is no variation in dispersal coefficients but in (c) there is. In going from (b) to (c), the bulk regions of the spectrum are elongated in the horizontal direction and the outlier is made more unstable.  The system parameters (after rescaling, see Sec.~\ref{section:scaling}) here are $D_u = 1$, $D_v = 5$, $NC\mu_{uu} = NC\mu_{vu} = -NC\mu_{uv} = 3$, $NC\mu_{vv} = -1.5$, $\gamma_u = 2/3$, $\gamma_v = 1/3$, $c = NC\sigma^2 = 0.375$, $d_u = d_v = 1$. The top two panels have $\Delta_0^u = \Delta_0^v = 0$ and the bottom have $\Delta_0^u = \Delta_0^v = 1$. Panels (a) and (c) $q = 0$, panels (b) and (d) $q =1$.   }\label{fig:varyingdispersal}
\end{figure}

Now we imagine that instead of the intra-species auto-repression rates $d^\alpha$, the dispersal rates $D^\alpha$ vary. We assume they are distributed uniformally between $D^\alpha_0 - \Delta_0^\alpha/2$ and $D^\alpha_0 + \Delta_0^\alpha/2$. This is a fairly minor change to make with respect to the preceding analysis, the only difference being that now the degree of diagonal variation depends on $q^2$. In Eqs.~(\ref{uniformdiagonal}), for example, one simply makes the replacement $d_0 \to d_0 + q^2 D_0$ and $\Delta_0 \to  q^2 \Delta_0$. In Fig.~\ref{fig:varyingdispersal} we show eigenvalue spectra for the case of varying dispersal rates. As can be seen variation in the dispersal rates can promote dispersal-induced instability.

\clearpage
\part{Turing instability and the stabilising effects of spatial heterogeneity}\label{part:heterogeneity}
In this part of the Supplement we investigate the effects of spatial heterogeneity. To do this, we allow the interaction matrix to vary across different points in space. Such a setup was considered by Gravel,  Massol and Leibold (`GML') in a model without trophic structure and with all-to-all dispersal between a set of discrete patches  \cite{gravel}. GML report that this results in a stabilising effect for the bulk eigenvalues. The focus of our work is on dispersal-induced instability in complex ecosystems with trophic structure. We therefore ask how spatial heterogeneity impacts such an instability. 

To answer this question we construct a model which combines the features described in the main text and those in GML. That is, we consider the combination of dispersal, spatial heterogeneity in the interaction coefficients, and trophic structure. We detail how the same analytical approach used to produce the results in the main text can be applied in this case. The definition of the combined model and the subsequent calculations can be found in Section \ref{sec:heterogeneity}. In Section~\ref{sec:previous} we demonstrate that this combined set-up contains the model in the main text and that of \cite{gravel} as limiting cases. In particular, we rederive the stability criteria reported in GML and observe the same stabilising effect when heterogeneity and dispersal are combined.

In Section~\ref{sec:gravelextensions} we use this set-up to show that the stabilising mechanism of GML is largely independent of the way dispersal is implemented; that is, whether dispersal is modelled by local or `all-to-all' hopping between patches. We also introduce a non-zero mean to the random matrix elements in GML's model and demonstrate that outlier eigenvalues are not necessarily subject to the stabilising mechanism reported by GML for the bulk eigenvalues.

Conversely, in Section~\ref{sec:all-to-all} we show that dispersal-induced instability persists if the dispersal in the model in the main text is changed from local diffusion in continuous space to non-local `all-to-all' hopping amongst a set of patches (as it is in GML). This further demonstrates that the dispersal-induced effects under consideration are robust under changes in the implementation of dispersal. 

Finally, in Section~\ref{sec:interplay}, we analyse a setup combining the features of the model in Section~\ref{sec:model} and that of GML, i.e., trophic structure, spatial heterogeneity and dispersal. We illustrate that both the stabilising and destabilising dispersal-induced effects can be seen for different sets of model parameters. That is, the two effects can coexist within the same model. We then proceed to identify the roles of the different model parameters in promoting dispersal-induced stabilisation or destabilisation. 

\section{Spatial heterogeneity in interaction coefficients}\label{sec:heterogeneity}

\subsection{Model setup}
In this section, we construct and analyse a model involving trophic structure, spatial heterogeneity of interactions and dispersal. This allows us to investigate the interplay between the Turing instability induced by outlier eigenvalues, and the stabilising mechanism for bulk eigenvalues reported in Ref. \cite{gravel}.

The model involves $n$ discrete `patches', indexed by their spatial coordinate $x$. The coefficient $m_{ij;xx'}^{\alpha\beta}$ encodes the linear response of the abundance of species $i$ belonging to trophic block $\alpha$ at patch $x$ due to a change in abundance of species $j$ belonging to trophic block $\beta$ at patch $x'$. We note that the only way the abundance of a species at one location is affected by the abundance at another location is through migration. So for $x \neq x'$, $m_{ij;xx'}^{\alpha\beta} \propto \delta_{ij}\delta_{\alpha\beta}$ (see Eqs.~(\ref{eq:mstattrophic}) below). 

The introduction of the patch coordinate $x$ in this way gives rise to a higher level of block structure: the matrix is split into $n^2$ blocks corresponding to pairs of patches, and each of these blocks is split into four sub-blocks reflecting the trophic structure of the ecosystem. 

More precisely, the interaction matrix has entries with the following statistics
\begin{align}
\overline{m^{\alpha \beta}_{ij;xx}} &= \frac{\mu_{\alpha\beta}}{N}, \,\,\,\,\, \mathrm{for} \,\,\,\,\, i\neq j, \nonumber \\
\overline{ (m^{\alpha \beta}_{ij;xx})^2}  &= \frac{\sigma^2}{N},   \nonumber \\
\overline{ m^{\alpha \beta}_{ij;xx} m^{\alpha \beta}_{ij;x'x'}} &= \frac{c_{x,x'} \sigma^2}{N}  , \nonumber \\
\overline{ m^{\alpha \beta}_{ij;xx} m^{\beta\alpha}_{ji;xx}} &= \frac{ \Gamma_{\alpha\beta}\sigma^2}{N}  , \nonumber \\
\overline{ m^{\alpha \beta}_{ij;xx} m^{\beta\alpha}_{ji;x'x'}} &= \frac{c_{x,x'} \Gamma_{\alpha\beta}\sigma^2}{N}  , \nonumber \\
m_{ii;xx}^{\alpha \alpha} &= - d_\alpha - D_\alpha, \nonumber \\
m_{ij;xx'}^{\alpha \beta} &= D_\alpha \phi_{x,x'} \delta_{ij}\delta_{\alpha\beta},\label{eq:mstattrophic}
\end{align}
where $x \neq x'$. We have introduced the spatial correlation kernel $c_{x,x'}$, which determines the degree to which interaction coefficients vary from patch to patch. For example, if $c_{x,x'}=0$ for all $x\neq x'$, then the interactions are entirely uncorrelated across patches. If on the other hand $c_{x,x'}=1$ for all $x, x'$ then interactions are the same everywhere (no spatial heterogeneity). When we refer to the matrix $\mathbf{m}$ in this section, we assume the matrix to have this finer block structure and the above statistics.

By taking $c_{x,x'} = 1$ for all $x,x'$, we recover the discrete-patch model discussed in Section \ref{sec:patches1}, which in turn reduces to the model in the main text in the continuum limit. If instead we set $\Gamma_{\alpha\beta} = 0$ and $\mu_{\alpha\beta} = 0$ (i.e. remove statistical correlations and set the mean of the random matrix elements to zero) and do not have any notion of trophic structure (by setting $\gamma_u = 1$ for instance), we recover the model in GML \cite{gravel}.
 
\subsection{Saddle-point calculation for the bulk spectrum}
\subsubsection*{Replica method and Hubbard-Stratononich transformation}
In a similar spirit to the earlier calculation in Section \ref{section:support}, in order to find the bulk of the eigenvalue spectrum, we set the mean of the random matrix elements to zero $\mu_{\alpha\beta} = 0$. Any non-zero mean is later restored in order to find the outlier eigenvalues. We evaluate a potential $\Phi(\omega)$ obeying Eq.~(\ref{potential}) in the limit $N\to \infty$. We begin with the Gaussian representation of the determinant [c.f. Eq.~(\ref{potentialdeterminant})]
\begin{align}\
&\det\left[ (\id_N\omega^\star - \mathbf{m}^T )(\id_N\omega - \mathbf{m} )\right]^{-1} = \int \prod_{i, \alpha,x} \left( \frac{d^2z^\alpha_{i,x}}{\pi}\right) \nonumber \\
&\times \exp\left[ -\epsilon \sum_{i,\alpha,x} \vert z^\alpha_{i,x} \vert^2 - \sum_{i,j,k, \alpha, \beta, \gamma,x,x',x''}z^{\alpha \star}_{i,x} (\omega^\star\delta_{ik}\delta_{\alpha \gamma}\delta_{xx''} - (m^T)_{ik;xx''}^{\alpha \gamma})(\omega\delta_{kj}\delta_{\gamma \beta}\delta_{x'x''} - m^{\gamma \beta}_{kj;x''x'})z^\beta_{j,x'}\right].
\end{align}
Next, we perform a Hubbard--Stratonovich transformation \cite{hubbard} in order to linearise the terms involving $m^{\alpha \beta}_{ij;x,x'}$ yielding
\begin{align}
&\det\left[ (\id_N\omega^\star - \mathbf{m}^T )( \id_N\omega - \mathbf{m} )\right]^{-1} = \int \prod_{i\alpha x} \left( \frac{d^2z^\alpha_{i,x} d^2y^\alpha_{i,x}}{2 \pi^2}\right) \exp\left[ - \sum_{i\alpha x} y^{\alpha\star}_{i,x} y^\alpha_{i,x} + \epsilon z^{\alpha\star}_{i,x} z^\alpha_{i,x} \right]\nonumber \\
\times &\exp\left[ i \sum_{ij\alpha \beta x x'} z^{\alpha\star}_{i,x} \left( m^{ \beta\alpha}_{ji;x'x} - \omega^\star\delta_{ij}\delta_{\alpha \beta} \delta_{xx'} \right) y^\beta_{j x'}\right] \exp\left[ i \sum_{ij\alpha \beta x x'} y^{\alpha\star}_{i,x} \left( m^{\alpha \beta}_{ij;xx'} - \omega  \delta_{x x'}\delta_{ij}\delta_{\alpha \beta}\right) z^\beta_{j,x'}\right]. \label{HStransformed2}
\end{align}
\subsubsection*{Average over the random matrix and introduction of order parameters}
\noindent Carrying out the average over the random matrix elements in Eq.~(\ref{HStransformed2}) then gives 
\begin{align}
\exp\left[ -N \Phi(\omega) \right] &= \int \prod_{i\alpha x} \left( \frac{d^2z^\alpha_{ix} d^2y^\alpha_{ix}}{2 \pi^2}\right) \exp\left[ - \sum_{i\alpha x} y^{\alpha\star}_{i,x} y^\alpha_{ix} + \epsilon z^{\alpha\star}_{i x} z^\alpha_{i x} \right]\nonumber \\
\times & \exp\left[ -i \sum_{i \alpha x} z^{\alpha\star}_{i x} y^\alpha_{i x}  (\omega^\star + d_\alpha + D_\alpha) + z^\alpha_{i x} y^{\alpha\star}_{i x} ( \omega + d_\alpha + D_\alpha)  \right]\nonumber \\
\times & \exp\left[ i \sum_{i \alpha x x'} z^{\alpha\star}_{i x} y^\alpha_{i x'}  D_\alpha \phi_{x,x'} + z^\alpha_{i x} y^{\alpha\star}_{i x'} D_\alpha \phi_{x,x'} \right]\nonumber \\
\times & \exp\left[ - \frac{\sigma^2}{2N}\sum_{i j \alpha \beta x } (z_{i x }^{\alpha\star} y^\beta_{j x} + z^\alpha_{i x} y_{j x}^{\beta\star})^2 + \Gamma_{\alpha \beta} (z_{ix}^{\alpha\star} y^\beta_{jx} + z^\alpha_{ix} y_{jx}^{\beta\star})(z_{jx}^{\beta\star} y^\alpha_{ix} + z^\beta_{jx} y_{ix}^{\alpha\star}) \right]\nonumber \\
\times & \exp\left[ - \frac{\sigma^2}{2N}\sum_{i j \alpha \beta x x' } c_{x,x'} (z_{ix}^{\alpha\star} y^\beta_{jx} + z^\alpha_{ix} y_{jx}^{\beta\star})(z_{ix'}^{\alpha\star} y^\beta_{jx'} + z^\alpha_{ix'} y_{jx'}^{\beta\star}) \right] \nonumber \\
\times & \exp\left[ - \frac{\sigma^2}{2N}\sum_{i j \alpha \beta x x' } c_{x,x'} \Gamma_{\alpha\beta} (z_{ix}^{\alpha\star} y^\beta_{jx} + z^\alpha_{ix} y_{jx}^{\beta\star})(z_{jx'}^{\beta\star} y^\alpha_{ix'} + z^\beta_{jx'} y_{ix'}^{\alpha\star}) \right]. \label{averaged2}
\end{align}
\noindent To capture the non-trivial spatial variation in the interaction matrix elements between patches, we define the following order parameters (c.f. Eq.~(\ref{orderparameters}))
\begin{align}
u_{x,x'} &= \frac{1}{N} \sum_{i \alpha } z^{\alpha\star}_{i ,x  } z^\alpha_{ix'}, \,\,\,\,\, v_{x,x'} = \frac{1}{N } \sum_{i \alpha } y^{\alpha\star}_{i,x } y^\alpha_{i x'}, \nonumber \\
w_{\alpha ; x,x'} &= \frac{1}{N_\alpha } \sum_{i } z^{\alpha\star}_{i x} y^\alpha_{i,x '}, \,\,\,\,\, w_{\alpha; x,x'}^\star = \frac{1}{N_\alpha } \sum_{i } y^{\alpha\star}_{i,x} z^\alpha_{i x'}, \label{orderparameters2}
\end{align}
which we impose in the integral Eq.~(\ref{averaged2}) using Dirac delta functions in their complex exponential representation. We can thus rewrite Eq.~(\ref{averaged2}) as 
\begin{align}
\exp\left[ -N \Phi(\omega) \right] &= \int \mathcal{D}\left[ \cdots \right]\exp\left[N n(\Psi + \Theta + \Omega)\right] ,\label{saddlesetup2}
\end{align}
where $\mathcal{D}\left[ \cdots \right]$ denotes integration over all of the order parameters and their conjugate (`hatted') variables, and where 
\begin{align}
\Psi =& i \sum_{x,x'}\left[ \hat u_{x,x'} u_{x,x'}  + \hat v_{x,x'}  v_{x,x'}  + \sum_{\alpha}\gamma_\alpha(\hat w_{\alpha;x,x'}  w^\star_{\alpha; x,x'} +   \hat w^\star_{\alpha; x,x'}  w_{\alpha ; x,x'}   ) \right], \nonumber \\
\Theta =& \sum_x\Bigg\{-\epsilon  u_{x,x} - v_{x,x} - \sigma^2 \sum_{x'}c_{x,x'}u_{x,x'} v_{x',x} + i \sum_{\alpha x'} \phi_{x,x'} D_\alpha \left[ w_{\alpha; x,x'} + w^\star_{\alpha; x,x'}\right] \nonumber \\
&\,\,\,\,\,-i \sum_\alpha \gamma_\alpha\left[     w_{\alpha;x,x} (\omega^\star + d_\alpha + D_\alpha)  + w^\star_{\alpha; x,x} (\omega + d_\alpha + D_\alpha)\right]  \nonumber \\
&\,\,\,\,\, - \frac{1}{2} \sigma^2 \sum_{\alpha \beta}\gamma_\alpha \gamma_\beta\Gamma_{\alpha \beta}c_{x,x'}[w_{\alpha; x,x'} w_{\beta; x',x}+ w_{\alpha; x',x}^\star w_{\beta; x,x'}^\star] \Bigg\} , \nonumber \\
\Omega =& \sum_\alpha \gamma_\alpha \ln  \Bigg[\int \prod_{x}  \left( \frac{d^2z^\alpha_x d^2y^\alpha_x}{2 \pi^2}\right)  \exp\bigg\{ -i \sum_{x x'} (\hat u_{x,x'} z^{\alpha\star}_{x} z^\alpha_{x'}  + \hat v_{x,x'} y^{\alpha\star}_{x} y^\alpha_{x'} \nonumber \\
& \,\,\,\,\, + \hat w_{\alpha;x,x'} y^{\alpha\star}_{x} z^\alpha_{x'}  + \hat w_{\alpha; x,x'}^\star z^{\alpha\star}_x y^\alpha_{x'}  ) \bigg\} \Bigg].
\end{align}
To proceed we now assume that the correlation kernel $c_{x,x'}$  is only a function of the distance $|x-x'|$. The order parameters ($w_{\alpha;x,x'}$ and so on) then also only depend $\vert x-x'\vert$. We can exploit this translational invariance to simplify the calculation through the use of discrete Fourier transforms.

We introduce the Fourier transformed variables $\tilde z_{iq}$ and $\tilde y_{iq}$ through
 \begin{align}
\tilde z_{iq} = \sum_x e^{iqx} z_{ix}, \,\,\,\,\, \tilde y_{iq} =  \sum_x e^{iqx} y_{ix}.
\end{align}
The inverse transform is given by
\begin{align}
z_{ix} = \frac{1}{n}\sum_q e^{-iqx}\tilde z_{iq}, \,\,\,\,\, y_{ix} = \frac{1}{n}\sum_q e^{-iqx}\tilde y_{iq}.
\end{align}
We drop the tilde from further notation for the sake of compactness (all variables with subscript $q$ or $q'$ are Fourier transforms). One then obtains 
\begin{align}
\Psi &= i \sum_{q}\left[ \hat u_{q} u_{-q}  + \hat v_{q}  v_{-q}  + \sum_{\alpha}\gamma_\alpha(\hat w_{\alpha;q}  w^\star_{\alpha; -q} +   \hat w^\star_{\alpha; q}  w_{\alpha ; -q}   ) \right], \nonumber \\
\Theta &= \sum_q\Bigg\{-\epsilon  u_{q} - v_{q} - \frac{\sigma^2}{n}  \sum_{q'}c_{q-q'}u_{q}v_{q'} -i \sum_\alpha \gamma_\alpha\left[     w_{\alpha;q} (\omega^\star + d_\alpha + D_\alpha)  + w^\star_{\alpha; q} (\omega + d_\alpha + D_\alpha)\right]  \nonumber \\
&+ i  \sum_{\alpha } \phi_{q} D_\alpha \left[ w_{\alpha; q} + w^\star_{\alpha; q}\right] - \frac{\sigma^2}{2 n}  \sum_{\alpha \beta; q'}\gamma_\alpha \gamma_\beta\Gamma_{\alpha \beta} c_{q-q'}[w_{\alpha; q}  w_{\beta; q'}+ w_{\alpha; q'}^\star  w_{\beta; q}^\star]\Bigg\} ,\nonumber \\
\Omega &= -\sum_\alpha \gamma_\alpha\ln(\hat w_{\alpha;q} \hat w_{\alpha;q}^\star - \hat u_q \hat v_q) + \mbox{const.}
\end{align}

\subsubsection*{Saddle-point integration and evaluation of order parameters}
We now take the limit $N\to\infty$, and carry out the integral in Eq.~(\ref{saddlesetup2}) using the saddle-point method. To do this, we extremise the expression $\Psi + \Theta + \Omega$ with respect to the order parameters and the conjugates. Extremising with respect to the conjugate variables $\hat u, \hat v, \hat w_\alpha$ and $\hat w_\alpha^\star$, we find
\begin{align}
i  u_{-q} &= \sum_\alpha \frac{ -\gamma_\alpha \hat v_q}{\hat w_{\alpha;q} \hat w_{\alpha;q}^\star - \hat u_q \hat v_q } , \,\,\,\,\, iv_{-q} = \sum_\alpha \frac{ -\gamma_\alpha \hat u_q}{\hat w_{\alpha;q} \hat w_{\alpha;q}^\star - \hat u_q \hat v_q} , \nonumber \\ 
i\gamma_\alpha w_{\alpha;-q} &= \frac{\gamma_\alpha\hat w_{\alpha;q}}{\hat w_{\alpha;q} \hat w_{\alpha;q}^\star - \hat u_q \hat v_q} , \,\,\,\,\,  i\gamma_\alpha w_{\alpha;-q}^\star = \frac{\gamma_\alpha \hat w_{\alpha;q}^\star}{\hat w_{\alpha;q} \hat w_{\alpha;q}^\star - \hat u_q \hat v_q } .\label{diffwrthatted2}
\end{align}
Similar to Section \ref{sec:av}, we simplify matters by defining $\hat f_{\alpha;q} = \frac{1}{(\hat w_{\alpha;q} \hat w_{\alpha;q}^\star - \hat u_q \hat v_q)}$, and we obtain an implicit equation for $\hat f_{\alpha;q}$ in terms of $r_{-q}$ and $w_{\alpha;-q} w^\star_{\alpha;-q}$:
\begin{align}
- w_{\alpha;-q} w^\star_{\alpha;-q} &= \hat f_{\alpha;q}-\hat f_{\alpha;q}^2 \frac{  r_{-q}}{\left[\sum_\beta \gamma_\beta \hat f_{\beta;q}\right]^2} .\label{fdefinition}
\end{align} 
We now extremise $\Psi + \Theta + \Omega$ in Eq.~(\ref{saddlesetup2}) with respect to  $u_{q}, v_{q}, w_{\alpha;q}$ and $w_{\alpha;q}^\star$. We find
\begin{align}\label{eq:aux1}
i\hat u_{-q} &= \epsilon + \frac{\sigma^2}{n}\sum_{q'} c_{q-q'}v_{q'} , \,\,\,\,\, i\hat v_{-q} = 1 + \frac{\sigma^2}{n} \sum_{q'} c_{q-q'}u_{q'} , \nonumber \\
i\hat w_{\alpha;-q} &=  i[\omega + d_\alpha + D_\alpha(1- \phi_{q})] +  \frac{\sigma^2}{n} \sum_{\beta}\Gamma_{\alpha \beta}\gamma_\beta \sum_{q'} c_{q-q'} w^\star_{\beta;q'} , \nonumber \\
i\hat w_{\alpha;-q}^\star &= i [\omega^\star + d_\alpha + D_\alpha(1- \phi_{q})] + \frac{\sigma^2}{n} \sum_{\beta}\Gamma_{\alpha \beta}\gamma_\beta \sum_{q'}c_{q-q'}w_{\beta;q'} .
\end{align}
Combining these with Eqs.~(\ref{diffwrthatted2}), one finally obtains
\begin{align}
v_q &= \left(\epsilon + \frac{\sigma^2}{n}\sum_{q'} c_{q-q'}v_{q'}\right)\sum_\alpha \gamma_\alpha \hat f_{\alpha;-q} , \nonumber \\
u_q &= \left(1 + \frac{\sigma^2}{n}\sum_{q'}c_{q-q'}u_{q'} \right)\sum_\alpha \gamma_\alpha \hat f_{\alpha;-q}  , \nonumber \\
w_{\alpha;q} &= -\left[  i(\omega + d_\alpha + D_\alpha (1- \phi_{q}))+  \frac{\sigma^2}{n} \sum_{\beta,q'} \Gamma_{\alpha \beta} \gamma_\beta c_{q-q'}w^\star_{\beta;q'}\right]\hat f_{\alpha;-q} , \nonumber \\
w^\star_{\alpha;q} &=  -\left[  i(\omega^\star + d_\alpha + D_\alpha (1- \phi_{q}))+  \frac{\sigma^2}{n} \sum_\beta \Gamma_{\alpha \beta} \gamma_\beta c_{q-q'}w_{\beta;q'}\right]\hat f_{\alpha;-q} . \label{unhatted2}
\end{align}
If we compare this result with the one in Eq.~(\ref{unhatted}), we observe two key differences: (1) The Fourier transform of the continuous Laplacian operator $-q^2$ is now replaced with its discrete counterpart $1-\phi_q$, (2) Because of the imperfect correlation between the interaction coefficients in different patches, the eigenvalue spectra for different Fourier modes $q$ are no longer independent. This interdependence is encoded in the Fourier transform of the correlation kernel $c_{q-q'}$. For perfect correlation ($c_{x,x'}= 1$ for $x$ and $x'$) this reduces to $c_{q-q'} \propto \delta_{q,q'}$ and the eigenvalue spectra for different modes $q$ once again decouple.

Although obtaining a general solution of Eqs.~(\ref{unhatted2}) is complicated (while technically possible), the support of the eigenvalue spectrum can be found more easily in various special cases. We explore some of these in the following sections.

\subsection{Outlier eigenvalues}\label{sec:outliers2}
\subsubsection*{Setup and general relations}
In a similar fashion to Section \ref{sec:outliers}, we now wish to find the outliers that arise when we perturb the random matrix $\mathbf{m}$ [whose statistics are given in Eq.~(\ref{eq:mstattrophic}) with $\mu_{\alpha\beta} = 0$] by introducing a finite mean. That is, we are now interested in the matrix with elements 
\be\label{eq:mshift2}
m'^{\alpha \beta}_{ij;x,x'} =m^{\alpha \beta}_{ij;x,x'} + \frac{\mu_{\alpha \beta}}{N} \delta_{x,x'},
\ee
and we define the matrix
\be
\left(\boldsymbol{\upmu}\right)^{\alpha\beta}_{ij;x,x'}=\mu_{\alpha \beta} \delta_{x,x'},
\ee
which has elements independent of $i$ and $j$. 

Similar to the calculation in Section~\ref{sec:outliers} any eigenvalue of the perturbed matrix  outside the bulk of the unperturbed eigenvalue spectrum must satisfy
\begin{align}
\det\left[ \id_{Nn}   - \frac{1}{N}(\lambda\id_{Nn}  - \mathbf{m})^{-1} \boldsymbol{\upmu} \right] &= 0 . \label{determinantsetup2}
\end{align}

\subsubsection*{Simplification to an effective 2n-species problem}
\noindent We once again consider the resolvent matrix $\mathbf{G}= (\lambda\id_{Nn}  - \mathbf{m})^{-1}$. Using reasoning along the lines of Section \ref{sec:outliers}, one arrives at a similar expression to Eq.~(\ref{reduceddimensiondet})
\begin{align}
\det\left[ \id_{2n}   -  \boldsymbol{\upmu}^{(2n)} \frac{1}{N}\sum_{ij} \mathbf{G}^{(ij)}(\lambda) \right] &= 0 ,\label{reduceddimensiondet2}
\end{align}
where here $\left(\mathbf{G}^{(ij)} \right)_{\alpha\beta;x,x'}= \left((\id_N \lambda - \mathbf{m})^{-1}\right)^{\alpha \beta}_{ij;x,x'}$ instead. 

\subsubsection*{Final expression for outliers}
\noindent Assuming the translational invariance of the averaged resolvent, one finds using Eq.~(\ref{blockmatrixtheorems}) in combination with the discrete Fourier transform (which can be represented as an orthonormal transformation)
\begin{align}
\det\left[ \id_{2n}   -  \boldsymbol{\upmu}^{(2n)} \frac{1}{N}\sum_{ij} \mathbf{G}^{(ij)}(\lambda) \right] = \prod_q \det\left[ \id_{2}   -  \boldsymbol{\upmu}^{(2)} \frac{1}{N}\sum_{ij} \tilde{\mathbf{G}}^{(ij)}_q(\lambda) \right] &= 0 ,
\end{align}

\noindent Given that the resolvent matrix is approximately diagonal for large $N$ (see Section~\ref{sec:outliers}), we only need to calculate the sum of diagonal elements $\frac{1}{N}\sum_{i} \tilde G_{ii;q}^{\alpha\alpha}$ in Eq.~(\ref{reduceddimensiondet2}). As in the calculation for the model used in the main text, we can identify these averaged resolvents with the order parameters $i \gamma_\alpha w^{\star}_{q;\alpha}$.

\noindent Our final equation for the outliers $\lambda$ is 
\begin{align}
\prod_q\det\left[ \id_2   -  \boldsymbol{\upmu}^{(2)}\,\boldsymbol{\upgamma}\, \boldsymbol{\upchi}_q(\lambda)\right] &= 0, 
\end{align}
with
\begin{align}  
\boldsymbol{\upgamma}(\lambda) = \begin{bmatrix}
\gamma_u & 0 \\
0 & \gamma_v
\end{bmatrix}, \,\,\,\,\,
\boldsymbol{\upchi}_q(\lambda) &=   \begin{bmatrix}
\chi_{u;q} & 0 \\
0 & \chi_{v;q}
\end{bmatrix} .
\end{align}
This can be written as
\begin{align}
\prod_q \left\{\left[ \gamma_u\mu_{uu} - \frac{1}{\chi_{u;q}(\lambda)}\right] \left[\gamma_v \mu_{vv} - \frac{1}{\chi_{v;q}(\lambda)}\right] - \gamma_u\gamma_v\mu_{uv}\mu_{vu}\right\} = 0. \label{productq}
\end{align}
Each value of $q$ thus yields possible values for $\lambda$. We consider each value of $q$ individually:
\begin{align}
\left[ \gamma_u\mu_{uu} - \frac{1}{\chi_{u;q}(\lambda)}\right] \left[\gamma_v \mu_{vv} - \frac{1}{\chi_{v;q}(\lambda)}\right] - \gamma_u\gamma_v\mu_{uv}\mu_{vu} = 0.\label{simult12}
\end{align}
We note that this takes the same form as in Eq.~(\ref{simult1}). From Eqs.~(\ref{unhatted2}), the quantities $\chi_{\alpha;q}(\lambda) = i w_{q;\alpha}^{ \star}$ satisfy 
\begin{align}
- 1 &=  - [\lambda + d_\alpha + D_\alpha(1-\phi_{q}) ]\chi_{\alpha;q} +  \frac{\sigma^2}{n}\sum_{\beta;q'}\gamma_{\beta}\Gamma_{\alpha\beta}  c_{q-q'} \chi_{\alpha;q}\chi_{\beta;q'} , \label{simult22}
\end{align}
and so the outliers for different values of $q$ remain coupled through these equations. 

\noindent As with the model used in the main text, not all solutions of this set of equations are valid. At the saddle point, the order parameter $u_q $ must satisfy $u_q\geq 0$. This gives an additional criterion which is important for eliminating invalid solutions for $\lambda$.

\section{Previous results as limiting cases}\label{sec:previous}
\subsection{Model in the main text}\label{sec:maintextmodel}
We now recover the model in the main text from Eqs.~(\ref{eq:mstattrophic}) by taking the continuum limit in a similar way to Section \ref{sec:patches1}, imposing that interactions do not vary in space, and making dispersal local. More precisely, we impose the following: (1) There are a large number patches, (2) these patches are spaced a small distance apart, (3) dispersal is only allowed between adjacent patches, (4) the interaction coefficients are the same in every patch (i.e. there is perfect correlation between the interaction coefficients in different patches).  

The reduction can then be seen as follows. We set the correlation kernel to be constant in space $c_{x,x'} = 1$. The Fourier transform of the correlation kernel is then given by $\tilde c_{q-q'} = n\delta_{q,q'}$. Immediately, we see that the first two equations in Eq.~(\ref{unhatted2}) decouple and we obtain
\begin{align}
- w_{\alpha;-q} w^\star_{\alpha;-q} &= \hat f_{\alpha;q}-\hat f_{\alpha;q}^2 \frac{  r_{-q}}{\left[\sum_\beta \gamma_\beta \hat f_{\beta;q}\right]^2} , \nonumber \\
v_q &= \left(\epsilon + \sigma^2 v_{q}\right)\sum_\alpha \gamma_\alpha \hat f_{\alpha;-q} , \nonumber \\
u_q &= \left(1 + \sigma^2 u_{q} \right)\sum_\alpha \gamma_\alpha \hat f_{\alpha;-q} , \label{uncoupled1}
\end{align}
which are equivalent to the first two of Eqs.~(\ref{unhatted}) and the first of Eqs.~(\ref{rneq0}). We now choose the following local dispersal kernel
\begin{align}
\phi(x,x') = \begin{cases}
\frac{1}{2} \,\,\,\,\,\, \mathrm{if} \,\,\,\,\, x = x' + \Delta x ,\\
\frac{1}{2} \,\,\,\,\,\, \mathrm{if} \,\,\,\,\, x = x' - \Delta x, \\
0 \,\,\,\,\,\, \mathrm{otherwise},
\end{cases} \label{localhoppingkernel}
\end{align}
where the separation between patches $\Delta x$ is assumed small. We take the continuum limit in a similar way to Section~\ref{sec:patches1}. Noting that once again $\phi_q = \cos(\Delta x\, q)$, the last two equations in Eq.~(\ref{unhatted2}) become
\begin{align}
w_{\alpha;q} &= -\left[  i(\omega + d_\alpha + D'_\alpha q^2)+  \sigma^2 \sum_{\beta} \Gamma_{\alpha \beta} \gamma_\beta w^\star_{\beta;q'}\right]\hat f_{\alpha;-q} , \nonumber \\
w^\star_{\alpha;q} &=  -\left[  i(\omega^\star + d_\alpha + D'_\alpha q^2)+  \sigma^2 \sum_\beta \Gamma_{\alpha \beta} \gamma_\beta w_{\beta;q'}\right]\hat f_{\alpha;-q} , 
\end{align}
and Eq.~(\ref{simult22}) for the outliers becomes
\begin{align}
- 1 &=  - [\lambda + d_\alpha + D'_\alpha q^2 ]\chi_{\alpha;q} +  \sigma^2 \gamma_{\beta}\Gamma_{\alpha\beta}   \chi_{\alpha;q}\chi_{\beta;q}  
\end{align}
where $D'_\alpha = D_\alpha (\Delta x)^2/2$. These take the same form as in Eqs.~(\ref{unhatted}) and Eqs.~(\ref{simult2}) respectively. Hence, with appropriate scaling of the hopping constant, we recover the continuum model used in the main text. 

\subsection{Stabilising mechanism reported by Gravel, Massol and Leibold (`GML') \cite{gravel}}\label{sec:recovergravel}
\subsubsection*{Specifications of the model}
We now show how the results by Gravel, Massol and Leibold (GML) can be recovered. The model by GML does not contain trophic block-structure, so we need to remove the distinction between the groups of predator and prey species. This can be achieved by setting $\gamma_u=1$ (implying $\gamma_v=0$). The group labelled `$v$' in our model is then no longer present. In the rest of this subsection (\ref{sec:recovergravel}) for simplicity write $D$ instead of $D_u$ and $d$ instead of $d_u$. We also replace $\Gamma_u$ by $\Gamma$, $\mu_{uu}$ by $\mu$ and the remove the subscript $u$ in the order parameters. 

Further, the interaction matrix entries for a single patch in GML \cite{gravel} are each drawn independently -- correlations between random matrix entries are not considered ($\Gamma = 0$). Finally, the mean of the interaction matrix elements is also zero ($\mu = 0$). As a result, there are no outlier eigenvalues.

With these simplifications, the model by GML can be obtained from Eqs.~(\ref{eq:mstattrophic}) by choosing the following dispersal kernel and correlations between interactions in different patches,
\begin{align}
c_{x,x'} &= \delta_{x,x'} + (1-\delta_{x,x'}) \rho , \nonumber \\
\phi_{x,x'} &= (1-\delta_{x,x'}) \frac{1}{n-1}, \label{alltoallkernels}
\end{align}
The degree to which there is spatial heterogeneity (or conversely, how much different patches are correlated) is quantified by the parameter $\rho$. This may vary between $0$ (patches are completely independent) and $1$ (where all patches have identical interaction coefficients). With the aforementioned restrictions in mind, we now show that the results for the eigenvalue spectra derived in \cite{gravel} also follow from Eqs.~(\ref{unhatted2}). From these, we then also demonstrate that stabilisation can be result from the combination of dispersal and spatial heterogeneity.

\subsubsection*{Eigenvalue spectra}
\noindent The Fourier transforms of the kernels in Eqs.~(\ref{alltoallkernels}) are
\begin{align}
c_{q} &= \left[1+ \rho(n\delta_{q,0}-1)\right] , \nonumber \\
\phi_{q} &=  \frac{(n \delta_{q,0} -1)}{n-1} . \label{transformedalltoall}
\end{align}
We observe that there are now only two relevant cases for the value of $q$. The first is $q=0$ and the second encompasses all values $q\neq 0$. In particular we have $v_q = v_{q'}$ (for $q$ and $q'$ both non-zero) and similarly for the other order parameters.

\noindent In the limit $\epsilon \to 0^+$ [c.f. Eq.~(\ref{eq:aux1})], the first of  Eqs.~(\ref{unhatted2}) becomes
\begin{align}
v_0 &= \frac{\sigma^2}{n} \left[[1+\rho(n-1)]v_0 + (n-1)(1-\rho)v_{q} \right] \hat f_{0} , \nonumber \\
v_q &= \frac{\sigma^2}{n} \left[(n-1+\rho)v_q + (1-\rho)v_{0} \right] \hat f_{-q} , \label{eigenvalueproblem}
\end{align}
where again $q\neq 0$. There are two solutions to these equations. One is the trivial solution $v_0=v_q=0$. The other is found by recognising Eqs.~(\ref{eigenvalueproblem}) as an eigenvector problem. Thus for non-trivial $v_0$ and $v_q$ we must have
\begin{align}
&\hat f_{0} \left[ 1 + \rho (n-1)\right] + \hat f_{-q} \left[ \rho + n -1\right] \nonumber \\
&+ \frac{\sigma^2}{n} \hat f_{0} \hat f_{-q} \left[ (1-\rho)^2 (n-1) - (1 + \rho(n-1))(\rho + n -1) \right] = \frac{n}{\sigma^2} . \label{boundcond}
\end{align}
The following conditions must be satisfied in order for the order parameters to have the correct properties (i.e. sign and complex conjugate relations)
\begin{gather}
- w_{0}^\star w_{0} > 0, \,\,\,\,\, - w_{0}^\star w_{0} \in \mathbb{R} , \,\,\,\,\,  - w_{q}^\star w_{q} > 0, \,\,\,\,\, - w_{q}^\star w_{q} \in \mathbb{R} , \nonumber \\
1 - \frac{\sigma^2}{n} [1 + \rho (n-1)]  \hat f_{0} >0 , \,\,\,\,\, 1 - \frac{\sigma^2}{n} [\rho + n-1] \hat f_{-q} >0  .
\end{gather}
When $v_0= v_q = 0$, we have $\hat f_{0} = - w^\star_{0}w_{0}$ and $\hat f_{-q} = - w^\star_{-q}w_{-q}$ [by definition -- c.f. Eq.~(\ref{fdefinition})]. Therefore, the last two of Eqs.~(\ref{unhatted2}) yield analytic expressions for $iw^\star_{0}$ and $iw^\star_{q}$ in $\omega$, meaning that there is zero eigenvalue density where this trivial solution is the only one possible. The boundary of the region of the complex plane where there is non-zero eigenvalue density is given by the values of $\omega$ for which Eq.~(\ref{boundcond}) and $\hat f_{0} = - w^\star_{0}w_{0}$ and $\hat f_{-q} = - w^\star_{-q}w_{-q}$ are simultaneously true. Together with Eqs.~(\ref{unhatted2}), we have
\begin{align}
\hat f_{0} = \frac{1}{\vert \omega + d \vert^2} , \,\,\,\,\, \hat f_{-q} = \frac{1}{\vert \omega + d + D n/(n-1)\vert^2} .
\end{align}
Combining this with Eq.~(\ref{boundcond}) yields for the boundary of the eigenvalue support
\begin{align}
&\frac{n}{\sigma^2} \vert \omega + d + D n/(n-1)\vert^2 \vert \omega + d\vert^2 - \left[ 1 + \rho (n-1)\right] \vert \omega + d + D n/(n-1)\vert^2 - \left[ \rho + n -1\right] \vert \omega + d \vert^2 \nonumber \\
&= \frac{\sigma^2}{n}\left[ (1-\rho)^2 (n-1) - (1 + \rho(n-1))(\rho + n -1) \right] . \label{alltoallsupportgravel}
\end{align}
In the case $\rho = 0$, this simplifies further and we obtain
\begin{align}\label{eq:aux2}
\frac{1}{\vert \omega + d \vert^2} + \frac{(n-1)}{\vert \omega + d + D n/(n-1)\vert^2} = \frac{n}{\sigma^2}.
\end{align}
Simplification is also possible in the case $\rho = 1$. The support of the eigenvalue spectrum here is given by the union of the area within two circles:
\begin{align}\label{eq:aux3}
\vert \omega + d \vert^2 = \frac{\sigma^2}{n}, \,\,\,\,\, \vert \omega + d + D n/(n-1)\vert^2 = \frac{\sigma^2}{n}.
\end{align}
The expressions in Eqs.~(\ref{eq:aux2}) and (\ref{eq:aux3}) were reported in Ref. \cite{gravel} [see the discussion in Section S1 of the  Supplementary Material of \cite{gravel} and in particular Eq.~(1.6)] but were arrived at using a different method. 

The result for general $0<\rho<1$ in Eq.~(\ref{alltoallsupportgravel}) was not explicitly written down in \cite{gravel}. Using our approach, we are able to extend the results to matrices with block structure and more complicated statistics ($\Gamma\neq 0$ and $\mu \neq 0$) and local dispersal (as opposed to all-to-all), as is discussed in more detail in Section \ref{sec:gravelextensions}.
\subsubsection*{Stabilising mechanism}
\begin{figure}[H]
	\centering
	\includegraphics[scale = 0.4]{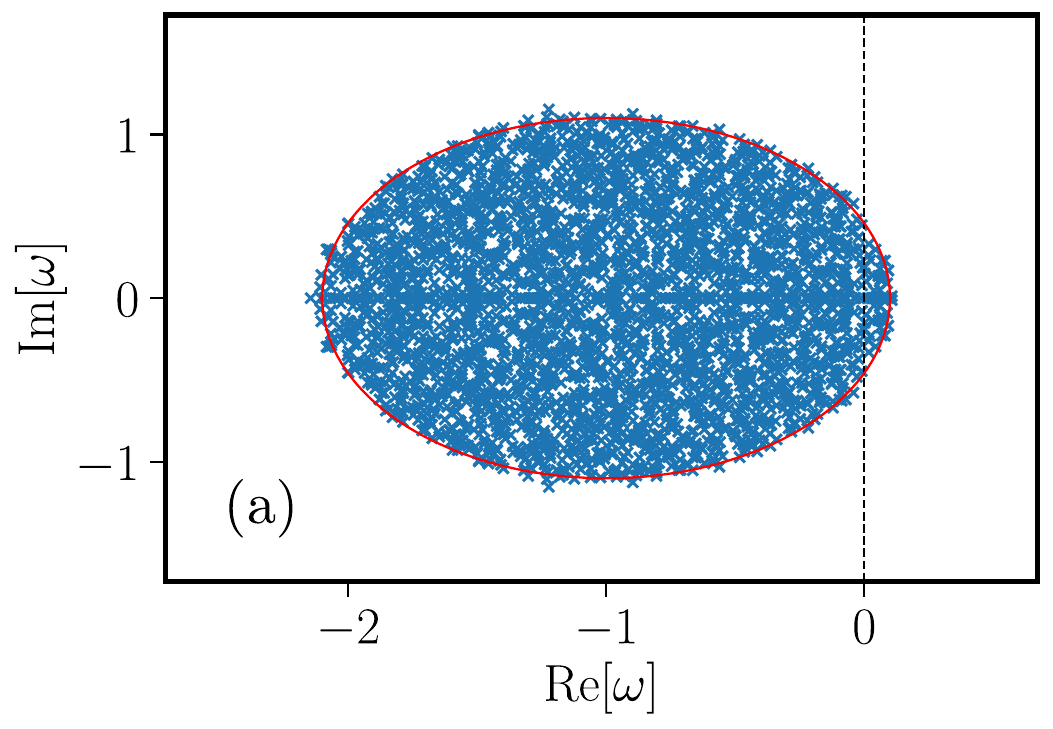} 
	\includegraphics[scale = 0.4]{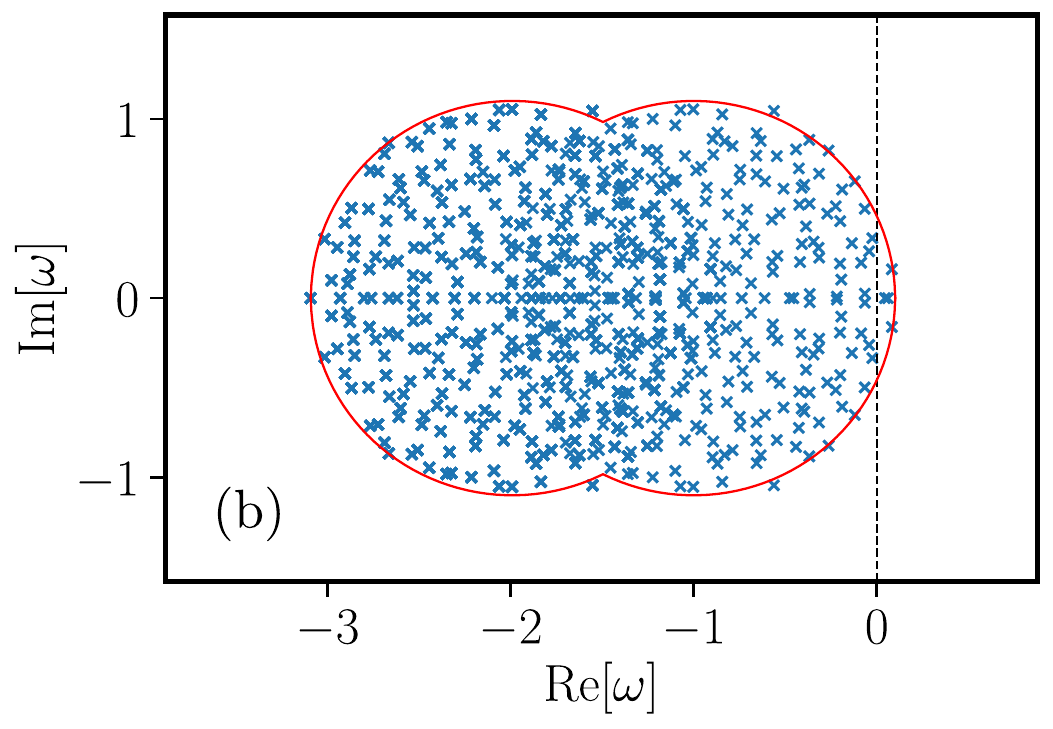}
	\includegraphics[scale = 0.4]{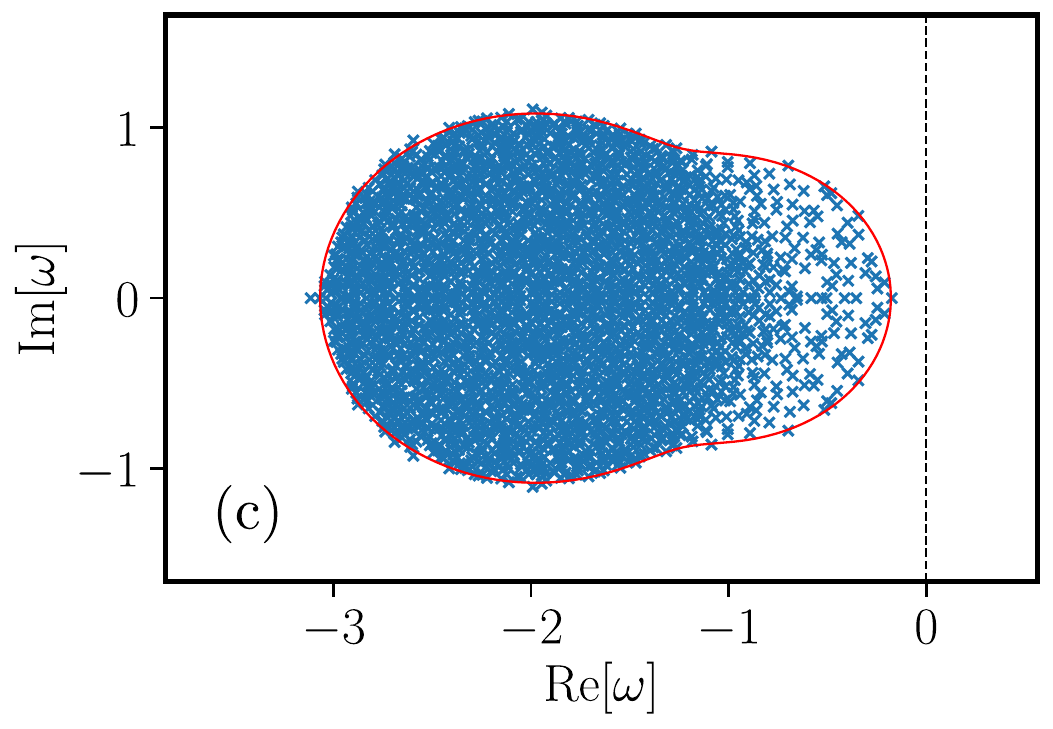}
	\includegraphics[scale = 0.4]{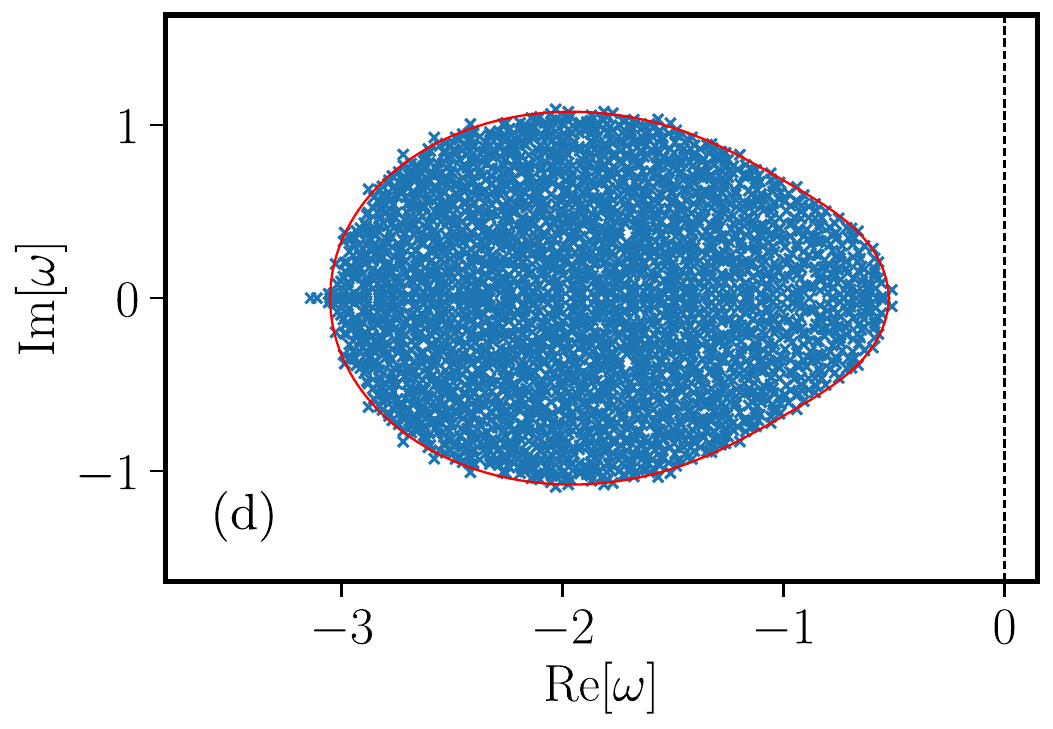}
	\caption{{\bf Dispersal can stabilise an equilibrium when interactions are heterogeneous in space.} The figure shows eigenvalue spectra for instances of the model in \cite{gravel}. Spectra of computer-generated random interaction matrices are shown as blue crosses. Red lines are the theoretical predictions for the boundary of the support from Eq.~(\ref{alltoallsupportgravel}). In all panels (after rescaling, see Sec.~\ref{section:scaling}), $d = 1$, $n=11$, $N = 300$, $C =1$, $c = NC\sigma^2 = 1.21$. In panel (a) there is no dispersal ($D=0$) and the equilibrium is unstable (data is for $\rho=1$). In panel (b) there is now dispersal ($D = 0.9$) but no spatial heterogeneity ($\rho = 1$). The equilibrium is still unstable. In panel (c) spatial heterogeneity and dispersal has been introduced ($\rho = 0.5,~D=0.9$); the disk in the spectrum corresponding to the $q=0$ mode has become smaller, and the equilibrium stabilised. In panel (d) the equilibrium is stabilised further by greater heterogeneity ($\rho = 0$ -- complete independence between patches, $D=0.9$).}\label{fig:gravelmatrices}
\end{figure}

Now that we have obtained analytical results for the eigenvalue spectra, we highlight the stabilising effect of the combination of spatial heterogeneity and dispersal in the model by GML. First we examine the case $D= 0$. Here, Eq.~(\ref{alltoallsupportgravel}) reduces to 
\begin{align}
\left\{ \vert \omega + d\vert^2 - \frac{\sigma^2}{n} \left[1 + \rho (n-1)\right]\right\} \left\{ \vert \omega + d\vert^2 - \frac{\sigma^2}{n} \left[\rho +  (n-1)\right]\right\} = \frac{\sigma^4}{n^2} (1-\rho)^2 (n-1).
		\end{align}
For reasonably large $n$ therefore, the boundary of the spectrum for $D=0$ is well-approximated by $\vert \omega +d\vert^2 = \sigma^2$ (this is verified in Fig. \ref{fig:gravelmatrices}(a)). This is the same result one would obtain without spatial heterogeneity. This means that spatial heterogeneity only has an appreciable effect on stability once dispersal is introduced.

In Fig.~\ref{fig:gravelmatrices} we demonstrate how the combination of dispersal and spatial heterogeneity can be stabilising. Panel (a) shows an example of a spectrum for an unstable equilibrium of a system without dispersal. The data shown is for $\rho=1$, but as one would expect from the above reasoning, we find that the spectrum does not depend substantially on $\rho$ for $D = 0$. When we introduce dispersal but no spatial heterogeneity [Fig. \ref{fig:gravelmatrices}(b)], we find no stabilising effect. This would be expected from our discussion in Section \ref{sec:dispersalalone}: dispersal alone cannot stabilise a system. However, when we introduce heterogeneity along with dispersal, the system is stabilised. This can be seen by comparing panel (c) [$D=0.9, \rho=0.5$] to panel (a) [$D=0, \rho=1$]. Fig. \ref{fig:gravelmatrices}(d) [$D=0.9, \rho=0$] shows that the stabilising effect is stronger as the interactions become more heterogeneous in space.

\begin{figure}[H]
	\centering
	\includegraphics[scale = 0.4]{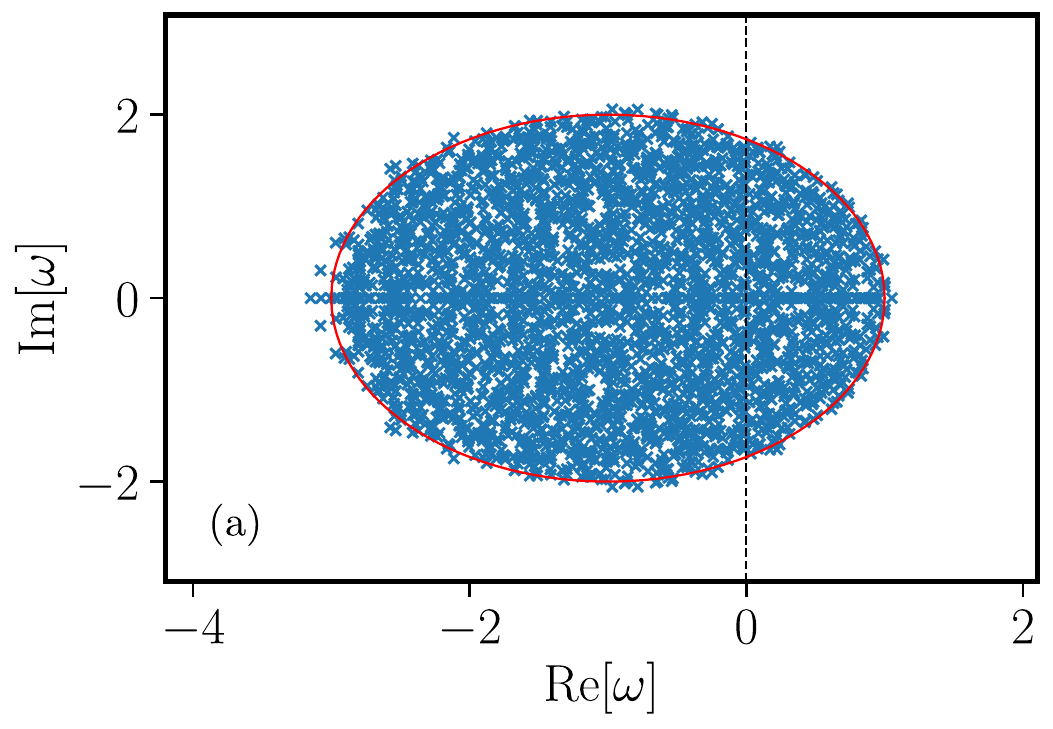} 
	\includegraphics[scale = 0.4]{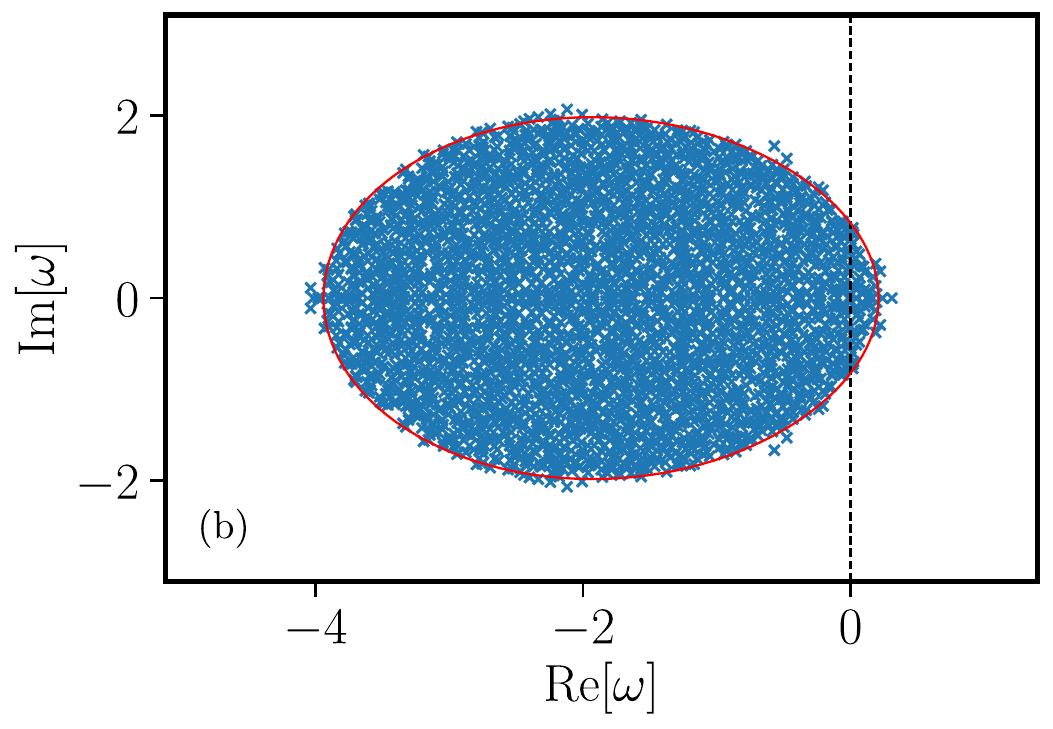}
	\includegraphics[scale = 0.4]{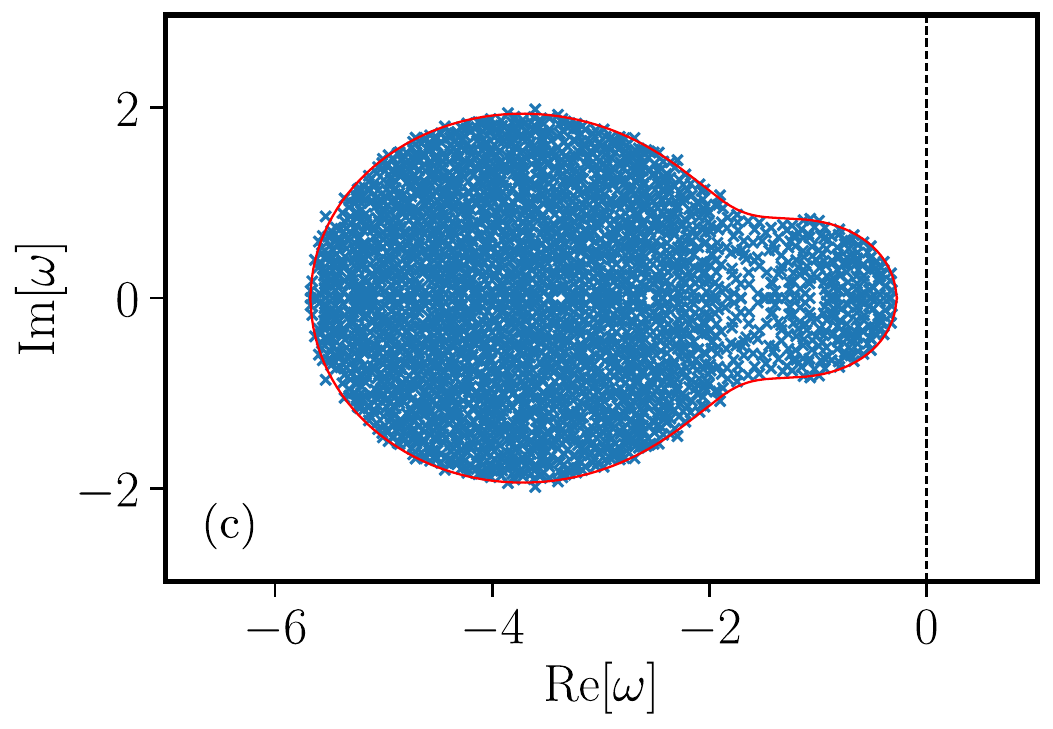}
	\includegraphics[scale = 0.4]{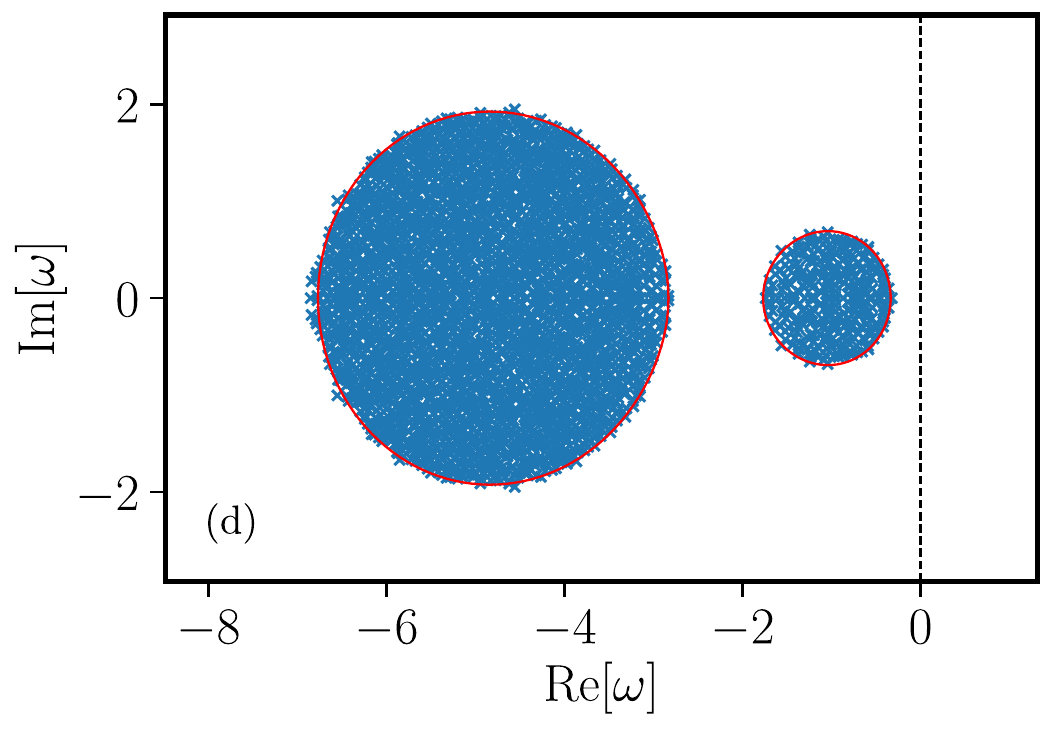}
	\caption{{\bf Dispersal must be sufficiently strong to produce stabilisation.} The figure shows eigenvalue spectra for instances of the model in \cite{gravel}. Spectra of computer-generated random interaction matrices are shown as blue crosses. Red lines are the theoretical predictions for the boundary of the support from Eq.~(\ref{alltoallsupportgravel}). In all panels (after rescaling, see Sec.~\ref{section:scaling}), $\rho = 0$, $d = 1$, $n=11$, $N = 300$, $C =1$, $c = NC\sigma^2 = 4$. In panel (a) there is no dispersal ($D=0$) and the equilibrium is unstable. In panel (b) there is now dispersal ($D = 0.9$). The equilibrium is still unstable. In panel (c) dispersal is increased ($D = 2.5$), the larger bulk region associated with the $q \neq 0$ modes is shifted leftwards and the equilibrium stabilised. In panel (d), the larger bulk region for $q \neq 0$ is shifted further leftwards, while the region associated with $ q =0$ remains in the same location (here $D=3.5$).}\label{fig:effectofdbar}
\end{figure}

Conversely, in Fig. \ref{fig:effectofdbar}, we illustrate the effect of increasing the rate of dispersal $D$. Panel (a) once again shows an eigenvalue spectrum for $D=0$, this time with maximum heterogeneity ($\rho = 0$). As we increase the rate of dispersal from zero (panels (b) $D= 0.9$, (c) $D = 2.5$, (d) $D = 3.5$), the larger part of the spectrum (associated with the $q\neq 0$ Fourier modes) moves leftwards. The part of the spectrum associated with the $q=0$ Fourier mode (the smaller disk in panels (c) and (d)) remains unmoved as the magnitude of the dispersal is increased. If this smaller disk were large enough that some of the eigenvalues comprising it crossed the imaginary axis, then dispersal would not be able to stabilise the equilibrium, regardless of the value of $D$.

\clearpage

\section{Additions to the model by GML}\label{sec:gravelextensions}
We now introduce some additional features to the model of GML \cite{gravel} using our general results derived in Section~\ref{sec:heterogeneity}. Specifically, we show in Section~\ref{sec:stabhop} that stabilisation not only occurs for all-to-all dispersal between patches, but also when dispersal is local (i.e., between nearest-neighbour patches). In Section~\ref{sec:stabcor} we then discuss the introduction of local interspecies correlations (setting $\Gamma \neq 0$). Section~\ref{sec:stabout} then contains a discussion of outlier eigenvalues. These are responsible for the Turing instability, which is the focus of our work. We show that, unlike bulk eigenvalues, they are not necessarily subject to the stabilisation mechanism.

For this section, as with subsection \ref{sec:recovergravel}, we take $\gamma_v = 0$ and suppress the subscript $u$ for the sake of brevity.

\subsection{The stabilising mechanism is preserved when dispersal is local}\label{sec:stabhop}

\subsubsection*{Eigenvalue spectra for local dispersal}
\noindent We now examine the impact that changing the dispersal kernel from all-to-all to local has on the results presented in Fig. \ref{fig:gravelmatrices} and in Eq.~(\ref{alltoallsupportgravel}). We now use the local dispersal kernel (see also Eq.~(\ref{localhoppingkernel}))
\begin{align}
\phi_{x,x'} = 
\begin{cases}
\frac{1}{2} \,\,\,\,\, \mathrm{if} \,\,\,\,\, x = x'-1 \\
\frac{1}{2} \,\,\,\,\, \mathrm{if} \,\,\,\,\, x = x'+1 \\
0 \,\,\,\,\, \mathrm{otherwise}
\end{cases}
.
\end{align}
This has the Fourier transform $\phi_q = \cos(q )$ (as in Section~\ref{sec:maintextmodel}, but setting $\Delta x=1$). We keep the same form for $c_{x,x'}$ as in Eq.~(\ref{alltoallkernels}). Eqs.~(\ref{unhatted2}) then reduce to
\begin{align}
v_q  &= \frac{1}{n}\left[\epsilon n  +\sigma^2 (1-\rho)\sum_{q' }  v_{q'} + n \rho \sigma^2 v_q \right] \hat f_{-q}, \nonumber \\
u_q  &= \frac{1}{n}\left[ n + \sigma^2 (1-\rho)\sum_{q' }  u_{q'} + n \rho \sigma^2 u_q \right] \hat f_{-q}, \nonumber \\
w_{q} &= -i\left[  \omega + d + D [1-\cos(q)] \right] \hat f_{-q}, \nonumber \\
w^\star_{q} &= -i\left[ \omega^\star + d + D [1-\cos(q)] \right] \hat f_{-q} , \label{ftsaddlechain}
\end{align}
bearing in mind the restrictions mentioned at the start of Section \ref{sec:recovergravel}. Due to the larger number of contributing Fourier modes, we cannot proceed as easily as in the case of all-to-all dispersal. However, we can still approximate the boundary of the eigenvalue support quite effectively. As we will see below, the equations approximately decouple for different values of $q$ when the diffusion coefficient is sufficiently large.

\noindent First we deduce from the last two of Eqs.~(\ref{ftsaddlechain}) that 
\begin{align}\label{eq:aux4}
\hat f_{-q} = \frac{1}{\vert \omega + d + D[1-\cos(q)] \vert^2}
\end{align}
on the boundary of the eigenvalue support, where $\hat f_{-q} = - w^\star_{q}w_{q}$.  This means that for large $D$, $\hat f_{-q}$ is small unless $\omega \approx -d -D[1-\cos(q)]$. This indicates that the eigenvalue support is divided into clumps, one for each value of $1-\cos(q)$. When $\hat f_{-q}$ is small, so is $v_q$. Consequently, we can proceed based on the approximation that only one value of $v_q$ is non-zero in any one region of the complex plane.

We wish to find the values of $\omega$ at which Eqs.~(\ref{ftsaddlechain}) first permit a non-zero solution. The solution only depends on the value of $\cos(q)$ and not on $q$ directly. This means that $v_q = v_{q'}$ when $\cos(q) = \cos(q')$. We first focus on the case in which $q$ does not take values $q=0$ or $q=\pi$. For these values of $q$, the non-trivial solution of the first of Eqs.~(\ref{ftsaddlechain}) is given by$\hat f_{-q} =\frac{n}{\sigma^2}\left[ 2(1-\rho) + n\rho \right]^{-1}$. The boundaries of the clumps of eigenvalues are therefore given by the solution of $\hat f_{-q} =\frac{n}{\sigma^2}\left[ 2(1-\rho) + n\rho \right]^{-1} = - w^\star_{q} w_{q}$, with $w^\star_{q}$ and $w_{q}$ being given by the last two of Eqs.~(\ref{ftsaddlechain}). This yields
\begin{align}
\vert \omega + d + D[1-\cos(q)] \vert^2 = \frac{\sigma^2}{n} \left[ 2 + \rho (n-2)\right] . \label{chain2}
\end{align} 
When $q=0$ or $q= \pi$, instead the the non-trivial solution of the first of Eqs.~(\ref{ftsaddlechain}) is given by $\hat f_{-q} =\frac{n}{\sigma^2}\left[ (1-\rho) + n\rho \right]^{-1}$. Hence, the boundaries of the clumps of eigenvalues corresponding to these values of $q$ are given by
\begin{align}
\vert \omega + d + D(1-\cos(q))  \vert^2 = \frac{\sigma^2}{n} \left[ 1 + \rho (n-1)\right] . \label{chain1}
\end{align}

We recall that we have assumed $D$ to be large so that the approximation after Eq.~(\ref{eq:aux4}) is valid. 

For $D=0$, we see that the boundary of the eigenvalue spectrum becomes independent of the dispersal kernel. For sufficiently large $n$ we can therefore proceed as in as in Section \ref{sec:recovergravel}, and approximate the equation characterising the boundary as $\vert \omega + d \vert^2 = \sigma^2$.

\begin{figure}[H]
	\centering
	\includegraphics[scale = 0.4]{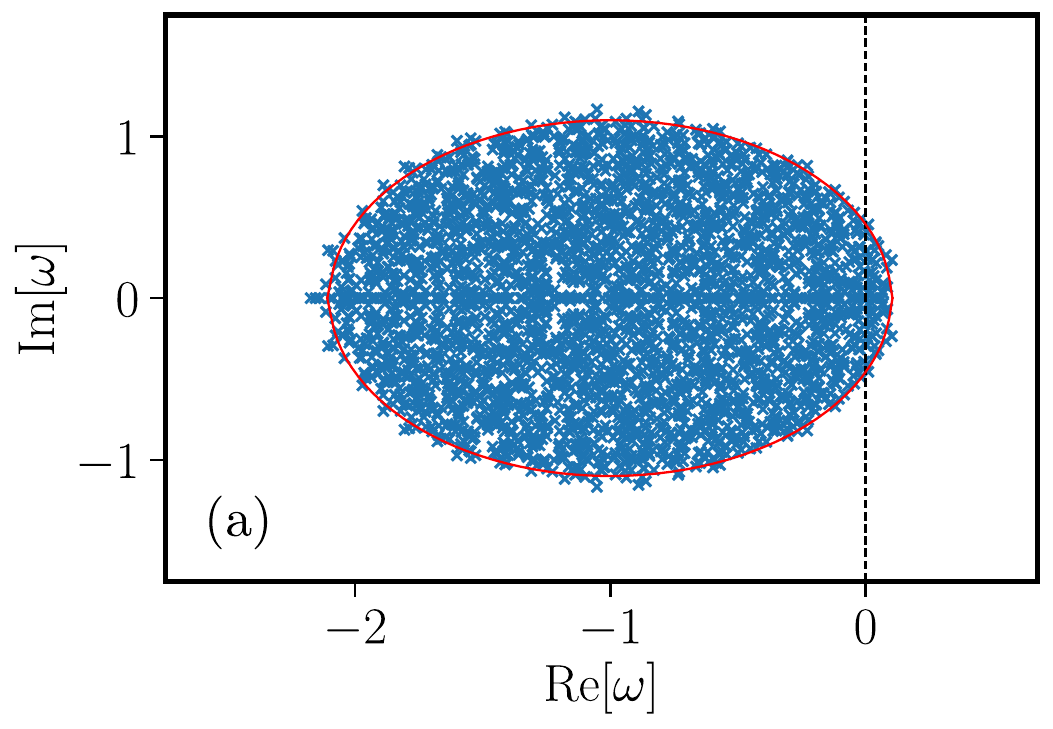} 
	\includegraphics[scale = 0.4]{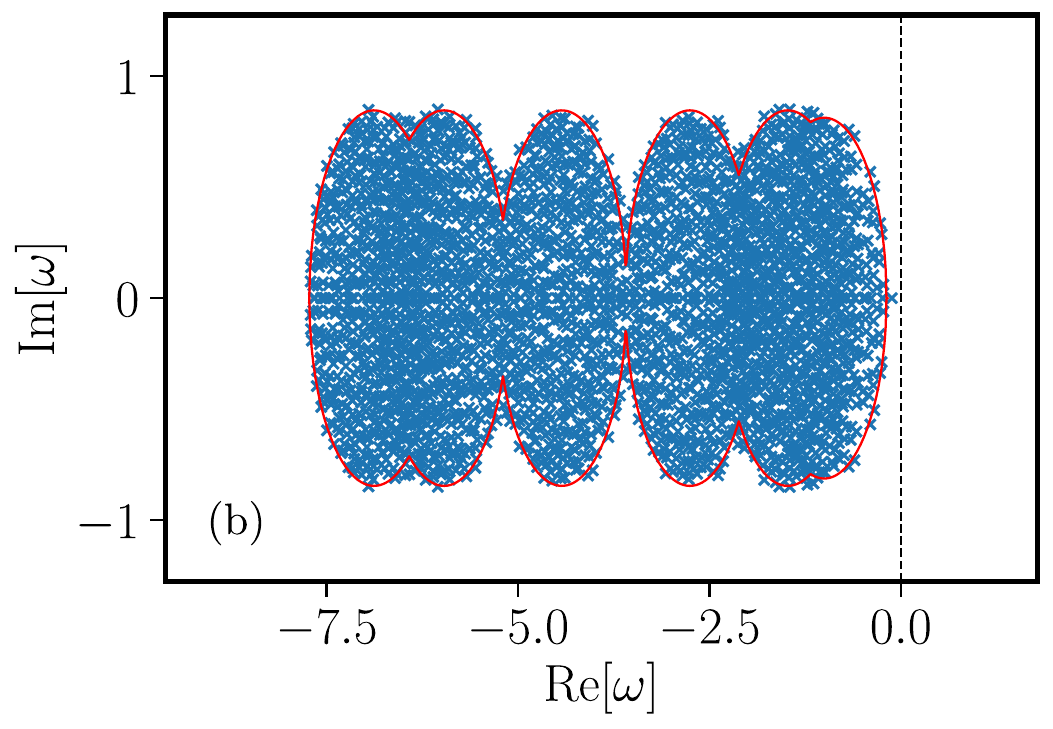} 
	\caption{{\bf Stabilising mechanism is preserved when dispersal is local.} The system parameters here are the same as in Fig. \ref{fig:gravelmatrices}, with the exception that dispersal is now local. In particular we have $\rho=0.5$. Panel (a) shows the case $D=0$ so that the spectrum is the same as Fig. \ref{fig:gravelmatrices}(a). The diffusion constant in panel is $D=3$. The example also demonstrates the approximation of the bulk regions as circles [see text after Eq.~(\ref{eq:aux4})] is reasonably accurate even when $D$ is only moderately large. The aspect ratio and size of the image for panel (a) was chosen in-line with that of panel (b). As a consequence the scaling on the axes in (a) stretches the circular disk in horizontal direction.}\label{fig:localhopping}
\end{figure}

We therefore see that the stabilising effect of the combination of dispersal and heterogeneity remains when the dispersal is local: For $D=0$, the support of the spectrum is approximately circular, with radius approximately $\sigma$ and centre $-d$. An example is shown in Fig.~\ref{fig:localhopping}, where we choose $\rho=0.5$. We note that the aspect ratio for the image in panel (a) was chosen in-line with that of panel (b), the scaling on the axes in (a) stretches the circular disk in horizontal direction. When we introduce dispersal ($D>0$), modes with non-zero values of $q$ become relevant. The circular patch of eigenvalues closest to the imaginary axis now has a smaller radius of approximately $\sigma \sqrt{[1 + \rho (n-1)]/n} < \sigma$ and the spectrum is now entirely in the negative half-plane.
\subsubsection*{Stability criteria for all-to-all and nearest-neighbour dispersal}
We first consider all-to-all dispersal. The boundary of the support of the eigenvalue spectrum in this case is given by Eq.~(\ref{alltoallsupportgravel}). When $D$ is large, this is well approximated by two circles, 
\begin{align}\label{eq:aux7}
\vert \omega + d  \vert^2 &= \frac{\sigma^2}{n} \left[ 1 + \rho (n-1)\right], \nonumber \\
\vert \omega + d + D n/(n-1) \vert^2 &= \frac{\sigma^2}{n} \left[ n-1 + \rho \right]  .
\end{align}
For large diffusion constants, the first of these disks is closer to the imaginary axis and so the criterion for stability is
\begin{align}\label{eq:aux6}
\sigma^2 <\frac{d^2n}{1 + \rho(n-1)}.
\end{align}
Keeping in mind that we have re-scaled the variance of interactions (see Section~\ref{section:scaling}), we need to carry out the replacement $\sigma^2\to \sigma^2 N$. In the limit of large $N$ (which we always assume), Eq.~(\ref{eq:aux6}) is then the same as the stability criterion in Eq.~(3) of \cite{gravel}. 

In the case of nearest-neighbour dispersal between patches, since the clump of eigenvalues corresponding to $q = 0$ is the same as in the all-to-all case [compare Eqs.~(\ref{eq:aux7}) and (\ref{chain1})], we obtain the same criterion for stability from Eqs.~(\ref{chain1}). As a consequence, the stability criteria for the model in \cite{gravel} remain unchanged if all-to-all dispersal between patches is replaced by local dispersal. The stabilising mechanism for bulk eigenvalues due to spatial heterogeneity and dispersal therefore remains.

\subsection{Inclusion of correlations: $\Gamma \neq 0$}\label{sec:stabcor}
The work in Ref. \cite{gravel} assumed that the interaction coefficients between species within a patch could all be drawn independently, in a similar style to May's original model \cite{may}. Our analytical approach allows one to include correlations. We deduce the boundary of the eigenvalue support in this case and show that $\Gamma <0$ can further stabilise the community.

\begin{figure}[H]
	\centering 
	\includegraphics[scale = 0.4]{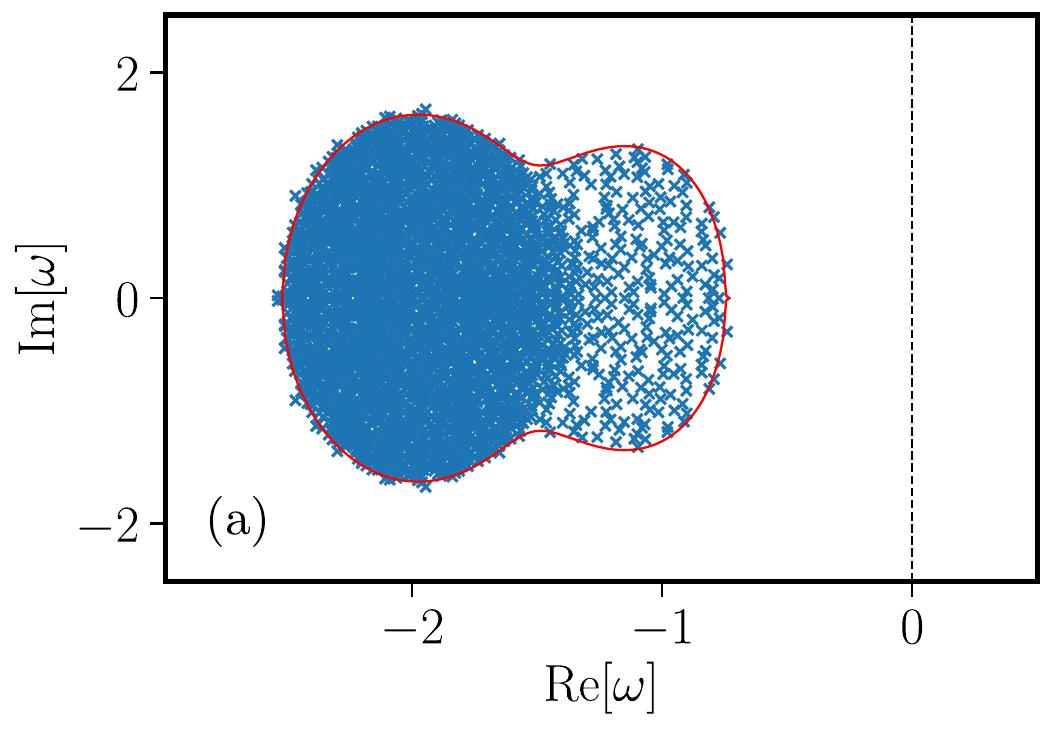}
	\includegraphics[scale = 0.4]{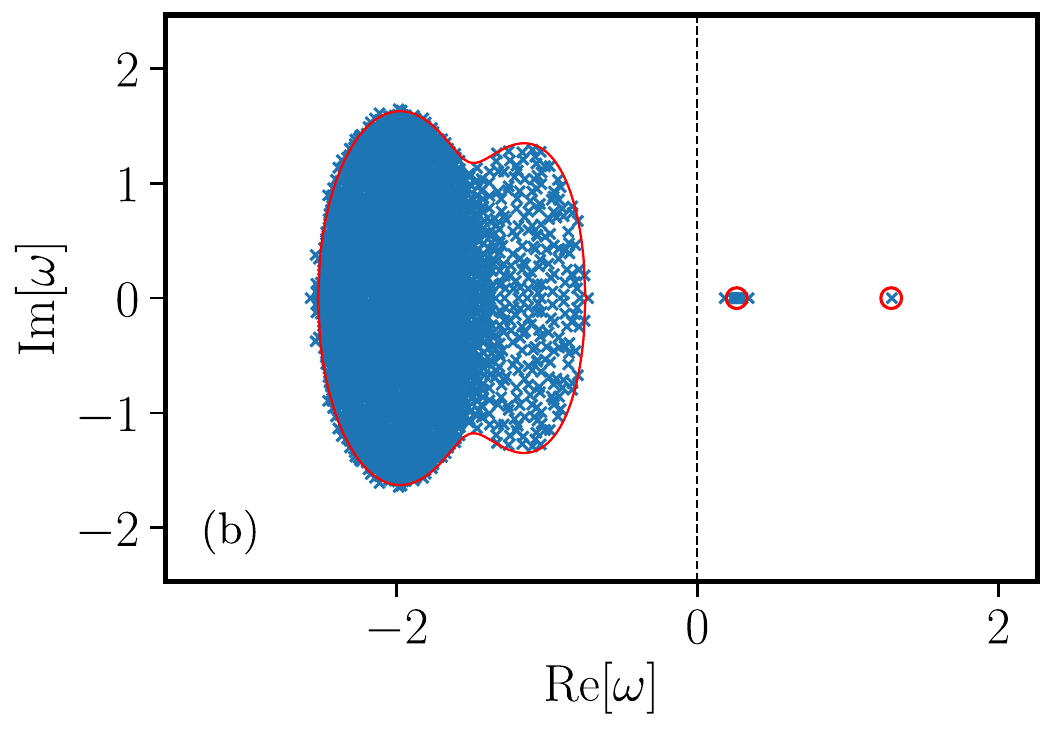}
	\caption{{\bf Verification of theory when interactions are locally correlated, but heterogeneous in space.} Panel (a):  The setup and parameters are identical to those in Fig.~\ref{fig:gravelmatrices}(c) with the sole adjustment $\Gamma = -0.5$. The boundary of the bulk region is predicted by Eqs.~(\ref{resolventsgammaneq0}) and the outlier eigenvalues are found using Eq.~(\ref{outliersgammamu}).  The equilibrium is more stable than that in Fig.  \ref{fig:gravelmatrices}(c), confirming that anti-correlation in the interaction matrix acts to stabilise. In panel (b) the interaction matrix elements have a finite mean $NC\mu = 2.5$ (after rescaling, see Sec.~\ref{section:scaling}) This produces two sets of outlier eigenvalues [one for $q=0$ and one for $q\neq 0$ in equations Eq.~(\ref{productq})]. As before blue crosses are from numerically diagonalising computer-generated random matrices, red lines and circles represent the analytical predictions for bulk and outlier eigenvalues respectively.}\label{fig:rho0point5gammaminus0point5}
\end{figure}

We again use the dispersal and correlation kernels in Eq.~(\ref{alltoallkernels}). From the first relation in Eq.~(\ref{unhatted2}), we once again obtain [c.f. Eq.~(\ref{boundcond})]
\begin{align}
\hat f_0 \left[ 1 + \rho (n-1)\right] + \hat f_{-q} \left[ \rho + n -1\right] + \frac{\sigma^2}{n}\hat f_0 \hat f_{-q}\left[ (1-\rho)^2 (n-1) - (1 + \rho(n-1))(\rho + n -1) \right] = \frac{n}{\sigma^2} . \label{hatscond}
\end{align}
Now, because of the new asymmetry $\Gamma \neq0$, the last two of Eqs.~(\ref{unhatted2}) for the order parameters $w^\star_{q}$ and $w_{q}$ no longer decouple for different values of $q$. We instead obtain 
\begin{align}
w_{0} &= -\left\{ i [\omega + d ] + \frac{\Gamma\sigma^2}{n}\left[ [1 + \rho(n-1)] w^\star_0 + (1-\rho)(n-1) w^\star_{q}\right] \right\} \hat f_{0}, \nonumber \\
w_{0}^\star &= -\left\{ i [\omega^\star + d ] + \frac{\Gamma\sigma^2}{n}\left[[1+\rho(n-1)] w_0 + (1-\rho)(n-1) w_{q}\right]\right\} \hat f_{0} , \nonumber \\
w_{q} &= -\left\{ i [\omega + d +Dn/(n-1)] + \frac{\Gamma\sigma^2}{n}\left[(\rho + n -1)  w^\star_q + (1-\rho) w^\star_{0}\right] \right\} \hat f_{-q}, \nonumber \\
w_{q}^\star &= -\left\{ i [\omega^\star + d +Dn/(n-1)] + \frac{\Gamma\sigma^2}{n}\left[(\rho + n -1)  w_q + (1-\rho) w_{0}\right]\right\} \hat f_{-q} , \label{resolventsgammaneq0}
\end{align}
For the boundary of the support of the eigenvalue spectrum, these must be solved simultaneously with $\hat f_{0} = - w_0 w_0^\star$ and $\hat f_{-q} = - w_q w_q^\star$ and Eq.~(\ref{hatscond}) subject to the following conditions
\begin{align}
- w_0^\star w_0 > 0, \,\,\,\,\, - w_0^\star w_0 \in \mathbb{R} ,& \,\,\,\,\,  - w_q^\star w_q > 0, \,\,\,\,\, - w_q^\star w_q \in \mathbb{R} , \nonumber \\
1 - \frac{\sigma^2}{n} \hat f_0[1 + \rho (n-1)] >0 ,& \,\,\,\,\, 1 - \frac{\sigma^2}{n} \hat f_{-q}[\rho + n-1] >0  .
\end{align}
Although these equations are difficult to handle analytically, they can be solved numerically. One finds that negative correlations $\Gamma<0$ between the interaction coefficients $m_{ij;xx}$ and $m_{ji;xx}$ of species $i$ and $j$ within the same patch $x$ further stabilises an equilibrium to an even greater extent than dispersal and heterogeneity on their own, an example is shown in Fig. \ref{fig:rho0point5gammaminus0point5}. Such negative correlation between pairs of interaction coefficients in matrices without block structure is tantamount to predation.

\subsection{Outlier eigenvalues are not necessarily subject to the same stabilising effect as the bulk}\label{sec:stabout}
As a precursor to the full inclusion of trophic structure in Section~\ref{sec:interplay}, we demonstrate the effect including a non-zero mean ($\mu \neq 0$) in the model of GML \cite{gravel}. In the original work in Ref. \cite{gravel}, the interaction matrix elements were assumed to be drawn from a Gaussian distribution with zero mean ($\mu = 0$). As we have previously discussed, the inclusion of a non-zero mean can have a profound effect on stability. We can examine this effect using Eq.~(\ref{simult22}), which becomes 
\begin{align}
- 1 &=  - [\lambda + d ]\chi_{0} +  \frac{\sigma^2\Gamma}{n} \chi_{0}\left\{ [1+\rho(n-1)]\chi_{0} + (1-\rho) (n-1) \chi_q \right\}, \nonumber \\
- 1 &=  - [\lambda + d + Dn/(n-1)]\chi_{q} +  \frac{\sigma^2\Gamma}{n} \chi_{q}\left\{ [\rho+(n-1)]\chi_{q} + (1-\rho)  \chi_0 \right\}.\label{outliersgammamu}
\end{align}
These are solved simultaneously [c.f. Eq.~(\ref{simult12})] first with 
\begin{align}
\chi_0(\lambda)  = \frac{1}{\mu},
\end{align}
to yield one set of outlier eigenvalues and then with 
\begin{align}
\chi_q(\lambda)  = \frac{1}{\mu},
\end{align}
to yield another set. Again, one must impose $u_0>0$ and $u_q>0$ [c.f. Eqs.~(\ref{unhatted2})] to eliminate invalid solutions. These expressions for the outliers are verified in Fig. \ref{fig:rho0point5gammaminus0point5}.

When $\Gamma = 0$ (as was the case in the model originally presented in Ref. \cite{gravel}), we obtain the remarkably simple result for the outliers
\begin{align}
\lambda_q = \mu - \left[ d + D (1-\phi_q)\right].
\end{align}
Crucially, when $\Gamma =0$, the outlier eigenvalues are independent of $\rho$ (and therefore heterogeneity). This example demonstrates that the outlier eigenvalues are not necessarily subject to the same stabilising effects as the bulk distribution. This will prove important when we discuss the interplay between the stabilising effect of spatial heterogeneity and our destabilising mechanism in Section \ref{sec:interplay}.

\section{Dispersal-induced instability persists on a discrete-patch landscape with all-to-all dispersal}\label{sec:all-to-all}
In Section~\ref{sec:stabhop} we showed that the stabilising mechanism reported by \cite{gravel} in a model without trophic structure persists when dispersal is local. Here, we now demonstrate a complementary statement: dispersal-induced instability in the model with trophic structure continues to be possible in case of all-to-all dispersal between a set of patches. 

Using the dispersal and correlation kernels in Eq.~(\ref{alltoallkernels}) with $\rho = 1$, the equations for the order parameters decouple for different values of $q$, and we obtain
\begin{align}
- w_{\alpha;-q} w^\star_{\alpha;-q} &= \hat f_{\alpha;q}-\hat f_{\alpha;q}^2 \frac{  r_{-q}}{\left[\sum_\beta \gamma_\beta \hat f_{\beta;q}\right]^2} , \nonumber \\
v_q &= \left(\epsilon + \sigma^2 v_{q}\right)\sum_\alpha \gamma_\alpha \hat f_{\alpha;-q} , \nonumber \\
u_q &= \left(1 + \sigma^2 u_{q} \right)\sum_\alpha \gamma_\alpha \hat f_{\alpha;-q} , \nonumber \\
w_{\alpha;q} &= -\left[  i\{\omega + d_\alpha + D_\alpha[1-(n\delta_{q,0}-1)/(n-1)]\}+  \sigma^2 \sum_{\beta} \Gamma_{\alpha \beta} \gamma_\beta w^\star_{\beta;q}\right]\hat f_{\alpha;-q} , \nonumber \\
w^\star_{\alpha;q} &=  -\left[  i\{\omega^\star + d_\alpha + D_\alpha[1-(n\delta_{q,0}-1)/(n-1)]\}+  \sigma^2 \sum_\beta \Gamma_{\alpha \beta} \gamma_\beta w_{\beta;q}\right]\hat f_{\alpha;-q} , \label{decoupledpatches}
\end{align}
These equations are almost identical to those derived for the model used in the main text, the only difference being the dispersal kernel. They can be obtained from Eqs.~(\ref{unhatted}) with the substitution $q^2 \to 1 - \phi_q$.

We expect to see two contributions to the eigenvalue spectra, one for $q=0$ and one for $q\neq 0$. The $q=0$ contribution is the same as that which is found from Eqs.~(\ref{unhatted}) since, due to total probability, $1- \phi_0 = 0$ (a statement which is independent of the hopping kernel). When $q\neq 0$, we have for the all-to-all hopping kernel that $1-\phi_q =\frac{n}{n-1}$ [c.f. Eqs.~(\ref{transformedalltoall})]. The second contribution to the spectrum can therefore be obtained by setting $q = \sqrt{\frac{n}{n-1}}$ in the model in Section \ref{sec:model}. 

\begin{figure}[H]
	\centering 
	\includegraphics[scale = 0.4]{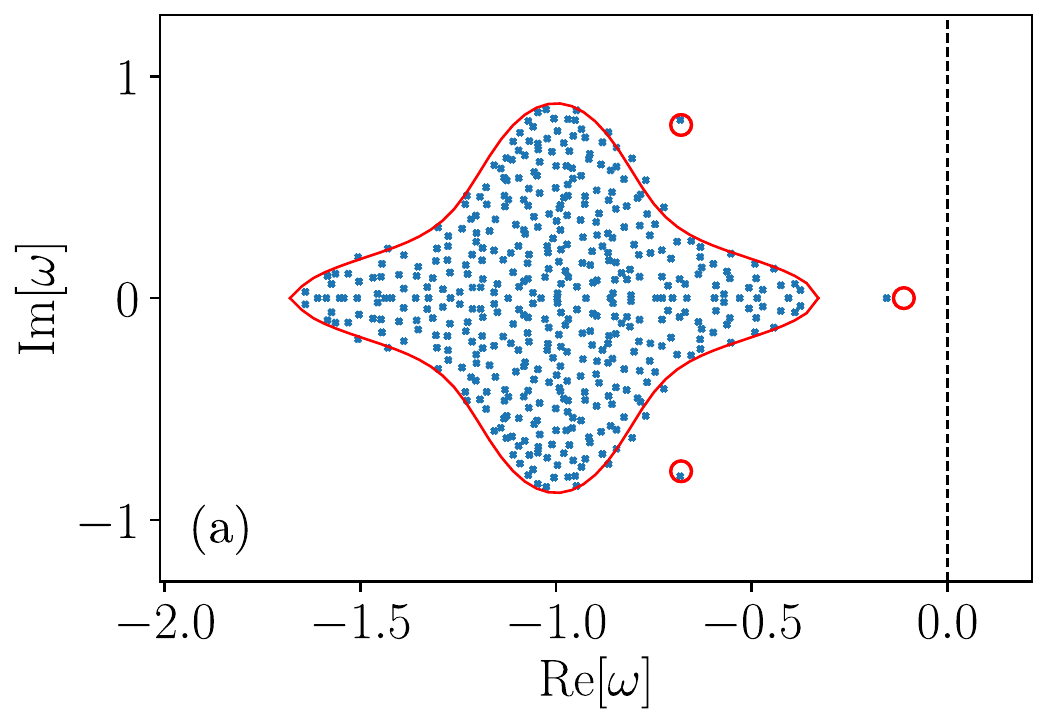}
	\includegraphics[scale = 0.4]{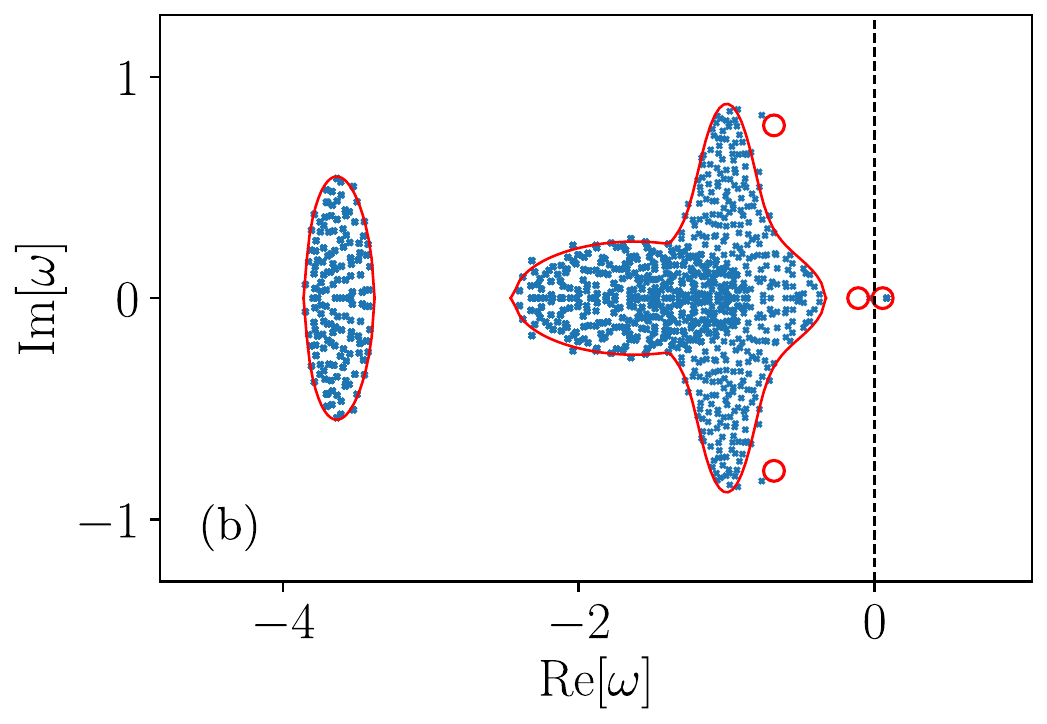}
	\caption{{\bf Dispersal-induced instability persists in discrete-patch landscape with all-to-all dispersal.}  Panel (a) shows the spectrum when there is no dispersal ($D_u = D_v = 0$). In panel (b) $D_u = 0.5$, $D_v = 2.5$. The spectrum in (a) has the same support and outliers as that in Fig. 2(a) in the main text. The spectrum in (b) demonstrates that dispersal-induced instability still occurs when dispersal is all-to-all between a discrete set of patches. As before, predictions from the analytical theory are shown as red lines and circles, and numerical results for computer-generated random matrices as blue crosses. Parameters are (after rescaling, see Sec.~\ref{section:scaling}) $NC\mu_{uu} = 3$, $NC\mu_{uv} = -3$, $NC\mu_{vu} = 3$, $NC\mu_{vv} = -1.5$, $c = NC\sigma^2 = 0.325$, $n = 11$, $\rho = 1$, $\Gamma_{uv} = -0.9$, $\Gamma_{uu} = 0.5$, $\Gamma_{vv} = -0.5$, $\gamma_u = 2/3$, $\gamma_v = 1/3$, $C = 1$ and $N = 450$. }\label{fig:turinginst}
\end{figure}

We can also calculate the outlier eigenvalues, whose locations are given by Eqs.~(\ref{simult12}) and (\ref{simult22}). Again, the equations decouple for different values of $q$ when $\rho =1$, and we obtain a result similar to those in Eqs.~(\ref{simult1}) and (\ref{simult2}), with the modification $q^2 \to 1-\phi_q$,
\begin{align}
&\left[ \gamma_u\mu_{uu} - \frac{1}{\chi_{u;q}(\lambda)}\right] \left[\gamma_v \mu_{vv} - \frac{1}{\chi_{v;q}(\lambda)}\right] - \gamma_u\gamma_v\mu_{uv}\mu_{vu} = 0, \nonumber \\
- 1 &=  - [\lambda + d_\alpha + D_\alpha(1-\phi_{q}) ]\chi_{\alpha;q} +  \sigma^2\sum_{\beta;q'}\gamma_{\beta}\Gamma_{\alpha\beta}   \chi_{\alpha;q}\chi_{\beta;q}.
\end{align}
These are to be solved subject to
\begin{align}
\sum_{\alpha}\gamma_\alpha \vert \chi_{\alpha;q}\vert^2 < \sigma^2.
\end{align}
We reach the conclusion that the system with all-to-all dispersal is equivalent to a system with continuous-space local diffusion restricted to Fourier modes $q=0$ and $q = \sqrt{\frac{n}{n-1}}$. Therefore, if the mode with $q=0$ is stable and the mode with $q = \sqrt{\frac{n}{n-1}}$ unstable in the main-text model (with local, diffusive dispersal), then we also observe a dispersal-induced instability in the model with discrete patches and all-to-all dispersal. An explicit example is given in Fig.~\ref{fig:turinginst}.

\section{Combining trophic structure, interaction heterogeneity and dispersal}\label{sec:interplay}
\subsection{Stabilising and destabilising mechanisms: explicit example}
 
The focus of our work is the possibility of dispersal-induced instability. Ref. \cite{gravel} argues that dispersal can act to stabilise an equilibrium. In this section we comment on these seemingly contrasting findings, and we show that there is no contradiction. 

The two mechanisms can in fact be thought of as quite separate. The stabilising effect reported in \cite{gravel} acts on bulk eigenvalues and is dependent upon spatial heterogeneity in the inter-species interactions. The destabilising effects of dispersal that we highlight, on the other hand, are due to outlier eigenvalues and require trophic structure in the community. As we will show, in a model which combines all of these features, both the stabilising and the destabilising effects of dispersal can be seen. Precisely which is observed depends on the circumstances.

\begin{figure}
	\centering
	\includegraphics[scale = 0.4]{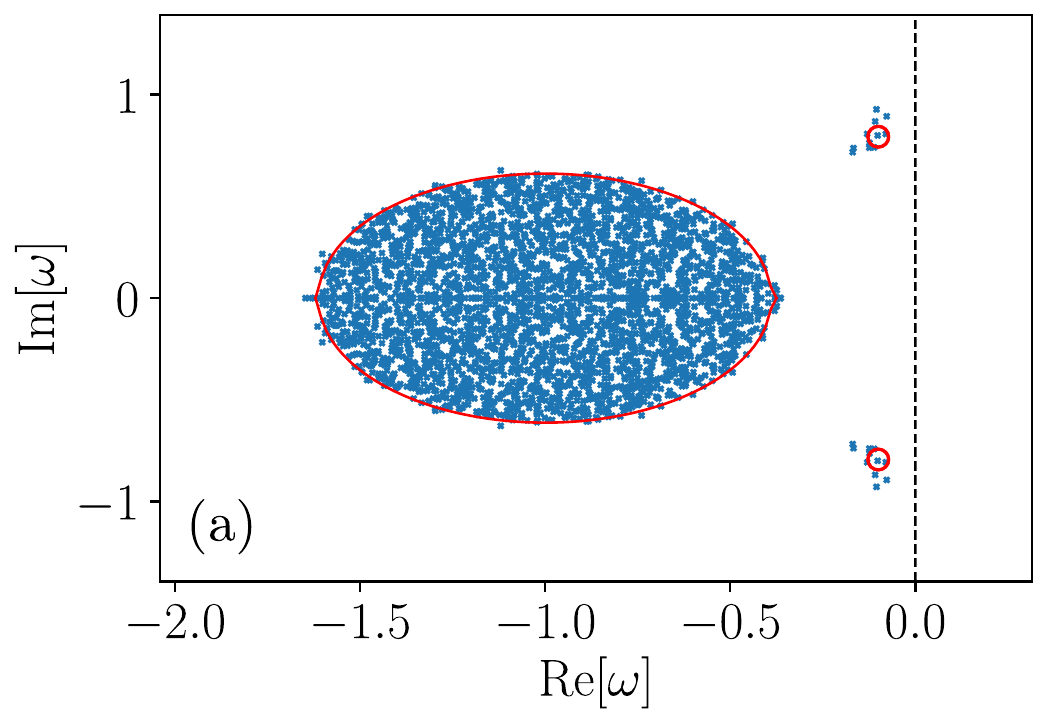} 
	\includegraphics[scale = 0.4]{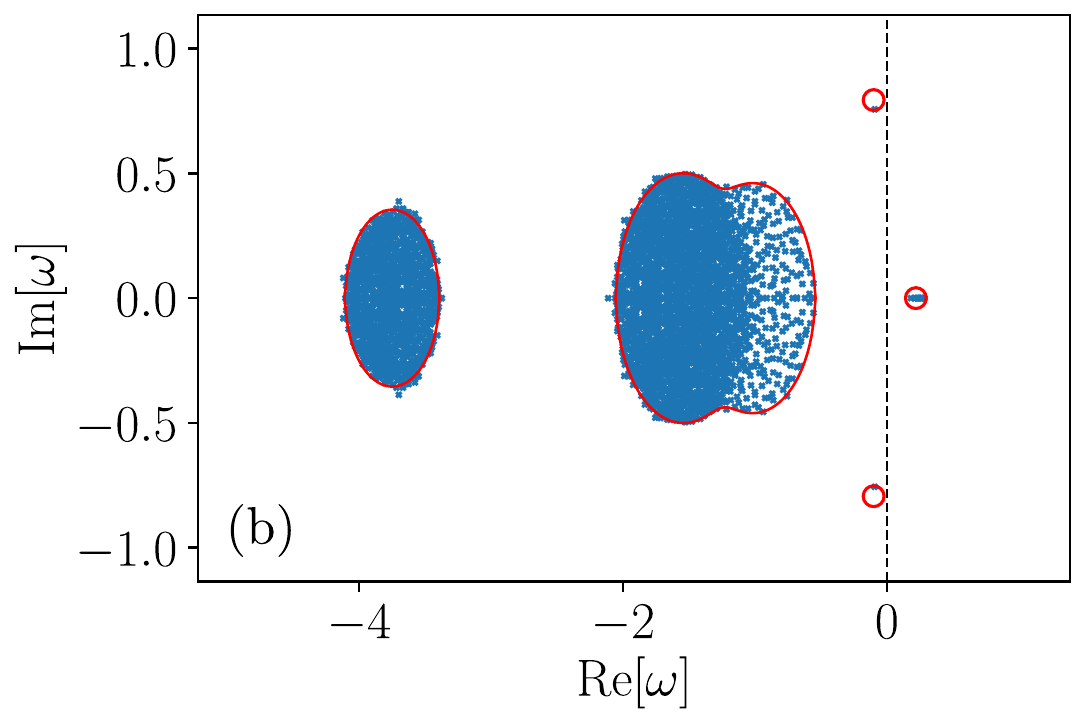}
	\includegraphics[scale = 0.4]{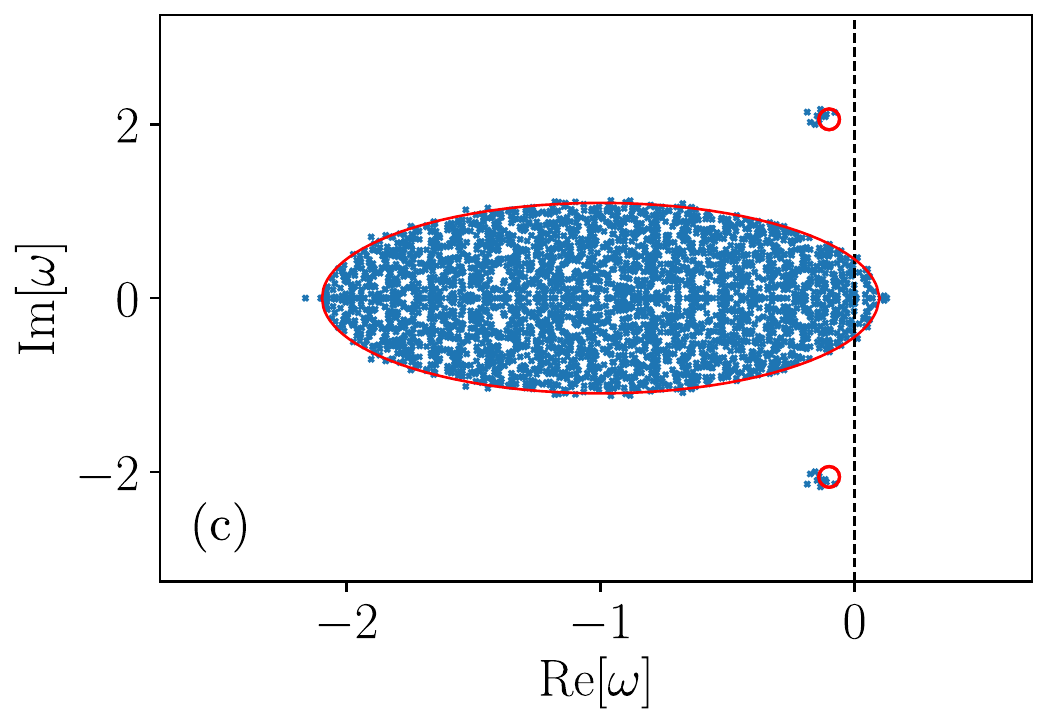} 
	\includegraphics[scale = 0.4]{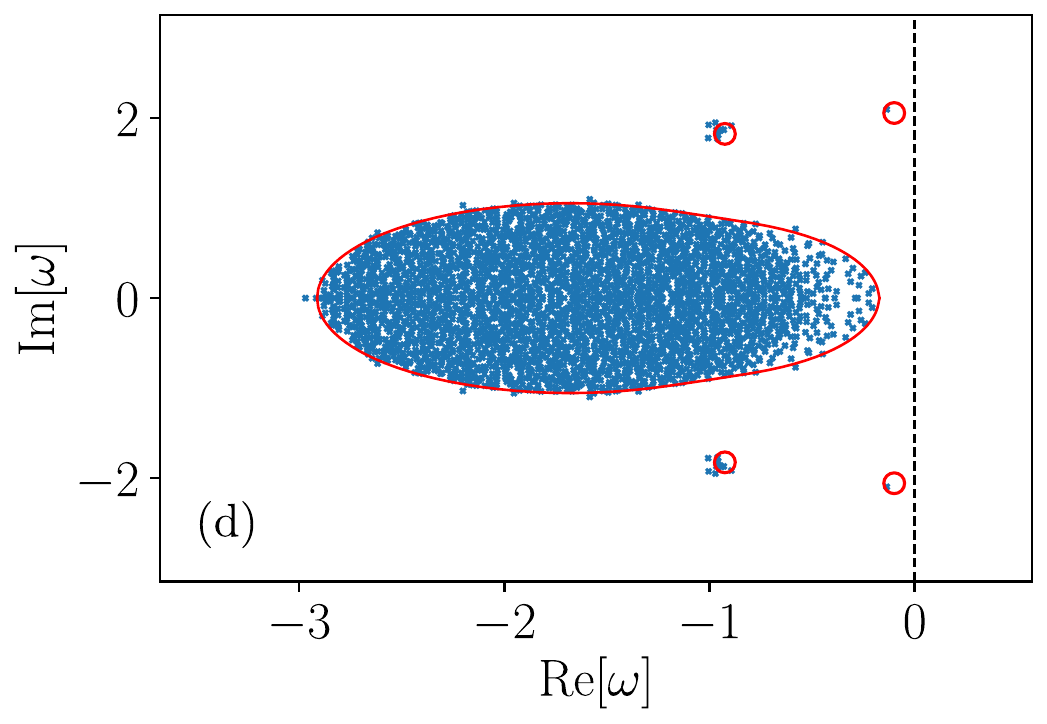}
	\caption{{\bf Demonstration of the stabilising and destabilising mechanisms in the same model.} Eigenvalue spectra demonstrating the different effects that result from the combination of trophic structure, spatial variation in interaction coefficients and dispersal. The combination of dispersal and spatial variation in interaction coefficients acts to stabilise whereas the combination of dispersal and trophic structure acts to destabilise. Depending on the system parameters, one effect takes precedence over the other. Panels (a) and (b) demonstrate the former effect and panels (c) and (d) demonstrate the latter. For the model parameters here (after rescaling, see Sec.~\ref{section:scaling}), we have (a)  $D_u = 0$, $D_v = 0$, $NC\mu_{vu} = - NC\mu_{uv} = 3.6$, $c = NC\sigma^2 = 0.325$ (b)  $D_u = 0.5$, $D_v = 2.5$, $NC\mu_{vu} = - NC\mu_{uv} = 3.6$, $c = NC\sigma^2 = 0.325$ (c) $D_u = 0$, $D_u = 0$, $NC\mu_{vu} = - NC\mu_{uv} = 5.4$, $c = NC\sigma^2 = 1.21$ (c) $D_u = 0.5$, $D_v = 1$, $NC\mu_{vu} = - NC\mu_{uv} = 5.4$, $c = NC\sigma^2 = 1.2$. The remaining model parameters are $NC\mu_{uu} = 3.6$, $NC\mu_{uv} = -3.6$, $NC\mu_{vu} = 3.6$, $NC\mu_{vv} = -1.8$, $n = 11$, $\rho = 0.5$, $\Gamma_{uv} = 0$, $\Gamma_{uu} = 0$, $\Gamma_{vv} = 0$, $\gamma_u = 2/3$, $\gamma_v = 1/3$, $C = 1$ and $N = 300$.  }\label{fig:gamma0trophic}
\end{figure}

We give an explicit demonstration of this in Fig.~\ref{fig:gamma0trophic}. The data in all four panels is obtained from a model combining spatial heterogeneity, trophic structure and dispersal. The different panels are for different choices of the model parameters. Panel (a) shows a spectrum for an equilibrium of a system with trophic structure, heterogeneity of interactions in space, but without dispersal. The example is such that the equilibrium is stable -- the bulk eigenvalues and the outliers all have negative real parts. In panel (b) we now introduce dispersal, keeping all other model parameters unchanged. As seen in the figure, there are now outlier eigenvalues in the positive half-plane. The equilibrium has become unstable. The bulk eigenvalues however remain to the left of the real axis [with an overall tendency to move to the left compared to panel (a)]. The comparison of panels (a) and (b) therefore shows the destabilising effect of dispersal on outlier eigenvalues. This is the mechanism highlighted in the main text (see Fig. 2).

In Fig.~\ref{fig:gamma0trophic}(c) we show again an eigenvalue spectrum for an equilibrium of a system without dispersal. Parameters are chosen such that the bulk spectrum crosses the imaginary axis into the positive half-plane. This makes the equilibrium unstable. We note that outlier eigenvalues in this example have negative real parts. In panel (d) we now allow for dispersal, with all other parameters as in panel (c). The stabilising effect on the bulk eigenvalues sets in, and the bulk spectrum is now strictly in the negative half-plane. At the same time, the outlier eigenvalues also remain in the negative half. The equilibrium is now stable. This is the mechanism highlighted by GML \cite{gravel} (see also Section \ref{sec:recovergravel}).
\clearpage

 \subsection{Analytical demonstration}\label{sec:analyticaldemonstration}
In order to take an analytically tractable example, we set $\Gamma_{\alpha\beta}=0$ in Eqs.~(\ref{unhatted2}) and once again use the dispersal and correlation kernels in Eq.~(\ref{alltoallkernels}). Similar to the previous examples in Section \ref{sec:recovergravel}, the first of Eqs.~(\ref{unhatted2}) yields
\begin{align}
&\sum_\alpha \gamma_\alpha \hat f_{\alpha;0} \left[ 1 + \rho (n-1)\right] + \sum_\alpha \gamma_\alpha \hat f_{\alpha;-q} \left[ \rho + n -1\right] \nonumber \\
&+ \frac{\sigma^2}{n}\sum_{\alpha,\beta} \gamma_\alpha \gamma_\beta\hat f_{\alpha;0} \hat f_{\beta;-q}\left[ (1-\rho)^2 (n-1) - (1 + \rho(n-1))(\rho + n -1) \right] = \frac{n}{\sigma^2} , 
\end{align}
subject to the following conditions
\begin{gather}
- w_{\alpha;0}^\star w_{\alpha;0} > 0, \,\,\,\,\, - w_{\alpha;0}^\star w_{\alpha;0} \in \mathbb{R} , \,\,\,\,\,  - w_{\alpha;q}^\star w_{\alpha;q} > 0, \,\,\,\,\, - w_{\alpha;q}^\star w_{\alpha;q} \in \mathbb{R} , \nonumber \\
1 - \frac{\sigma^2}{n} [1 + \rho (n-1)] \sum_\alpha \gamma_\alpha \hat f_{\alpha;0} >0 , \,\,\,\,\, 1 - \frac{\sigma^2}{n} [\rho + n-1] \sum_\alpha \gamma_\alpha \hat f_{\alpha;-q} >0 .
\end{gather}
Because $\Gamma_{\alpha\beta}=0$, the equations for $w_{\alpha;q}$ and $w_{\alpha;q}^\star$ in Eq.~(\ref{unhatted2}) decouple. When we solve these simultaneously with $\hat f_{\alpha;q} = -w^\star_{\alpha;q}w_{\alpha;q}$, we find
\begin{align}
\hat f_{\alpha;-q} = \frac{1}{\vert \omega + d_\alpha + D_{\alpha}( 1- \phi_{q})\vert^2} .
\end{align}
We thus obtain the following closed-form expression for the boundary of the eigenvalue support 
\begin{align}\label{eq:aux10}
&\sum_\alpha \gamma_\alpha \left[\frac{1 + \rho(n-1)}{\vert \omega + d_\alpha \vert^2} +\frac{\rho + n-1}{\vert \omega + d_\alpha + \frac{D_{\alpha} n}{n-1}\vert^2}\right] \nonumber \\
&+ \frac{\sigma^2}{n}\sum_{\alpha,\beta} \gamma_\alpha \gamma_\beta \frac{\left[ (1-\rho)^2 (n-1) - (1 + \rho(n-1))(\rho + n -1) \right]}{\vert \omega + d_\alpha \vert^2 \vert \omega + d_\beta + \frac{D_{\beta} n}{n-1}\vert^2 }  = \frac{n}{\sigma^2} .
\end{align}
For this simple example with $\Gamma_{\alpha\beta} =0$, separate expressions for outlier eigenvalues associated with different values of $q$ can be found from Eqs.~(\ref{simult12}) and (\ref{simult22}). One arrives at the compact result
\begin{align}\label{eq:aux11}
\left[ \gamma_u\mu_{uu} - \lambda_q - d_u -D_u (1-\phi_q)\right] \left[\gamma_v \mu_{vv} - \lambda_q - d_v -D_v (1-\phi_q)\right] - \gamma_u\gamma_v\mu_{uv}\mu_{vu} = 0.
\end{align}

This special case demonstrates that the stabilising and the destabilising mechanisms are rather separate. The region of the complex plane containing the bulk eigenvalues is determined by Eq.~(\ref{eq:aux10}). This relation has an explicit dependence on the level of spatial heterogeneity (parametrised by $\rho$), making heterogeneity a key factor for the stabilising influence of dispersal on the bulk eigenvalues. The outlier eigenvalues for this example on the other hand are determined by Eq.~(\ref{eq:aux11}), with no dependence on $\rho$ (in a similar way to the outliers obtained in the single-trophic-level case in Section \ref{sec:stabout}). This means that the outlier eigenvalues are not subject to the stabilising mechanism -- the dispersal-induced instability can not be mitigated by spatial heterogeneity.

The picture becomes more complicated when there are correlations ($\Gamma_{\alpha\beta} \neq 0$). However, this example serves to illustrate our point which is that one must be wary of both the stabilising and destabilising effects. Dispersal is therefore not a unilaterally stabilising or destabilising force by itself. Rather, the picture is more subtle; in combination with spatial heterogeneity dispersal may be stabilising, given the correct circumstances, and in combination with trophic structure it may be destabilising.  

\subsection{Coexistence of the Turing instability and the heterogeneity-associated stabilisation}

\begin{figure}[t]
	\centering
	\includegraphics[scale = 0.45]{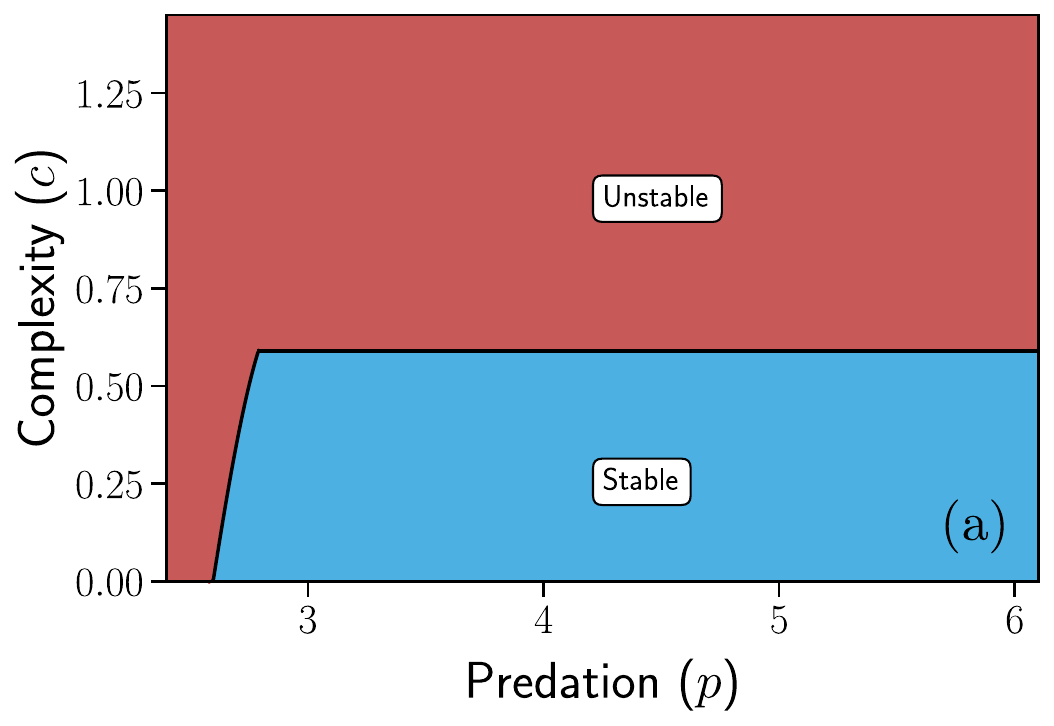}
	\includegraphics[scale = 0.45]{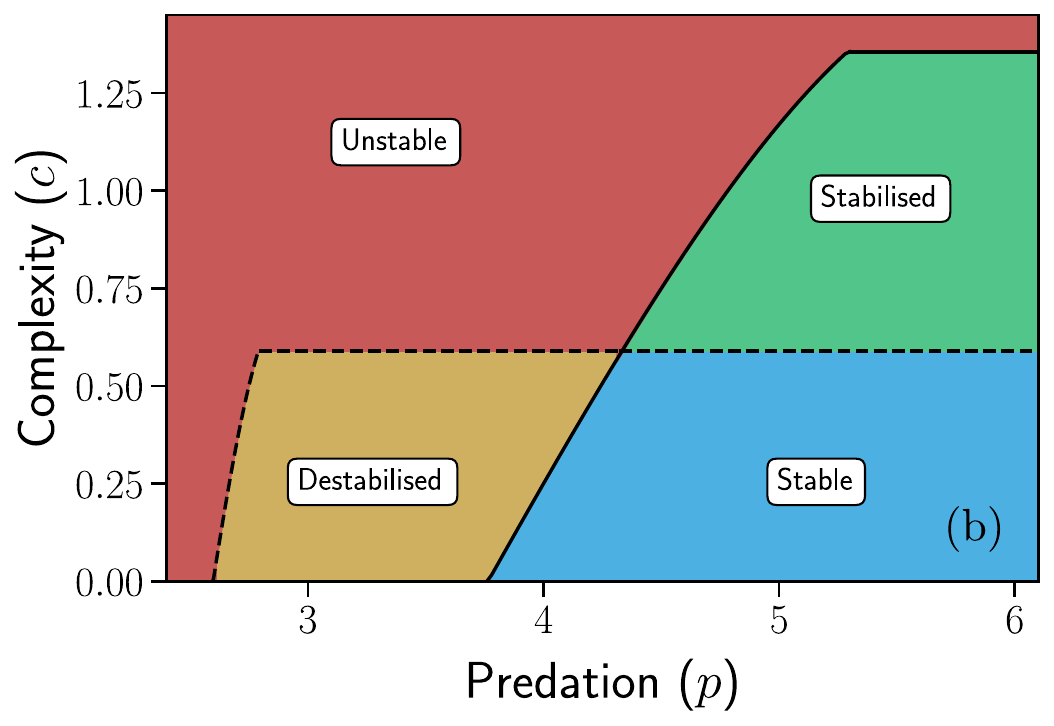}
	\caption{ If the interactions are spatially heterogeneous (species do not interact in the same way in one place as they do in another), stabilisation and destabilisation are both possible when dispersal is introduced. In a similar way to Fig. 3 of the main text, panel (a) shows sets of system parameters for which the model ecosystem is stable or unstable without taking dispersal into account. Panel (b) shows which sets of parameters are stable/unstable once dispersal is introduced. We have chosen a set of system parameters which accentuates both the stabilising and the destabilising mechanisms. The amount of spatial heterogeneity here is maximal ($\rho = 0$) and the ratio of the predator to the prey dispersal coefficients is quite large ($D_v/D_u = 10$). The remaining system parameters here are $n = 11$ (number of patches), $d_u = d_v = 1$, $\gamma_u = 2/3$, $\gamma_v = 1/3$, $\Gamma_u = 0.5$, $\Gamma_v = 0.6$,  $\Gamma_{uv} = 0$, $D_u = 0.5$ and $D_v = 5$ (in panel (b)), $NC\mu_{uu} = 3$, $NC\mu_{vv} = -1.5$, and like Fig. 3 in the main text $NC\mu_{vu} =-NC\mu_{uv}$. }
	\label{fig:combinedtrophicandheterogeneity}
\end{figure}

We have verified that the heterogeneity-associated stabilising mechanism can survive with the inclusion of trophic structure and that the Turing instability can survive with the inclusion of spatial heterogeneity in Fig. \ref{fig:gamma0trophic}. Now, we investigate in a more general fashion under what conditions dispersal is a stabilising or a destabilising influence. This can be achieved by constructing a stability diagram similar to Fig. 3 in the main text (see Fig. \ref{fig:combinedtrophicandheterogeneity}) for the case where spatial heterogeneity is included. In general, the lines which divide the stable region of parameter space from the unstable region can be found by identifying sets of parameters where (a) the bulk region crosses the imaginary axis or (b) an outlier crosses the imaginary axis. 

To find the horizontal lines in Fig. \ref{fig:combinedtrophicandheterogeneity}, one solves Eqs.~(\ref{unhatted2}) for the value of $\sigma^2$ where the boundary of the bulk eigenvalue support first touches the imaginary axis. Due to the symmetry of the support, this must occur at $\omega = 0$. That is, one solves the following equations for $\sigma^2$ [using the hopping and correlation kernels in Eqs.~(\ref{alltoallkernels})]

\begin{align}
\det\left[ V_{q,q'} \right] &= 0, \,\,\,\,V_{q,q'} = \delta_{q,q'} + \frac{\sigma^2}{n} c_{q-q'} \sum_\alpha \gamma_\alpha w_{\alpha;q}^{\star}w_{\alpha;q} , \nonumber \\
0 &= -1+ \left[  i( d_\alpha + D_\alpha (1- \phi_{q}))+  \frac{\sigma^2}{n} \sum_{\beta,q'} \Gamma_{\alpha \beta} \gamma_\beta c_{q-q'}w^\star_{\beta;q'}\right] w_{\alpha;q}^{\star} , \nonumber \\
0 &=  -1 +\left[  i( d_\alpha + D_\alpha (1- \phi_{q}))+  \frac{\sigma^2}{n} \sum_\beta \Gamma_{\alpha \beta} \gamma_\beta c_{q-q'}w_{\beta;q'}\right]w_{\alpha;q} . \label{criticalsigma}
\end{align}

To find the sloped lines in Fig. \ref{fig:combinedtrophicandheterogeneity}, one instead solves Eqs.~(\ref{productq}) to (\ref{simult22}) for the value of $\sigma^2$, as a function of $\mu_{vu}$, where an outlier first touches the imaginary axis. That is, one sets $\mathrm{Re}[\lambda] = 0$ in Eq.~(\ref{simult22}). We leave $\mathrm{Im}[\lambda]$ as a free parameter but we note that for the Turing instability the imaginary part of the most unstable eigenvalue turns out to be zero in all cases that we tried. We note that after rescaling we have $\sigma^2 \to NC \sigma^2 = c$ and $\mu_{vu} \to NC \mu_{vu} = p$ (see Section~\ref{section:scaling}).

Fig. \ref{fig:combinedtrophicandheterogeneity} highlights which regions of parameter space are subject to the stabilising and the destabilising effects when dispersal is `switched on'. As with Fig. 3 of the main text, there remains a lower bound on the amount of predation required for stability -- this lower bound is increased by the introduction of dispersal in a similar way to Fig. 3, hence destabilising certain parameters sets. There also remains an upper bound on the amount of complexity required for stability. Now, because of the inclusion of spatial heterogeneity, this upper bound is increased with dispersal and certain unstable parameter sets begin to exhibit stability. So when both trophic structure and spatial heterogeneity are included in the model, both the destabilising mechanism discussed in the main text and the stabilising mechanism discussed above play a role in determining how dispersal affects stability. Importantly, neither effect necessarily precludes nor diminishes the other.

\begin{figure}
	\centering
	\includegraphics[scale = 0.4]{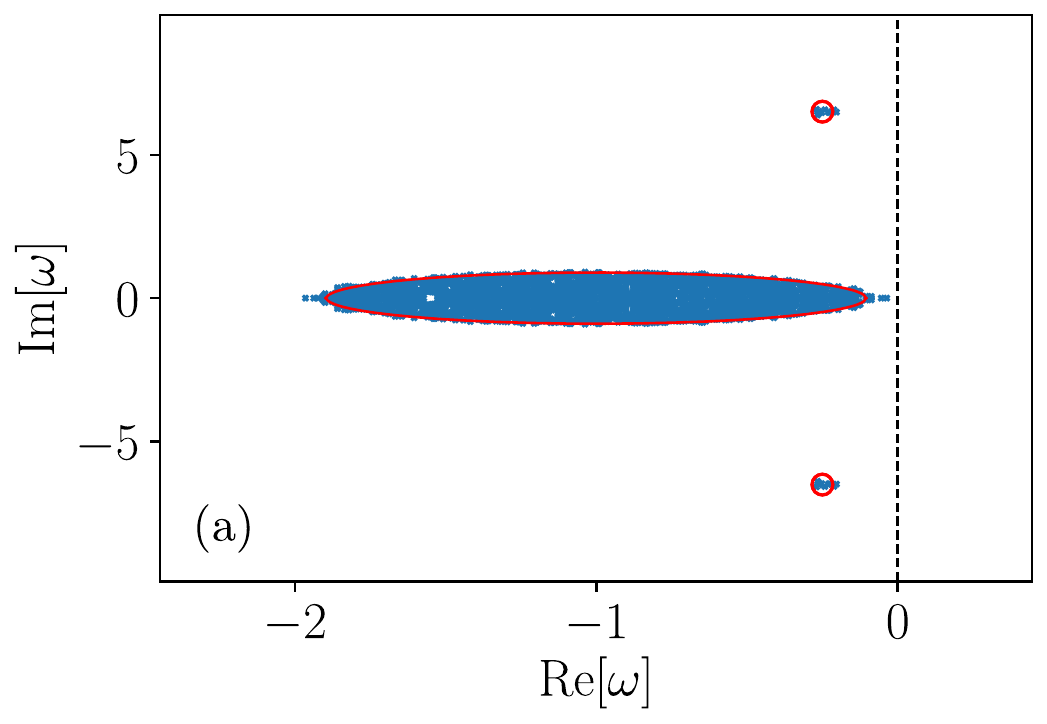} 
	\includegraphics[scale = 0.4]{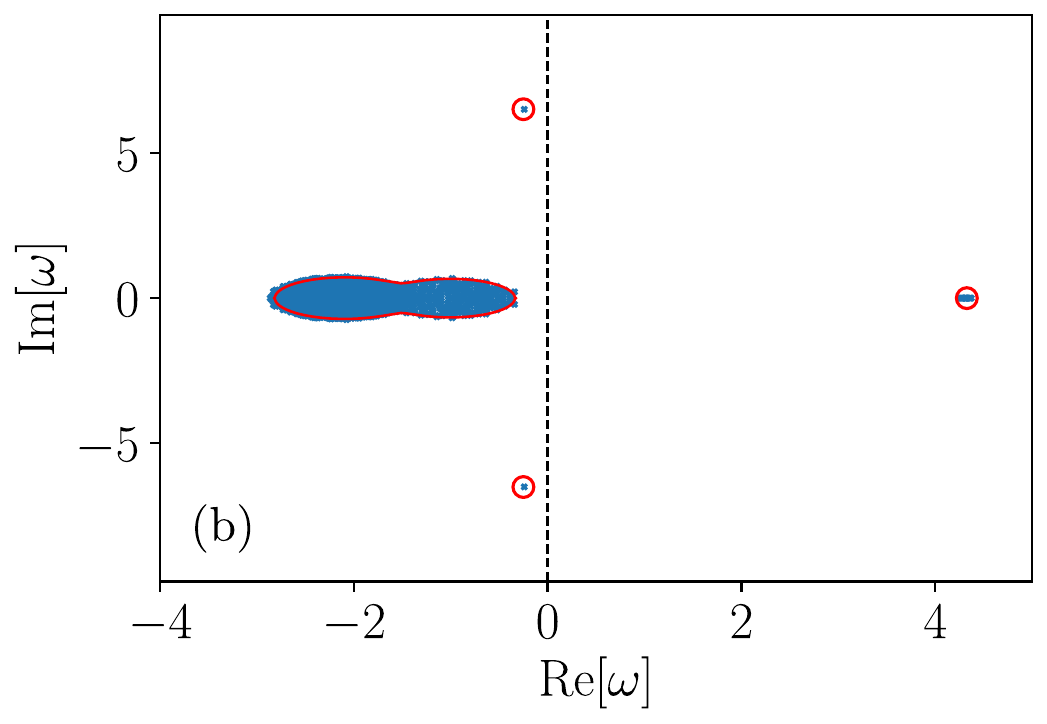}
	\includegraphics[scale = 0.4]{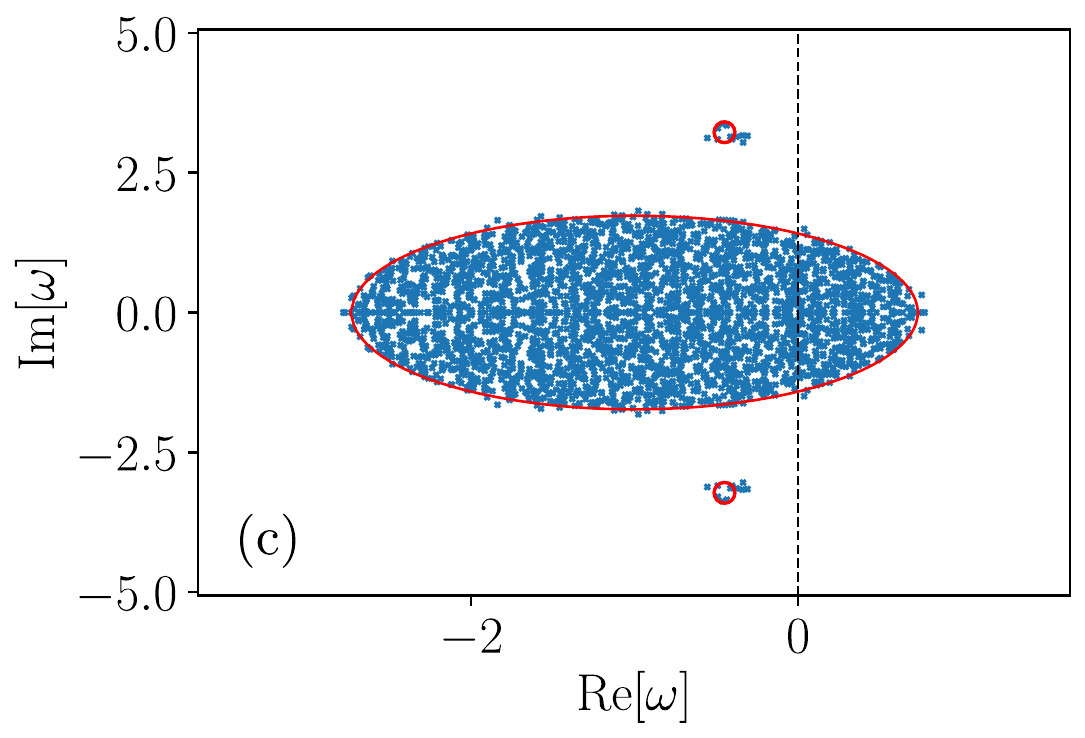} 
	\includegraphics[scale = 0.4]{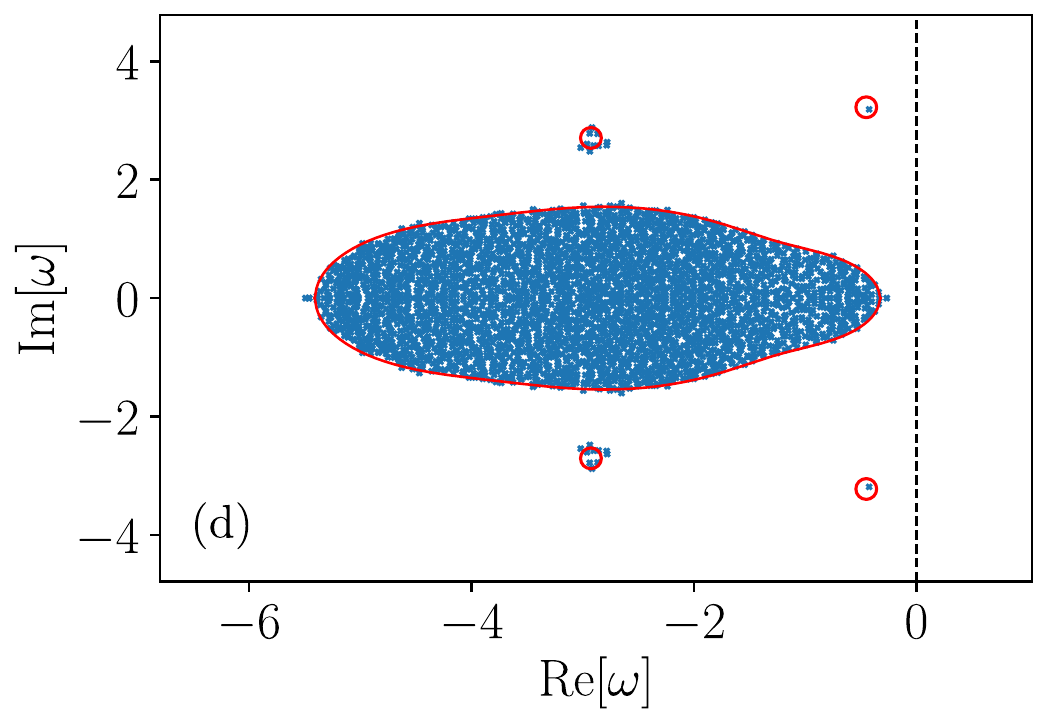}
	\caption{{\bf Exaggerated stabilising and destabilising mechanisms.}  In comparison to Fig. \ref{fig:gamma0trophic} (a) and (b), panels (a) and (b) here demonstrate that an increased ratio $D_v/D_u$, increased trophic structure (reflected in $\mu_{\alpha\beta}$) and moderate complexity gives rise to a strong destabilising effect for dispersal. In comparison to Fig. \ref{fig:gamma0trophic} (c) and (d), panels (c) and (d) here demonstrate that a larger degree of spatial heterogeneity coupled with a sufficiently large typical dispersal rate $\bar D$ and large enough predation gives rise to a strong stabilising effect. For the model parameters here (after rescaling, see Sec.~\ref{section:scaling}), we have (a)  $D_u = 0$, $D_v = 0$, $c = NC\sigma^2 = 0.8$, $\rho = 0.5$, $NC\mu_{uu} = 15$, $NC\mu_{uv} = -24$, $NC\mu_{vu} = 24$, $NC\mu_{vv} = -25.5$; (b) the same but with $D_u = 1$, $D_v = 20$; (c) $D_u = 0$, $D_u = 0$, $c = NC\sigma^2 = 3$, $\rho = 0$, $NC\mu_{uu} = 3$, $NC\mu_{uv} = -7.5$, $NC\mu_{vu} = 7.5$, $NC\mu_{vv} =-2.4$ (d) the same but with $D_u = 1.5$, $D_v = 3$. The remaining model parameters are (in all cases) $n = 11$, $\Gamma_{uv} = 0$, $\Gamma_{uu} = 0$, $\Gamma_{vv} = 0$, $\gamma_u = 2/3$, $\gamma_v = 1/3$, $C= 1$ and $N = 300$.  }\label{fig:exaggerated}
\end{figure}
\subsection{What determines if dispersal is stabilising or destabilising?}
We now explore the roles that the different components of the model play in determining whether one observes the stabilising or destabilising effects of dispersal. Because of the number of model parameters, it is difficult to be exhaustive. We therefore highlight the influence of the most salient components and focus on general tendencies.

The principal elements determining whether dispersal acts to stabilise or destabilise are: (i) the presence of a pronounced trophic structure; (ii) the ratio of the diffusion coefficients for predator and prey species; (iii) the overall magnitude of the dispersal; and (iv) spatial heterogeneity of the inter-species interactions. These different elements do not always act independently, and their interplay depends on other parameters, for example the complexity $c$. However, broadly speaking, elements (i) and (ii) are relevant mostly for the the destabilising mechanism presented in the main text, whereas (iii) and (iv) relate more to the stabilising mechanism.

Elements (i) -- (iv) listed above each correspond to a subset of the model parameters. For example, several parameters contribute to the degree of trophic structure in the population [item (i) above]. Among these are the relative sizes of the prey and predator blocks ($\gamma_u,\gamma_v$), the correlation parameters $\Gamma_{\alpha\beta}$, and the mean coefficients in each of the four blocks of the interaction matrix (i.e., the parameters $\mu_{\alpha\beta}$).

Elements (ii) and (iii) describe properties of the dispersal in the model. To discuss these quantitatively, it is useful to define $\theta = D_v/D_u$ as the predator-to-prey ratio of diffusion coefficients and $\bar D = \sqrt{D_u D_v}$ as the magnitude of dispersal. We then have $D_u =\bar D/\sqrt{\theta}$ and $D_v = \sqrt{\theta} \bar D$. 

Finally, correlation (or, conversely, heterogeneity) of the interactions in space finally [item (iv)] is indicated by the model parameter $\rho$, with $\rho=0$ representing the absence of any correlations (full heterogeneity), and $\rho=1$ a situation in which there is no heterogeneity at all.

\subsubsection*{What promotes the dispersal-induced instability?}

Similar to the simpler two-component systems in Turing's original work, an important characteristic for dispersal-induced instability in complex ecosystems is the ratio of the dispersal coefficients for predator and prey species. As shown in Fig. 4 of the main text, greater values of the predator-to-prey ratio of diffusion coefficients ($\theta$) promote dispersal-induced instability. This can also be seen from Eq.~(\ref{criticalratio}). Provided that there is dispersal, its magnitude $\bar D$ is not immediately important for the instability in the mathematical model. The quantity $\bar D$ only appears as a factor multiplying $q$, so its main role is to control the wavelength associated with the instability.

As a second main component, trophic structure is required to see dispersal-induced instability (see `Eigenvalue spectra' in Results in the main text). The structure of the community is determined by multiple parameters as discussed above. The role of each of these specific parameters is elucidated by the special cases in Section \ref{section:stability}, but we here focus on predation $p$. In order for genuine dispersal-induced instability to occur the equilibrium must be stable when there is no dispersal. Fig. \ref{fig:combinedtrophicandheterogeneity} (a) shows that this is only the case if $p$ exceeds a threshold. Predation is also a stabilising influence in the system with dispersal, and dispersal-induced instability is not seen when predation is too high. As a result predation must be within a limited range in order for dispersal to destabilise equilibria which would otherwise be stable [see Fig.~\ref{fig:combinedtrophicandheterogeneity} (b)]. A competition between the ratio of diffusion coefficients ($\theta$) and predation ($p$) then emerges. High values of $\theta$ promote dispersal-induced instability and increase the range of $p$ for which such instability is possible.

The statements in the previous two paragraphs apply -- in principle -- to Turing's original two-species system \cite{turing}. The parameter which best encapsulates how our model differs to a simple two-species system is the complexity $c = NC\sigma^2$. Complexity generally acts to destabilise equilibria in ecosystems with or without dispersal (as per May's original findings \cite{may}). Too high a complexity can therefore make the ecosystem unstable even if there is no dispersal [see Fig.~\ref{fig:combinedtrophicandheterogeneity}(a)]. Therefore, complexity must be sufficiently low in order for dispersal to be the factor which destabilises the equilibrium  [see Fig.~\ref{fig:combinedtrophicandheterogeneity} (b)].  After all, if the complexity is so large that the system without dispersal is unstable, then, of course, dispersal only has the potential to act as a stabilising influence. However, with this caveat in mind, an increase in complexity can also reduce the ratio of diffusion coefficients $\theta$ required for the Turing instability (Fig. 4 in the main text). Complexity then increases the propensity for dispersal to be destabilising. 
	
In Fig. \ref{fig:exaggerated} (a) and (b) we exaggerate the destabilising effect [compared to Fig. \ref{fig:gamma0trophic} (a) and (b)] by increasing $D_{v}/D_u$, increasing the relative sizes of the trophic parameters $\mu_{\alpha\beta}$ and choosing a moderate complexity. We note that neither the presence nor the complete absence of spatial heterogeneity are necessary for dispersal-induced instability. The instability shown in Fig.~2 of the main paper for example is for a situation with identical interactions across space (no heterogeneity). Figs.~\ref{fig:gamma0trophic}, \ref{fig:combinedtrophicandheterogeneity} and \ref{fig:exaggerated} demonstrate that dispersal-induced instability can persist in the face of heterogeneity.
\vspace{-1em}
\subsubsection*{What promotes the dispersal as a stabilising influence?}
May's original observation \cite{may} was that for sufficiently high complexity $c$, the eigenvalues of a random Jacobian strayed into the real half-plane, signifying instability. The primary driving factor behind the stabilising mechanism first presented in Ref. \cite{gravel} is spatial heterogeneity of the interactions. The main mathematical effect of the combination of heterogeneity and dispersal is to compress a part of the bulk region of the eigenvalue spectrum, bringing the eigenvalues into the negative half-plane (see Fig. \ref{fig:gravelmatrices}). Greater spatial heterogeneity gives rise to a greater amount of compression. This means that an unstable equilibrium can be stabilised by an increase of spatial heterogeneity.

In order to see this stabilising effect, spatial heterogeneity must be combined with a sufficiently large typical dispersal $\bar D$ (see Fig. \ref{fig:effectofdbar}). This is so that the part of the eigenvalue spectrum that is not subject to the shrinking effects of heterogeneity is moved sufficiently leftward. In a sense, dispersal in this context aids in allowing heterogeneity to be stabilising. We also note that the ratio of dispersal coefficients $\theta$ must not be too large so that both predator and prey species are subject to the stabilisation. 

In order for an equilibrium to be stable, both the bulk of the eigenvalue spectrum and any outlier eigenvalues must be entirely in the negative half-plane. The stabilising effect of dispersal with heterogeneity acts primarily on the bulk eigenvalues. However, it does not apply in the same way to any outliers, which arise from the inclusion of trophic structure. It is these outliers which are responsible for the dispersal-induced instability. This means that dispersal can only stabilise an unstable equilibrium if no Turing instability is introduced by the dispersal. 

In Fig. \ref{fig:exaggerated} (c) and (d), we demonstrate that the combination of increased spatial heterogeneity, a sufficiently large $\bar D$ and a large enough predation $p$ (so as to preclude the excursion of positive eigenvalues into the positive half-plane, see Fig. \ref{fig:combinedtrophicandheterogeneity}) acts to exaggerate the stabilising effect in comparison to Fig. \ref{fig:gamma0trophic} (c) and (d).